\documentclass[11pt]{article}
\usepackage[margin=1.15in]{geometry}
\usepackage{graphicx}
\usepackage{amsmath,amssymb,amsthm,color}
\usepackage{booktabs}
\usepackage{tikz}
\usepackage{authblk}
\usepackage[numbers]{natbib}
\usepackage[colorlinks=true,linkcolor=blue,citecolor=blue,urlcolor=blue]{hyperref}

\newcommand{\tr}{\operatorname{tr}}
\newcommand{\sgn}{\operatorname{sgn}}
\newcommand{\Part}{\operatorname{Part}}
\newcommand{\no}[1]{{:}\,#1\,{:}}   
\newcommand{\He}{\operatorname{He}}
\newcommand{\beq}{\begin{equation}}
\newcommand{\eeq}{\end{equation}}

\title{Single-Instance Observables in the Sachdev-Ye-Kitaev Model}
\author{Brian Swingle \\ \small \href{mailto:bswingle@brandeis.edu}{bswingle@brandeis.edu}}
\affil{Department of Physics, Brandeis University, Waltham, Massachusetts, 02453}
\date{September 14, 2026\\[0.5em]}

\begin{document}

\maketitle

\begin{abstract}
    We present a diagrammatic expansion for thermal expectation values of low-weight Majorana strings in a fixed realization of the $N$-fermion Sachdev--Ye--Kitaev (SYK) model. At each order in the expansion, the observable is expressed as a finite sum of polynomials in the couplings multiplied by temperature-dependent kernels built from the large-$N$ melonic propagator. At fixed $\beta J$, the expansion is organized in powers of $N^{-1/2}$, and each order can be efficiently evaluated with classical resources polynomial in $N$. The expansion also extends to correlations at arbitrary imaginary times and sufficiently short real times. For the four-body SYK model, we show that the diagrammatic predictions give an excellent account of exact diagonalization results at sizes up to $N=24$. We also apply the method to the Maldacena--Qi (MQ) model formed from two identical copies of a fixed SYK realization. This enables the identification of a nearby SYK-like Hamiltonian whose thermofield double state has higher fidelity with the Maldacena--Qi ground state than the thermofield double of the original SYK Hamiltonian.
\end{abstract}

\tableofcontents


\section{Introduction}
\label{sec:intro}

The Sachdev--Ye--Kitaev (SYK) model~\cite{SachdevYe1993,KitaevKITP,MaldacenaStanford2016} is a strongly interacting quantum many-body system consisting of $N$ Majorana fermions with random $p$-body interactions that is solvable at large $N$. A single instance of the model is determined by a sample of Gaussian random couplings $J_I$, also known as a disorder realization, where $I$ runs over collections of $p$ fermions. Most analytic results for the model concern disorder-averaged quantities at large $N$, for which the melonic diagram technology~\cite{PolchinskiRosenhaus2016,GrossRosenhaus2017gen,GrossRosenhaus2017,KitaevSuh2018,BonzomNadorTanasa2019,Rosenhaus2019,Chowdhury2022} gives a closed set of Schwinger--Dyson (SD) equations. In this paper we are interested in a different class of observables: thermal expectation values in a \emph{single disorder realization} of the couplings,
\begin{equation}
    \xi_X = \tr\!\big(\mu_X\, \rho(\beta)\big), \qquad \rho(\beta) = \frac{e^{-\beta H}}{Z},
\end{equation}
where $\mu_X$ is a low-weight Hermitian Majorana string built from the fermions in set $X$ (the \emph{weight} of a string is the number of Majorana fermions it contains; precise conventions are given in Section~\ref{sec:model}). The disorder average of $\xi_X$ typically vanishes, but for a fixed realization $\xi_X$ is generically a nonzero number, and one can ask how to compute it.

The answer we develop is a diagrammatic framework that can be viewed as a dressed high-temperature expansion. The structure of the result is
\begin{equation}
    \xi_X \;=\; \sum_{\text{diagrams } D} \big[\text{coupling polynomial}\big]_D(J) \times \big[\text{kernel}\big]_D(\beta) ,
    \label{eq:expansion}
\end{equation}
where each coupling polynomial is an explicit contraction of the couplings $J_I$ of the given sample, and each kernel is an imaginary-time integral of products of the large-$N$ melonic propagator $G(\tau)$, which is the dressed two-point function that solves the large-$N$ Schwinger--Dyson equations. Essentially the method is a Taylor expansion of $e^{-\beta H}$, resummed by dressing every line with the melonic propagator so that it remains sensible at any fixed $\beta J$ as $N \to \infty$. After this melonic dressing, the remaining skeleton diagrams and higher-order coupling polynomials are suppressed by inverse powers of $N$ relative to the leading prediction.

\begin{figure}[t]
\centering
\includegraphics[width=\textwidth]{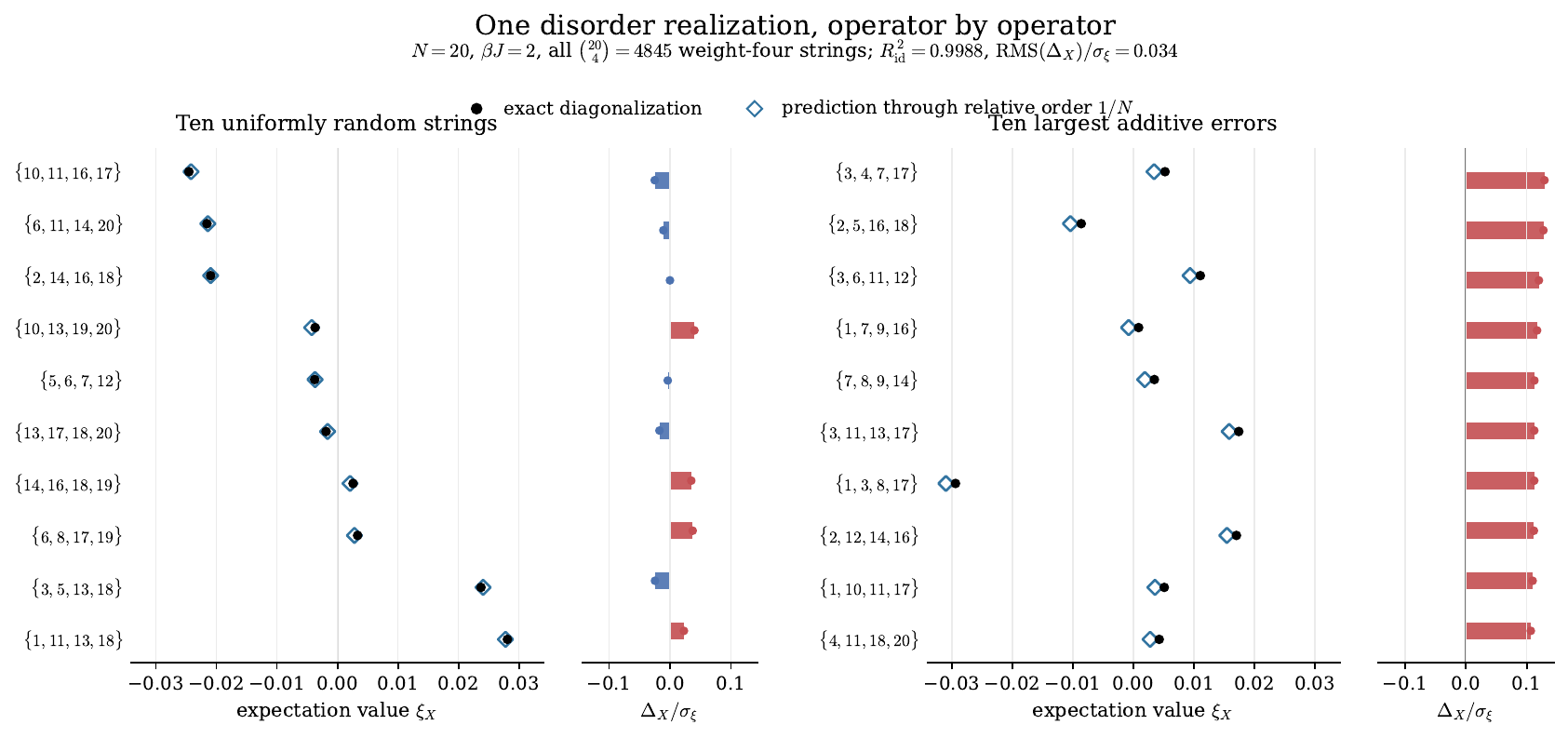}
\caption{An operator-by-operator preview for one fixed disorder realization at $N=20$ and $\beta J=2$. Black circles are exact-diagonalization values and open diamonds are the parameter-free predictions through relative order $1/N$: the leading term, the degree-two and Wick-ordered degree-three fluctuation tensors, and the one-loop correction to the leading coefficient (Section~\ref{sec:corrections}). The left panel shows ten strings selected uniformly with a random seed fixed independently of the coupling values; the right shows the ten largest additive errors among all $4845$ strings. Here ``largest'' means largest $|\Delta_X|$, where $\Delta_X=\xi_X^{\rm ED}-\xi_X^{\rm pred}$; the narrow axes display the signed residual in units of the standard deviation $\sigma_\xi$ over all exact values. Across the full realization, the identity-line coefficient of determination is $R^2_{\rm id}=0.9988$ and ${\rm RMS}(\Delta_X)/\sigma_\xi=0.034$. No coefficient is fitted to the displayed data.}
\label{fig:w4intro}
\end{figure}

At any fixed truncation order and fixed $\beta J$, the resulting expression can be evaluated classically in polynomial time. The truncated expansion is found to be in excellent agreement with exact results at small $N$, with Figure~\ref{fig:w4intro} previewing the output for one preselected disorder realization at $N=20$ and $\beta J=2$. We evaluate all $\binom{20}{4}=4845$ weight-$4$ expectation values and compare exact diagonalization with the parameter-free prediction through relative order $1/N$, assembled in Section~\ref{sec:corrections}. The prediction tracks both a uniformly random sample of strings and a stress-test sample consisting of the ten strings with the largest additive errors.\footnote{The ten largest outliers happen to share a sign in the realization displayed; the full residual distribution is nearly sign-balanced, and other disorder seeds have both signs among their largest outliers.}

In spite of this clear success, there are two qualifications worth emphasizing. First, the convergence is \emph{not uniform} in $\beta J$: higher-order diagrams carry more time integrals and hence typically more powers of $\beta J$, so at fixed $N$ they matter more as $\beta J$ grows, and a truncation at fixed order degrades at low temperature. Indeed, if $N$ is fixed and $\beta J$ is taken to infinity, one moves well outside the regime of validity of the standard melonic resummation. Second, many questions about the expansion, such as whether it is convergent or asymptotic, remain open. These issues are subtle because the large-$N$ power counting is inherently statistical, so when comparing the expansion at a given order to the exact $\xi_X$, the most natural target is a bound on the mean-square-error. We return to these questions in Section~\ref{sec:outlook}.

This approximation scheme is interesting in part because the SYK thermal state at $\beta J \gtrsim 1$ is highly entangled and highly magical (far from stabilizer states), as quantified for instance by the stabilizer R\'enyi entropy~\cite{LeoneOlivieroHamma2022,HaugPiroli2023}, which has been studied in SYK and its variants in~\cite{BeraSchiro2025,JasserOdavicHamma2025,RussomannoPassarelliRossiniLucignano2025,SantraWindeyBandyopadhyayLegramandiHauke2025,zhang2026stabilizerrenyientropytransition,SunZhang2026,MalvimatSarkisSukYoon2026,GarciaGarciaLiuZheng2026,BettaqueSwingle2026}. So while averaged properties are known to be efficiently accessible classically at sufficiently large $N$ at fixed $\beta J$, one might nevertheless suspect that quantum simulation would be required to efficiently obtain single-instance properties of the model. However, the formulas derived here show that an interesting class of observables --- low-weight string expectation values, and more generally low-weight correlators at arbitrary imaginary-time separations and sufficiently short real-time separations --- admit polynomial-time classical approximations at any fixed truncation order: one solves the large-$N$ SD equations once, and then evaluates explicit low-degree polynomials in the couplings of the instance at hand.

\paragraph{Background.} The statistical moments of $\xi_X$ were studied in~\cite{Berkooz2021,BettaqueSwingle2026}, although formulas for a single instance were left implicit. Part of our motivation to develop this approach arose from a desire to see the statistical results about thermal expectation values in \cite{BettaqueSwingle2026} in a different, diagrammatic, light. Another key motivation is recent work giving a rigorous quasi-polynomial-time classical algorithm for thermal expectation values in SYK at sufficiently high constant temperature~\cite{zlokapa2026sykthermalexpectationsclassically,Zlokapa2026}. In contrast, the framework discussed here is not rigorously established, but we demonstrate the effectiveness of the method by extensive comparisons with exact diagonalization (ED). Moreover, the classical computational cost of our approach grows only weakly with $\beta J$ when $N$ is large and we also access real-time observables at sufficiently short times.

The framework has a number of precursors. In classical mean-field spin glasses, the TAP equations and the Plefka and Georges--Yedidia expansions combine the couplings of a particular sample with systematic corrections to naive mean field~\cite{ThoulessAndersonPalmer1977,Plefka1982,GeorgesYedidia1991}. Adaptive TAP and later cavity/TAP iterations make this fixed-instance and algorithmic character explicit~\cite{OpperWinther2001,Bolthausen2014,ChenTang2021}. Related approaches for quantum Hamiltonians include quantum TAP equations~\cite{Biroli_2001}, quantum belief propagation~\cite{Hastings2007QBP}, and quantum cavity methods~\cite{LaumannScardicchioSondhi2008}, as well as rigorous high-temperature algorithms based on locality, bounded degree, or cluster expansions~\cite{KuwaharaKatoBrandao2020,KuwaharaAlhambraAnshu2021,HarrowMehrabanSoleimanifar2020,MannHelmuth2021,YinLucas2023}. Those latter methods have a rather different regime of control from the dense large-$N$ organization used here.

In SYK itself, collective-field descriptions at fixed couplings were developed in the half-wormhole and factorization literature~\cite{SaadShenkerStanfordYao2024,GotoSuzukiUgajin2022,Mukhametzhanov2022HalfWormholes,Mukhametzhanov2023Factorization}; related extensions include~\cite{PengTianYang2023}. These works are most directly connected to the collective-field discussion in Appendix~\ref{sec:rules}, but primarily concern partition functions, factorization, and spectral observables rather than thermal one-point functions.

Complementary systematic diagrammatics exist for the disorder-averaged theory. The bilocal collective action admits perturbative and semiclassical $1/N$ expansions~\cite{JevickiSuzuki2016,BenedettiGurau2018,ArefevaKhramtsovTikhanovskaya2018}, while combinatorial methods classify the leading and next-to-leading graphs in colored SYK and related tensor models~\cite{BonzomLionniTanasa2017}. These constructions underlie parts of our collective-kernel expansion, but do not retain the explicit coupling-polynomial dependence of a fixed realization. At a broader level, the use of dressed skeletons and bilocal effective actions descends from the Luttinger--Ward and two-particle-irreducible traditions~\cite{LuttingerWard1960,CornwallJackiwTomboulis1974}. In the double-scaled limit $p\sim\sqrt N$, disorder-averaged moments and correlators admit an exact chord-diagram organization~\cite{ErdosSchroder2014,BerkoozNarayanSimon2018,BerkoozIsachenkovNarovlanskyTorrents2019,Lin2022}.

Other discussions of realization-dependent structure in and around SYK include multi-trace correlators and non-geometric wormholes~\cite{Berkooz2021}, long-wavelength spectral fluctuations~\cite{JiaVerbaarschot2020}, fluctuations in sparse SYK~\cite{XuSusskindSuSwingle2020}, and the distinction between quenched and annealed descriptions~\cite{Baldwin_2020}. There has also been progress in the rigorous analysis of the SYK free energy~\cite{gamarnik2026freeenergylimitsyk}. The broader separation between universal statistics and reproducible sample-specific fingerprints is also familiar from mesoscopic conductance fluctuations~\cite{Stone1985,LeeStone1985}. To the best of our knowledge, however, these works do not give a systematic fixed-instance expansion for low-weight thermal observables in terms of explicit coupling polynomials and melonic kernels, together with extensive finite-$N$ numerical tests.

\paragraph{Paper organization.} After defining the model and setting conventions in Section \ref{sec:model}, there are three main components.
\begin{itemize}
    \item First, in Section~\ref{sec:leading} we discuss the diagrammatic expansion including the melonic resummation, and test the leading diagrams against ED at modest $N \leq 24$.
    \item Second, in Section~\ref{sec:corrections} we discuss corrections to the leading diagrams and present a detailed study of weight four showing how the ED results can be well captured by subleading diagrams. Along the way we exhibit an exact identity that turns the aggregate of all weight-four one-point functions into the thermal energy; because it needs only the spectrum, it gives a sharp test of the corrected coefficients with no operator-sampling noise.
    \item Third, in Section~\ref{sec:time-dep} we extend the rules to correlations at separated imaginary and real times, and in Section~\ref{sec:mq} we apply the method to the Maldacena--Qi (MQ) model~\cite{MaldacenaQi2018}, showing how the nature of the MQ ground state can be understood as a thermofield double state of a ``nearby'' SYK-like Hamiltonian.
\end{itemize}
Following this, we present an outlook in Section \ref{sec:outlook} that sketches some open questions. The appendices, in order, contain collected conventions, including the orientation signs carried by the coupling polynomials (Appendix~\ref{app:conventions}); a more systematic form of the diagram method which helps to efficiently organize higher-order corrections (Appendix \ref{sec:rules}); the derivation of the one-loop coefficient correction used in Section~\ref{sec:corrections} (Appendix~\ref{app:w4-oneloop}); additional information about weights $4$ (Appendix \ref{app:w4-extra}) and $8$ (Appendix \ref{app:w8}); and details of the numerical methods (Appendix~\ref{app:numerics}).

\section{Model}
\label{sec:model}

Consider $N$ Majorana fermions with the normalization convention
\begin{equation}
    \{ \chi_i , \chi_j \} = \delta_{ij}, \qquad \chi_i^2 = \tfrac{1}{2},
    \label{eq:chialg}
\end{equation}
and the $p$-body SYK Hamiltonian ($p$ even),
\begin{equation}
    H = i^{p/2} \sum_{i_1 < \cdots < i_p} J_{i_1 \cdots i_p}\, \chi_{i_1} \cdots \chi_{i_p},
    \qquad
    \operatorname{var}(J_I) = \frac{J^2}{\binom{N-1}{p-1}} .
    \label{eq:H}
\end{equation}
We focus on $p=4$ and comment on other cases. For $p=4$ the Hamiltonian reads
\begin{equation}
    H = -\sum_{i<j<k<l} J_{ijkl}\, \chi_i \chi_j \chi_k \chi_l,
    \qquad
    \operatorname{var}(J_{ijkl}) = \frac{J^2}{\binom{N-1}{3}} = \frac{6 J^2}{(N-1)(N-2)(N-3)} .
    \label{eq:H4}
\end{equation}
Throughout, the couplings $J_I$ are regarded as \emph{fixed known numbers}, and we only disorder average if explicitly stated. The variance \eqref{eq:H}, which differs from the more standard choice of $(p-1)! J^2 / N^{p-1}$, is chosen so that the disorder-averaged self-energy melon has a particularly simple normalization at finite $N$ as discussed in Section~\ref{sec:leading-melonic}.

It is convenient to introduce a basis of Hermitian Majorana strings. For a subset $X = \{x_1 < \cdots < x_W\} \subseteq \{1,\ldots,N\}$ of size $W = |X|$ (the \emph{weight}), define
\begin{equation}
    \mu_X = i^{W(W-1)/2}\, 2^{W/2}\, \chi_{x_1} \cdots \chi_{x_W} .
    \label{eq:mu}
\end{equation}
These operators are Hermitian, square to one, and form an orthonormal basis with respect to the normalized trace,
\begin{equation}
    \mu_X^\dagger = \mu_X, \qquad \mu_X^2 = 1, \qquad 2^{-N/2} \tr\!\big(\mu_X \mu_Y\big) = \delta_{XY},
\end{equation}
for even $N$ in the irreducible Hilbert space of dimension $2^{N/2}$. In terms of the strings, the $p=4$ Hamiltonian is
\begin{equation}
    H = \sum_{A} \frac{J_A}{4}\, \mu_A,
    \label{eq:Hmu}
\end{equation}
with $A$ running over sorted quartets, since $\mu_A = -4 \chi_{a_1}\chi_{a_2}\chi_{a_3}\chi_{a_4}$.

With the normalization \eqref{eq:chialg}, the free Euclidean Green function is
\begin{equation}
    G_0(\tau) = \big\langle T_\tau\, \chi_i(\tau) \chi_i(0) \big\rangle_0 = \frac{\sgn(\tau)}{2},
\end{equation}
and the large-$N$ melonic two-point function $G(\tau)$ solves the standard SD equations
\begin{equation}
    \Sigma(\tau) = J^2 G(\tau)^{p-1},
    \qquad
    G(i\omega_n)^{-1} = -i\omega_n - \Sigma(i\omega_n),
    \label{eq:SD}
\end{equation}
with fermionic Matsubara frequencies $\omega_n = 2\pi (n + \tfrac12)/\beta$ and antiperiodic continuation $G(\tau + \beta) = -G(\tau)$. Solving \eqref{eq:SD} numerically on a fine imaginary-time grid is inexpensive (Appendix~\ref{app:numerics}); it is the only self-consistent many-body input the method requires.

The observable of interest is the single-instance thermal one-point function
\begin{equation}
    \xi_X = \tr\!\big( \mu_X\, \rho(\beta) \big),
    \qquad
    \rho(\beta) = \frac{e^{-\beta H}}{Z},
    \qquad
    Z = \tr\, e^{-\beta H} ,
\end{equation}
together with its generalization in which the Majoranas in the string are placed at separated imaginary times,
\begin{equation}
    \xi_X(\tau_1, \ldots, \tau_W) = \tr\!\Big( \rho(\beta)\; T_\tau\big[\, i^{W(W-1)/2}\, 2^{W/2}\, \chi_{x_1}(\tau_1) \cdots \chi_{x_W}(\tau_W)\big] \Big),
    \quad
    \chi_i(\tau) = e^{\tau H}\, \chi_i\, e^{-\tau H},
    \label{eq:bilocaldef}
\end{equation}
with analytic continuation and the corresponding real-time observables introduced separately in Section~\ref{sec:time-dep}.

The scaling of temperature with $N$ distinguishes several physically different regimes. For $\beta J \lesssim 1$, the model is in its ultraviolet or high-temperature regime, while $1 \ll \beta J \ll N$ is the large-$N$ conformal regime. The Schwarzian dynamics that emerges in this limit also describes a 2d model of quantum gravity known as Jackiw--Teitelboim (JT) gravity~\cite{almheiri2015modelsads2backreactionholography,KitaevKITP,MaldacenaStanford2016,KitaevSuh2018}. We focus on $\beta J \ll N$ in this paper. Within this regime, it is straightforward to obtain the SD solution even at large $\beta J$ (for example, by using the fast Fourier transform), making it feasible to go to large $\beta J$ at any fixed order in the expansion.

It is an interesting open question to extend our results to larger imaginary time at fixed $N$. At $\beta J \sim N$, fluctuations of the reparametrization mode become important~\cite{MaldacenaStanford2016,KitaevSuh2018}. A further distinction separates inverse temperatures that grow only polynomially with $N$ from those of order $e^{aN}$. At polynomially large $\beta J$, the thermal window still contains exponentially many many-body levels, whereas at exponentially large inverse temperature spectral discreteness and eventually individual low-energy eigenstates become important~\cite{Gur_Ari_2018,Cotler2017}.

\subsection{Symmetries and selection rules}
\label{sec:symmetries}

There are symmetries that constrain the $\xi_X$ observables, including those that apply for a single instance and those that apply only to the ensemble as a whole.

\paragraph{Fermion parity.} Since $p$ is even, $H$ commutes with the fermion parity
\begin{equation}
    P = i^{N/2}\, 2^{N/2}\, \chi_1 \chi_2 \cdots \chi_N , \qquad P^2 = 1,
\end{equation}
(for even $N$, up to an overall sign convention), and
\begin{equation}
    P \mu_X P^{-1} = (-1)^{W} \mu_X .
\end{equation}
Since $\rho(\beta)$ commutes with $P$, every odd-weight one-point function vanishes exactly at finite $N$:
\begin{equation}
    W \text{ odd} \quad \Longrightarrow \quad \xi_X = 0 .
\end{equation}

\paragraph{Antiunitary symmetry for $p = 0 \bmod 4$.} Let $\mathsf{T}$ be the antiunitary operation that complex-conjugates numbers and leaves each $\chi_i$ fixed. The phase $i^{p/2}$ in \eqref{eq:H} is real when $p = 0 \bmod 4$, so
\begin{equation}
    \mathsf{T} H \mathsf{T}^{-1} = H \qquad (p = 0 \bmod 4),
\end{equation}
while the strings transform as
\begin{equation}
    \mathsf{T} \mu_X \mathsf{T}^{-1} = (-1)^{W(W-1)/2}\, \mu_X .
\end{equation}
The sign is $-1$ exactly when $W = 2 \bmod 4$ or $W = 3 \bmod 4$. Since $\mu_X$ is Hermitian, $\xi_X$ is real, while antiunitary invariance maps it to $(-1)^{W(W-1)/2}\xi_X^*$. Combined with the odd-weight parity rule,
\begin{equation}
    p = 0 \bmod 4, \quad W = 2 \bmod 4 \quad \Longrightarrow \quad \xi_X = 0 .
\end{equation}
For $p=4$ this kills $W = 2, 6, 10, \ldots$, so the first weights not excluded by exact symmetries are
\begin{equation}
    W = 0, 4, 8, 12, \ldots
\end{equation}
For $p = 2 \bmod 4$, the same antiunitary sends $H \mapsto -H$ and relates positive to negative temperature instead; it is not a fixed-$\beta$ selection rule.

\paragraph{Ensemble symmetries.} The coupling \emph{distribution} is invariant under index permutations and under sign flips $\chi_i \mapsto s_i \chi_i$, $J_I \mapsto (\prod_{i \in I} s_i) J_I$. These ensemble symmetries constrain disorder-averaged quantities and, importantly for us, the allowed \emph{coupling polynomials} that can appear in single-instance formulas; but they are not symmetries of a generic fixed realization. More generally, the ensemble has a statistical $O(N)$ invariance under rotations $\chi_i \mapsto \sum_j O_{ij} \chi_j$. One can analogously transform the couplings, $J_{i_1 \cdots i_p} \mapsto \sum_{j_1 \cdots j_p} O_{i_1 j_1} \cdots O_{i_p j_p} J_{j_1 \cdots j_p}$, and these transformed couplings are equiprobable in the ensemble.

\section{Leading diagrams}
\label{sec:leading}

We now introduce the diagrammatic computation of $\xi_X$ focusing on the leading diagrams. Corrections to the leading diagrams are discussed in Section \ref{sec:corrections}, and the extension to time-dependent insertions in Section \ref{sec:time-dep}.

\subsection{Taylor expansion of Gibbs weight}
\label{sec:leading-taylor}

The starting point is elementary: the Gibbs weight is an entire function of $H$, so we may expand it in a power series and evaluate the resulting traces term by term. Writing the one-point function as a ratio of traces,
\begin{equation}
    \xi_X = \frac{\tr\!\big(\mu_X\, e^{-\beta H}\big)}{\tr\, e^{-\beta H}},
    \qquad
    e^{-\beta H} = \sum_{r=0}^\infty \frac{(-\beta)^r}{r!}\, H^r ,
    \label{eq:taylor}
\end{equation}
produces at each order a trace of $\mu_X$ (or the identity, in the case of the denominator) against a product of $r$ Hamiltonians. The expansion \eqref{eq:taylor} is a simple high-temperature series which converges at any finite $N$; related large-$N$ expansions of SYK moments and the free energy can be organized through intersection graphs~\cite{JiaVerbaarschot2018}. For a typical disorder realization, however, the number of terms required here grows with $\beta\|H\|=O(\beta JN)$; it is only after the linked-cluster cancellation described below that $\beta J$ becomes the effective expansion parameter. The construction of this paper is a reorganization and partial resummation of \eqref{eq:taylor} to work with dressed propagators; in this subsection we set up its bare form, with no dressing yet.

Each factor of $H$ is a sum over couplings, $H = \tfrac14 \sum_A J_A\, \mu_A$ for $p = 4$ \eqref{eq:Hmu}, so $H^r$ is a sum of coupling monomials $J_{A_1}\cdots J_{A_r}$ weighting products of Majorana strings. These products of strings are combined with the chosen external string $\mu_X$ and traced. This is equivalent to evaluating the expectation in the infinite temperature state,
\begin{equation}
    \langle\,\cdot\,\rangle_0 = 2^{-N/2}\, \tr(\,\cdot\,) ,
\end{equation}
in which the fermions obey Wick's theorem. To keep track of the ordering of coincident operators it is convenient to assign the $r$ vertices distinct imaginary times and time-order them: placing the external string at $\tau = 0$ and the vertices at $0 < \tau_1 < \cdots < \tau_r < \beta$, the ordered integral reproduces the combinatorial factor $\beta^r/r!$, while each Wick contraction of two Majoranas sharing an index is weighted by the free Euclidean propagator
\begin{equation}
    \big\langle T_\tau\, \chi_i(\tau)\, \chi_j(0) \big\rangle_0 = \delta_{ij}\, G_0(\tau) ,
    \qquad
    G_0(\tau) = \tfrac12 \sgn(\tau) .
\end{equation}
Because $G_0$ is odd, distinct time orderings of the vertices contribute differently, and integrating over the simplex turns each term of \eqref{eq:taylor} into a sum of \emph{bare diagrams}.

Equivalently, one may restore the full integration cube together with the $1/r!$ from the exponential and sum over assignments of the $r$ labeled vertices. Our convention below packages those assignments and time orderings into a diagram class: their permutation multiplicity cancels the $1/r!$, while any residual symmetry factor is included in the coupling polynomial and its index sums. This is why the kernels below are written as integrals over $[0,\beta]^r$ with no additional factorial.

A bare diagram is a collection of the following data:
\begin{itemize}
    \item $r$ \emph{vertices}, each a single coupling $J_A$ supplying $p$ legs, with an imaginary time integrated over $[0,\beta]$;
    \item the \emph{external string} $\mu_X$, anchored at $\tau = 0$ and supplying its $W$ legs;
    \item a complete pairing of all legs into \emph{internal lines}, each carrying a propagator $G_0(\tau_a - \tau_b)$ and forcing the two paired indices to be equal; the shared index of every line is summed over $1,\ldots,N$.
\end{itemize}
Its value is
\begin{equation}
    (-1)^r\, 2^{W/2}\; P_X(J)\; \mathcal{K}(\beta) ,
    \qquad
    \mathcal{K}(\beta) = \int_0^\beta \Big(\prod_{\text{vertices}} d\tau\Big) \prod_{\text{lines }(a,b)} G_0(\tau_a - \tau_b) ,
    \label{eq:barevalue}
\end{equation}
where $P_X(J)$ is the coupling polynomial carried by the diagram, including its index sums, residual symmetry factor, and the orientation sign from antisymmetrizing the Majoranas into sorted order (Appendix~\ref{app:conventions}). The overall prefactor is the \emph{coefficient rule}
\begin{equation}
    (-1)^r\, 2^{W/2} :
    \label{eq:coeffrule}
\end{equation}
the $(-1)^r$ comes from expanding $e^{-\beta H}$, and the $2^{W/2}$ from the normalization of $\mu_X$ in \eqref{eq:mu}.

The denominator $Z = \tr\, e^{-\beta H}$ is the same expansion with no external string. Dividing by it enforces the linked-cluster theorem: every vacuum subdiagram disconnected from $\mu_X$ cancels between numerator and denominator, leaving only diagrams in which each vertex is tied through internal lines to the external string. From here on diagram means a connected one.

The leading contribution at a given weight is the diagram with the fewest vertices whose legs can absorb the $W$ external Majoranas. For $W = 4$ in the $p = 4$ model a single vertex already suffices: its four legs are identified with the four external legs, which forces $A = X$, so $P_X = J_X$ with no internal lines left to sum. The corresponding diagram is
\begin{center}
\begin{tikzpicture}[baseline={(v.base)}, line width=0.8pt]
  \coordinate (v) at (0,0);
  \coordinate (a1) at (135:1.3);
  \coordinate (a2) at (45:1.3);
  \coordinate (a3) at (-45:1.3);
  \coordinate (a4) at (-135:1.3);
  \draw (v) -- (a1);
  \draw (v) -- (a2);
  \draw (v) -- (a3);
  \draw (v) -- (a4);
  \fill[black] (v) circle (2.2pt);
  \node[right=2pt] at (v) {$J_{X}$};
  \node[left=2pt] at (v) {$\tau$};
  \node[above left]  at (a1) {$x_1$};
  \node[above right] at (a2) {$x_2$};
  \node[below right] at (a3) {$x_3$};
  \node[below left]  at (a4) {$x_4$};
\end{tikzpicture}
\end{center}
and its kernel is the integral
\begin{equation}
    \int_0^\beta G_0(\tau)^4\, d\tau = \frac{\beta}{16} .
\end{equation}
The coefficient rule at $(W, r) = (4, 1)$ gives $(-1)^1\, 2^2 = -4$, so
\begin{equation}
    \xi_X = -4 \cdot \frac{\beta}{16}\, J_X + O(\beta^2) = -\frac{\beta}{4}\, J_X + \cdots ,
    \label{eq:w4bare}
\end{equation}
in agreement with the direct first-order expansion $\xi_X \sim 2^{-N/2}\tr\big(\mu_X(1 - \beta H)\big) = -\beta J_X/4$. We emphasize that $J_X$ is typically small, of order $J N^{-3/2}$ for $p=4$, so $\xi_X$ is small but nonzero.

Two features of \eqref{eq:w4bare} already show why the bare series cannot be left as is. First, it is a power series in $\beta J$: the order-$r$ term carries $r$ time integrals and hence grows like $\beta^r$, so at fixed $N$ a truncation at fixed order degrades once $\beta J \gtrsim 1$. Second, and more consequentially, not every higher-order diagram is small at fixed $\beta J$ as $N \to \infty$. A pair of vertices joined by $p - 1$ internal lines --- the self-energy melon, see Figure \ref{fig:melon} --- is a coherent sum over the $\sim N^{p-1}$ couplings sharing an index. This sum has the form $\sum_{i_2 \cdots  i_p} J_{i_1 i_2 \cdots i_p}^2$ and we call such sums coherent because the summands are positive, or more generally have the same phase, so that there is no destructive interference between different terms. The practical effect is that the sum is $O(1)$: more precisely, it has an $O(1)$ ensemble mean and, by the central limit theorem, approximately Gaussian relative fluctuations of order $N^{-(p-1)/2}$.\footnote{In contrast, $\sum_{i_2 \cdots  i_p} J_{i_1 i_2 \cdots i_p} J_{i_1' i_2 \cdots i_p}$ is an incoherent sum when $i_1 \neq i_1'$, and its typical size is $\sqrt{N^{p-1}}/N^{p-1} \sim N^{-(p-1)/2}$ arising from random-walk scaling.} Such coherent contributions are not suppressed at large $N$ and must be summed to all orders. Doing so dresses each bare propagator $G_0$ into the melonic propagator $G$, which is the subject of the next subsection.

\subsection{Melonic dressing}
\label{sec:leading-melonic}

The bare series of the previous subsection contains one class of correction that does not become small at large $N$, and we now review how to sum it to all orders. Consider a single propagator line of a bare diagram and dress it with a \emph{self-energy insertion}: a pair of vertices whose remaining $p-1$ legs join together, so that two vertices are joined by $p-1$ parallel internal lines. For $p=4$ this is the sunset melon of Figure~\ref{fig:melon}(a). Because the two vertices share $p-1$ index lines, the coupling monomial is $J_{i\,\cdots}J_{i\,\cdots}$ with $p-1$ indices summed in common: this is a coherent, Wick-paired sum whose disorder mean is $O(1)$. Crucially, while the precise coupling sum for fermion line $i$ depends on $i$, its mean does not, and its relative flavor-to-flavor fluctuations are of order $N^{-(p-1)/2}$. Replacing the coupling sum by its disorder mean gives the leading large-$N$ resummation; the sample-dependent remainder enters at subleading order.

\paragraph{Index counting and the variance.} The averaged melonic coefficient is fixed by an exact index count. Attach a melon to a fixed external leg carrying index $i$: the inserted vertex is any coupling $J_I$ containing $i$, its other $p-1$ legs propagating to the neighbouring vertex, so its disorder average evaluates to
\begin{equation}
    \Sigma(\tau) = \sum_{I \ni i} \operatorname{var}(J_I)\, G(\tau)^{p-1}
    = \binom{N-1}{p-1}\operatorname{var}(J_I)\, G(\tau)^{p-1}
    = J^2\, G(\tau)^{p-1},
    \label{eq:leadingsigma}
\end{equation}
where the last equality uses the exact variance $\operatorname{var}(J_I) = J^2/\binom{N-1}{p-1}$ of \eqref{eq:H}. There are $\binom{N-1}{p-1}$ couplings sharing the index $i$, so the $J^2$ prefactor is exact at finite $N$ with no $1/N$ correction. There are of course still instance-to-instance fluctuations in the actual coupling sum $\sum_{B} J_{iB}^2$.

\paragraph{Schwinger--Dyson resummation.} The bare propagator can be conveniently represented in frequency space as
\beq
G_0(\tau) = \frac{1}{2} \text{sgn}(\tau) =  \frac{1}{\beta}\sum_n e^{- i \omega_n \tau} \frac{1}{- i\omega_n}
\eeq
where the sum runs over the fermionic Matsubara frequencies, $\omega_n = \frac{(2n+1)\pi}{\beta}$. Summing a single line's bare propagator, one melon, two nested melons, and so on is the geometric (Dyson) series of Figure~\ref{fig:melon}(b), whose sum is the dressed propagator $G$ obeying
\begin{equation}
    G(i\omega_n)^{-1} = -i\omega_n - \Sigma(i\omega_n),
    \qquad
    \Sigma(\tau) = J^2\, G(\tau)^{p-1},
    \label{eq:leadingSD}
\end{equation}
the Schwinger--Dyson equations \eqref{eq:SD}, now read as the self-consistency condition that resums the melon on every line. Solving \eqref{eq:leadingSD} once for each $\beta$, on an imaginary-time grid, is part of the classical computation required to obtain the expectation values. In practice, the solution on the thermal circle is straightforward to obtain to high precision using fast Fourier transforms, but the equations are non-linear and it is possible that in some circumstances obtaining a solution might be computationally challenging.

\begin{figure}[t]
\begin{center}
\begin{tikzpicture}[line width=0.9pt, every node/.style={font=\small}]
  \node at (-3.8,0) {(a)};
  \node at (-2.9,0) {$\Sigma(\tau)\;=$};
  \coordinate (l) at (-1.3,0);
  \coordinate (r) at (1.3,0);
  \draw (l) -- (-2.0,0);
  \draw (r) -- (2.0,0);
  \draw (l) .. controls (-0.65,0.8) and (0.65,0.8) .. (r);
  \draw (l) -- (r);
  \draw (l) .. controls (-0.65,-0.8) and (0.65,-0.8) .. (r);
  \fill (l) circle (2.4pt);
  \fill (r) circle (2.4pt);
  \node at (0,1.05) {$p-1$};
  \node at (3.4,0) {$=\;J^2\, G(\tau)^{p-1}$};
\end{tikzpicture}

\vspace{1.1em}

\begin{tikzpicture}[line width=0.9pt, every node/.style={font=\small}]
  \node at (-4.3,0) {(b)};
  \draw[line width=1.9pt] (-3.7,0) -- (-2.5,0);
  \node[above=2pt] at (-3.1,0) {$G$};
  \node at (-2.05,0) {$=$};
  \draw (-1.5,0) -- (-0.2,0);
  \node[above=2pt] at (-0.85,0) {$G_0$};
  \node at (0.35,0) {$+$};
  \draw (0.9,0) -- (2.0,0);
  \fill[gray!25] (2.35,0) circle (0.30);
  \draw (2.35,0) circle (0.30);
  \node at (2.35,0) {$\Sigma$};
  \draw[line width=1.9pt] (2.7,0) -- (4.0,0);
\end{tikzpicture}
\end{center}
\caption{Melonic dressing for $p = 4$. (a) The self-energy insertion is a melon: two vertices (dots) joined by $p-1 = 3$ internal propagators, evaluating to $\Sigma(\tau) = J^2 G(\tau)^{p-1}$ by the exact index count \eqref{eq:leadingsigma}. (b) The Schwinger--Dyson equation \eqref{eq:leadingSD} resumming the melon on every line: the dressed propagator $G$ (thick) equals the bare propagator $G_0$ (thin) plus one self-energy insertion followed by a dressed propagator.}
\label{fig:melon}
\end{figure}

\paragraph{The skeleton prescription.} Dressing amounts to the following rule at leading order:
\begin{quote}
Draw only \emph{skeleton} diagrams --- those with no self-energy insertion on any internal line --- put the dressed propagator $G$ on every line, and drop from each coupling polynomial the melonic piece already absorbed into $G$.
\end{quote}
In the bare expansion the melon is an ordinary fixed-order term and no subtraction is needed; in the dressed expansion the same piece would be double-counted, so it is removed once. Only the disorder-mean part of a self-energy insertion is resummed into $G$; its mean-zero remainder is a separate, higher-order structure, handled by the last of the rules collected below and in Section~\ref{sec:corrections}. The combinatorial content of a diagram --- its index-contraction pattern, coupling monomial, orientation sign, and coefficient $(-1)^r 2^{W/2}$ --- is untouched by dressing; only the kernel changes, each bare $G_0$ becoming a dressed $G$ in the time integral \eqref{eq:barevalue}.

\paragraph{Dressed leading order.} Applying this to the leading $W = 4$ diagram of the previous subsection, the single-vertex kernel $\int_0^\beta G_0^4 = \beta/16$ is replaced by its dressed counterpart
\begin{equation}
    I_4(\beta) = \int_0^\beta G(\tau)^4\, d\tau ,
    \label{eq:I4}
\end{equation}
so that
\begin{equation}
    \xi_X = -4\, I_4(\beta)\, J_X + \cdots .
    \label{eq:w4dressed}
\end{equation}
This is the same tensor $J_X$ with the same coefficient $-4$; only the temperature kernel has changed. At high temperature $G \to G_0$ and $I_4 \to \beta/16$, recovering the bare result \eqref{eq:w4bare}; but \eqref{eq:w4dressed} is now sensible at any fixed $\beta J$ as $N \to \infty$, because the melons that would otherwise spoil the naive series are resummed into $G$.

Indeed, at fixed but large $\beta J$, the dressed propagator approaches the \emph{conformal limit}
\beq
G_c(\tau) \propto
\left[\frac{\pi}{\beta J\,\sin\!\left(\pi\tau/\beta\right)}\right]^{2/p},
\qquad 0<\tau<\beta,
\eeq
when $J \tau$ is far from both endpoints. Plugging this into the leading diagram,
\beq
\int d\tau\, G_c(\tau)^p
\propto \frac{1}{(\beta J)^2}\int_0^\beta d\tau\,
\frac{1}{\sin^2\!\left(\pi\tau/\beta\right)},
\eeq
shows that the large-$\beta$ growth of the naive high-temperature expansion is cut off. The short-time divergences of the conformal form around $\tau=0,\beta$ are regulated by the regular behavior of the full $G(\tau)$ and contribute a constant rather than a term growing with $\beta$. More precisely, at the large-$N$ melonic saddle the thermal energy $E=\langle H\rangle_\beta$ obeys
\beq
\frac{E}{N}=-\frac{J^2}{p}\int_0^\beta d\tau\,G(\tau)^p.
\eeq
For $p=4$, this gives $I_4(\beta)=-4E/(NJ^2)$, so $I_4$ is proportional to minus the energy density and approaches a constant as $\beta\to\infty$.

\paragraph{The rules, collected.} It is convenient to gather the prescription in one place, since the remainder of the paper is an application of it. To compute $\xi_X$ at a given order:
\begin{enumerate}
    \item Draw every connected \emph{skeleton} diagram with $r$ interaction vertices: each vertex is one coupling $J_A$ supplying $p$ legs at an imaginary time integrated over $[0,\beta]$, the external string $\mu_X$ is pinned at $\tau = 0$ and supplies its $W$ legs, and all legs are paired into internal lines. Skeleton means no self-energy insertion on any internal line: the disorder-mean part of every such insertion is already resummed into $G$, and its mean-zero remainder is handled by rule~5.
    \item Give each internal line the dressed propagator $G(\tau_a - \tau_b)$ of \eqref{eq:leadingSD}. A line forces its two indices to be equal, and the shared index is summed over $1, \ldots, N$ unless it is pinned by the external string.
    \item Read off the coupling polynomial $P_X(J)$: the index sum just described, with every coupling written with sorted indices and the antisymmetric extension of Appendix~\ref{app:conventions} understood. For even external blocks its orientation sign is the unordered-partition parity $\eta(P)$. For odd blocks the vertex order must be retained and the sign is fixed by the full ordered Wick contraction, as explained in Appendix~\ref{app:conventions-eta} and illustrated explicitly by the cubic weight-four formulas below.
    \item Multiply by the kernel $\mathcal{K}(\beta)$ of \eqref{eq:barevalue}, now with dressed propagators, and by the coefficient $(-1)^r 2^{W/2}$ of \eqref{eq:coeffrule}.
    \item Where two vertices carry the \emph{same} coupling, keep only what the dressing has not already taken: the coherent part of a repeated coupling lives in $G$, and only its mean-zero remainder belongs to the diagram. Section~\ref{sec:corrections} does this subtraction by hand at the orders we need; Appendix~\ref{sec:rules} gives the rule that extends this bookkeeping to all orders.
\end{enumerate}
The rules for insertions at separated times are identical except for the external legs, and are given in Section~\ref{sec:time-dep-rules}.

\subsection[Example: weight 4]{Example: weight $4$}
\label{sec:leading-w4}

We now put the dressed leading prediction to a quantitative test. Since odd weights and weight $2$ vanish for $p=4$, the first interesting case is weight $4$. For $X = \{a<b<c<d\}$ the single diagram of the two previous subsections gives
\begin{equation}
    \xi_X = -4\, I_4(\beta)\, J_X + \cdots ,
    \qquad
    I_4(\beta) = \int_0^\beta G(\tau)^4\, d\tau.
    \label{eq:w4pred}
\end{equation}
We emphasize again that this is a parameter-free prediction: the propagator $G$ comes from solving the Schwinger--Dyson equation \eqref{eq:leadingSD} once at the given $\beta J$ (as described in Appendix~\ref{app:numerics}) and $J_X$ is read directly off the instance.

\begin{figure}[t]
\begin{center}
\includegraphics[width=0.98\textwidth]{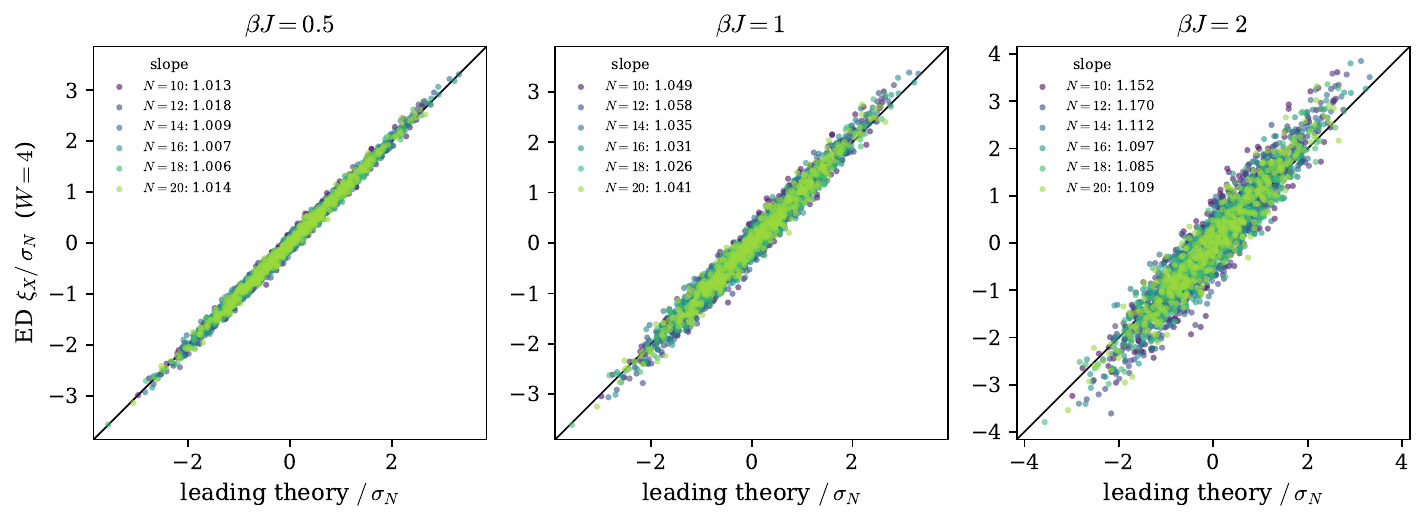}
\end{center}
\caption{ED weight-4 one-point values $\xi_X$ versus the leading prediction $-4 I_4(\beta) J_X$ of \eqref{eq:w4pred}, at $\beta J = 0.5, 1, 2$ (left to right), for $N = 10$--$20$ (color). Each point is one weight-4 string of one disorder realization; both axes are normalized per $N$ by the standard deviation of the prediction. Legends give the per-$N$ through-origin slope; the diagonal is unit slope. The prediction is nearly exact at high temperature (slopes ${\approx}1.01$ at $\beta J = 0.5$) and degrades smoothly as $\beta J$ grows, both the slope and the scatter increasing.}
\label{fig:w4betas}
\end{figure}

\paragraph{Temperature dependence at finite $N$.} Figure~\ref{fig:w4betas} plots the prediction \eqref{eq:w4pred} against exact diagonalization across sizes $N = 10$--$20$ and three temperatures $\beta J = 0.5, 1, 2$, each point a single weight-4 string of a single realization (values normalized per $N$ by the standard deviation of the prediction so the panels are comparable). At $\beta J = 0.5$ the agreement is essentially perfect: the cloud hugs the diagonal and the through-origin slopes sit at ${\approx}1.01$ for every $N$, a direct confirmation of the coefficient rule and the kernel $I_4$ at the percent level. As $\beta J$ increases the cloud both tilts and fattens --- the slope climbs to ${\approx} 1.1$ at $\beta J = 2$ and the operator-to-operator scatter about it grows. This is the non-uniformity announced in the introduction: at fixed $N$, a fixed-order truncation becomes less accurate as the neglected higher-order diagrams acquire larger kernels, so both the slope mismatch and the scatter worsen as the temperature is lowered.

As we discuss in more detail in Section \ref{sec:corrections}, once the melonic resummation is taken into account, corrections to the leading diagrams should vanish as $N$ is taken to infinity at fixed $\beta J$. Hence, one predicts that the deviation of the slope and the increased scatter are finite-$N$ effects. We can provide some independent evidence for this claim by studying exact diagonalization at fixed $\beta J$ and increasing $N$. We focus on $\beta J = 2$ in what follows, as this choice clearly demonstrates the finite $N$ effects of interest while also remaining below the $\beta J \sim N$ scale at which one expects more severe deviations from the leading large $N$ theory.

\paragraph{Finite-$N$ convergence.} We diagonalize the $p = 4$ Hamiltonian in fermion-parity blocks for $N = 16, 18, 20, 22, 24$ (Appendix~\ref{app:numerics}). For each of $160$ disorder realizations we measure $\xi_X$ for $48$ distinct weight-$4$ strings $X$ and compare with \eqref{eq:w4pred}, pooling all $160 \times 48 = 7680$ operator values at each $N$. Two summaries track the two trends of Figure~\ref{fig:w4betas}: the \emph{through-origin slope} of ED against prediction, which measures whether the leading coefficient is right on average, and the \emph{relative residual scatter} $\operatorname{rms}(\xi^{\rm ED} - s\,\xi^{\rm pred})/\operatorname{rms}(\xi^{\rm pred})$ about that slope, which measures how well the single tensor $J_X$ captures the operator-to-operator variation.

\begin{figure}[t]
\begin{center}
\includegraphics[width=0.72\textwidth]{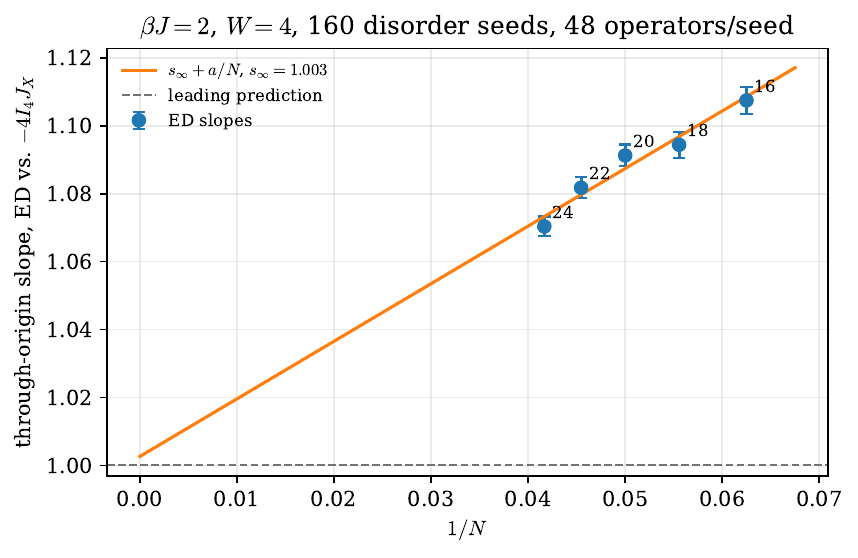}
\end{center}
\caption{Through-origin slope of the ED weight-4 one-point values against the leading prediction $-4 I_4(\beta) J_X$ at $\beta J = 2$, versus $1/N$ ($N = 16$--$24$; $160$ disorder seeds, $48$ operators per seed). The slope exceeds unity at finite $N$ --- about $1.11$ at $N = 16$ --- and falls almost linearly in $1/N$ toward the ideal value $1$ (dashed); a fit $s(N) = s_\infty + a/N$ gives $s_\infty \approx 1.00$. The residual excess at accessible $N$ is a deterministic coefficient correction analyzed in Section~\ref{sec:corrections}.}
\label{fig:w4slope}
\end{figure}

\begin{figure}[t]
\begin{center}
\includegraphics[width=0.72\textwidth]{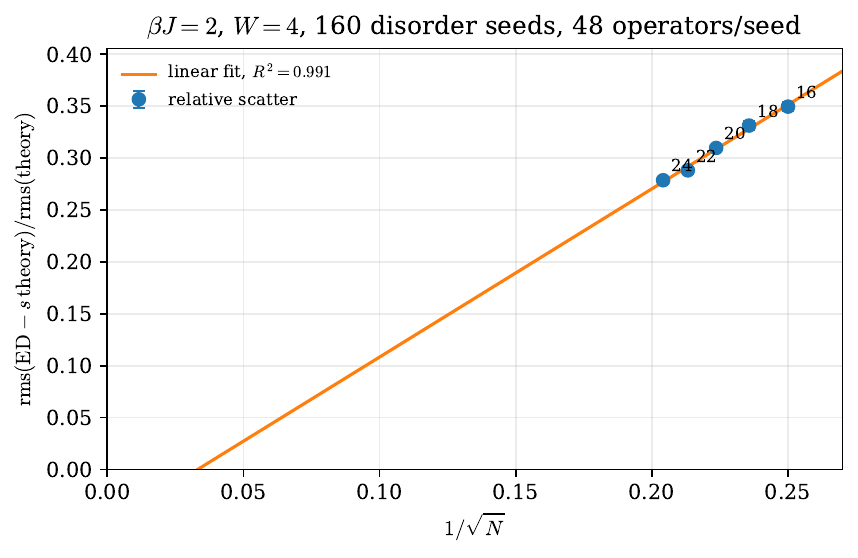}
\end{center}
\caption{Relative residual scatter $\operatorname{rms}(\xi^{\rm ED} - s\,\xi^{\rm pred})/\operatorname{rms}(\xi^{\rm pred})$ of the weight-4 one-point values about the fitted slope at $\beta J = 2$, versus $1/\sqrt{N}$. The scatter falls from ${\approx}0.35$ at $N = 16$ to ${\approx}0.28$ at $N = 24$ and extrapolates linearly in $1/\sqrt{N}$ to an intercept consistent with zero (${\approx}{-}0.05$, weighted $R^2 = 0.99$) --- the random-walk size of the leading fluctuation tensor omitted from \eqref{eq:w4pred}.}
\label{fig:w4scatter}
\end{figure}

Two features stand out in the data, both consistent with the leading formula becoming exact as $N \to \infty$, and they are of different kinds. The slope (Figure~\ref{fig:w4slope}) falls toward unity linearly in $1/N$. It is a property of the \emph{coefficient} of $J_X$: no tensor orthogonal to $J_X$ can move it, so the residual excess must be a deterministic correction to that coefficient, traced in Section~\ref{sec:corrections} to a connected four-point rung. The relative scatter (Figure~\ref{fig:w4scatter}) instead falls as $1/\sqrt{N}$ --- the size expected of the first fluctuation tensor beyond $J_X$ (the two-vertex structure of Section~\ref{sec:corrections}), which is statistically orthogonal to $J_X$ and so feeds the scatter but not the slope.


\subsection[Example: weight 8]{Example: weight $8$}
\label{sec:leading-w8}

The next weight allowed by the selection rules is $W = 8$, and it is the first case whose leading diagram has more than one vertex. Let $X = \{x_1 < \cdots < x_8\}$, so $\mu_X = 16\,\chi_{x_1}\cdots\chi_{x_8}$. A single quartic vertex supplies only four legs, so the fewest vertices that can absorb the eight external Majoranas is two, each carrying a quartet of the external legs. The two quartets must partition $X$: writing
\begin{align}
    \Part_4(X)
    &= \big\{\, P = \{A, B\} : |A| = |B| = 4,\ A \cup B = X,\
       A \cap B = \emptyset \,\big\}, \nonumber\\
    |\Part_4(X)| &= \tfrac12\tbinom{8}{4} = 35 .
\end{align}
the leading skeleton is one term per partition, with the two vertices attached to the external string but not to each other:
\begin{center}
\begin{tikzpicture}[line width=0.8pt, every node/.style={font=\small}]
  \coordinate (va) at (-1.9,0);
  \coordinate (vb) at (1.9,0);
  \foreach \ang/\lab in {120/$a_1$, 160/$a_2$, 200/$a_3$, 240/$a_4$} {
    \draw (va) -- ++(\ang:1.25) node[shift={(\ang:0.3)}] {\lab};
  }
  \foreach \ang/\lab in {60/$b_1$, 20/$b_2$, -20/$b_3$, -60/$b_4$} {
    \draw (vb) -- ++(\ang:1.25) node[shift={(\ang:0.3)}] {\lab};
  }
  \fill (va) circle (2.2pt);
  \fill (vb) circle (2.2pt);
  \node[right=3pt] at (va) {$J_A,\ \tau_1$};
  \node[left=3pt] at (vb) {$J_B,\ \tau_2$};
\end{tikzpicture}
\end{center}
Each vertex is exactly the leading $W = 4$ diagram of Section~\ref{sec:leading-w4}: its four dressed legs run to the external string at $\tau = 0$ and its time is integrated, contributing a factor $I_4(\beta)$ apiece. With the two vertices integrated independently the kernel is therefore $I_4(\beta)^2$, and the coefficient rule \eqref{eq:coeffrule} at $(W, r) = (8, 2)$ gives $(-1)^2\, 2^{4} = +16$:
\begin{equation}
    \xi_X = 16\, I_4(\beta)^2 \sum_{P = \{A, B\} \in \Part_4(X)} \eta(P)\, J_A J_B + \cdots ,
    \label{eq:w8pred}
\end{equation}
where $\eta(P) = \pm 1$ is the parity of the permutation carrying the concatenation of the sorted quartets $(a_1\cdots a_4\, b_1\cdots b_4)$ to sorted $X$ (Appendix~\ref{app:conventions}); because both blocks are even it is well defined on unordered partitions, and the two orderings of the distinct vertices cancel the $1/2!$ of the exponential. At high temperature $I_4 \to \beta/16$ and \eqref{eq:w8pred} reduces to $\tfrac{\beta^2}{16}\sum_P \eta(P) J_A J_B$, the second-order term of the bare expansion. As an aside, the structure generalizes to weight $W = 4m$ where the leading contribution is $\big(-4 I_4(\beta)\big)^m$ times the signed sum over partitions of $X$ into $m$ quartets.

\begin{figure}[t]
\begin{center}
\includegraphics[width=0.98\textwidth]{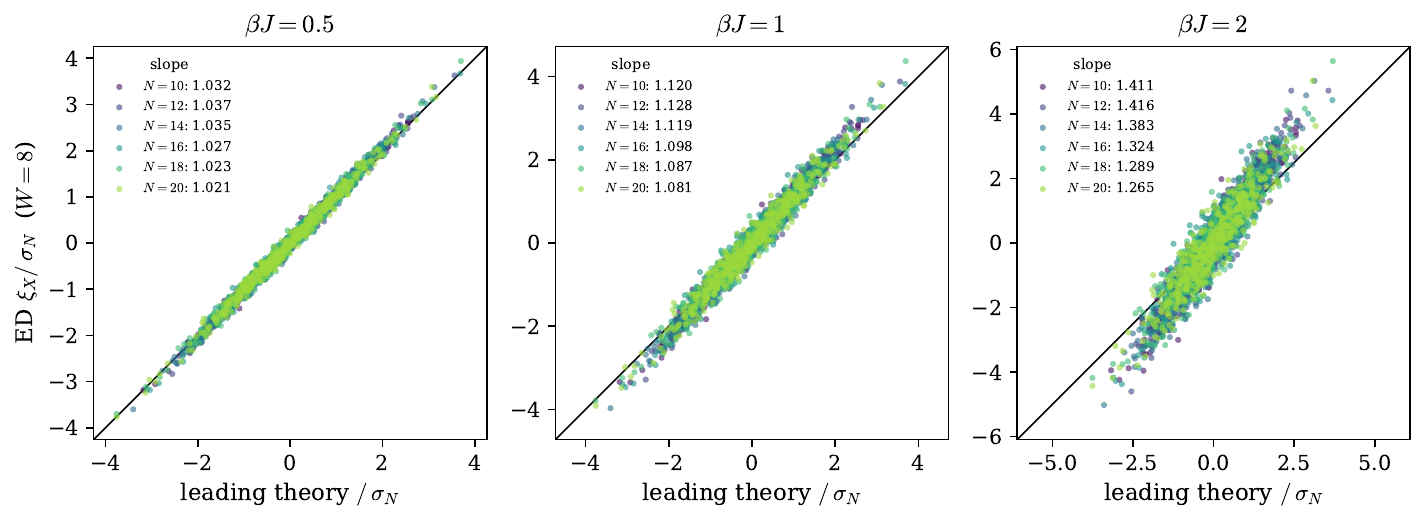}
\end{center}
\caption{ED weight-8 one-point values $\xi_X$ versus the leading prediction \eqref{eq:w8pred}, at $\beta J = 0.5, 1, 2$ (left to right), for $N = 10$--$20$ (color), normalized per $N$ as in Figure~\ref{fig:w4betas}. The prediction is nearly exact at high temperature (slopes ${\approx}1.02$ at $\beta J = 0.5$) and degrades as $\beta J$ grows, with a slope excess roughly three times that of $W = 4$ at the same temperature.}
\label{fig:w8betas}
\end{figure}

\paragraph{Temperature dependence at finite $N$.} Figure~\ref{fig:w8betas} tests \eqref{eq:w8pred} against ED at $\beta J = 0.5, 1, 2$ for $N = 10$--$20$, exactly as in Figure~\ref{fig:w4betas} for $W = 4$. The pattern is the same: near-perfect agreement at $\beta J = 0.5$ (slopes ${\approx}1.02$), with the cloud tilting and broadening as $\beta J$ grows. The slope excess is markedly larger than at $W = 4$ --- reaching ${\approx}1.3$--$1.4$ at $\beta J = 2$ against ${\approx}1.1$ there.

\begin{figure}[t]
\begin{center}
\includegraphics[width=0.72\textwidth]{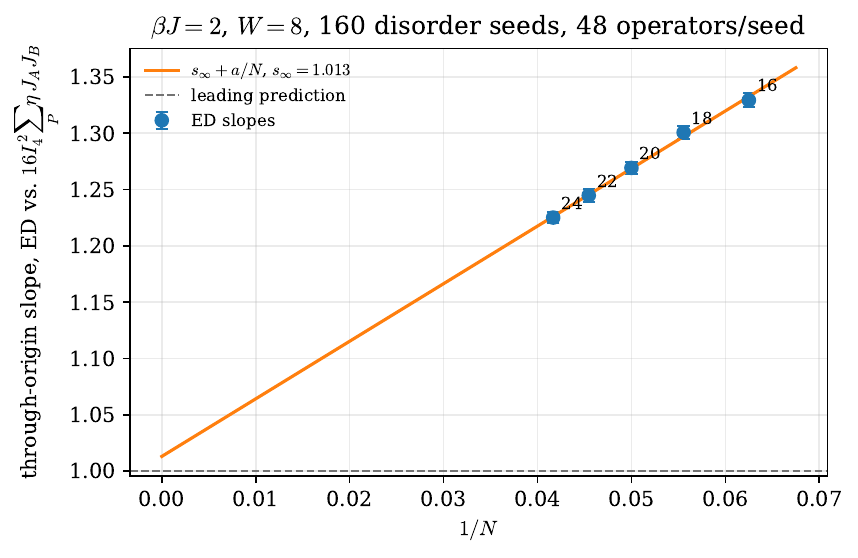}
\end{center}
\caption{Through-origin slope of the ED weight-8 one-point values against the leading prediction \eqref{eq:w8pred} at $\beta J = 2$, versus $1/N$ ($N = 16$--$24$; $160$ seeds, $48$ operators per seed). The slope decreases almost linearly in $1/N$; the fit $s(N) = s_\infty + a/N$ gives $s_\infty \approx 1.00$.}
\label{fig:w8slope}
\end{figure}

\begin{figure}[t]
\begin{center}
\includegraphics[width=0.72\textwidth]{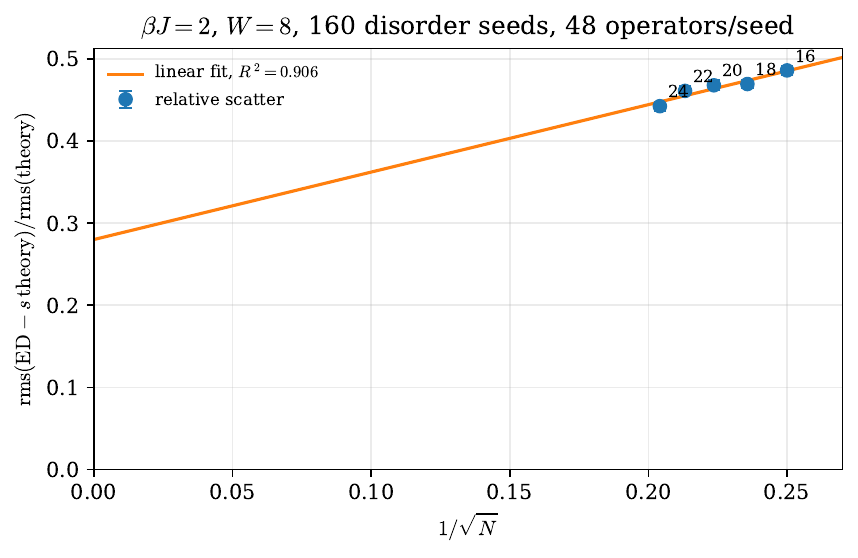}
\end{center}
\caption{Relative residual scatter of the weight-8 one-point values about the fitted slope at $\beta J = 2$, versus $1/\sqrt{N}$, as in Figure~\ref{fig:w4scatter} for $W = 4$. The scatter decreases with $N$ (${\approx}0.49 \to 0.45$ across $N = 16$--$24$) but much more slowly than at $W = 4$, and the linear fit leaves a sizeable intercept (${\approx}0.28$, weighted $R^2 = 0.91$): at weight $8$ several subleading structures of comparable size compete, and no single $1/\sqrt{N}$ term dominates at these sizes.}
\label{fig:w8scatter}
\end{figure}

\paragraph{Finite-$N$ convergence.} Pushing to $N = 16$--$24$ in parity-block ED at $\beta J = 2$ ($160$ seeds, $48$ weight-8 strings each), the through-origin slope again falls almost linearly in $1/N$ (Figure~\ref{fig:w8slope}), from ${\approx}1.33$ at $N = 16$ to ${\approx}1.23$ at $N = 24$, with a descriptive fit $s_\infty + a/N$ giving $s_\infty \approx 1.00$, consistent with the two-vertex leading formula being asymptotically exact. The excess itself is roughly three times the $W = 4$ excess at the same $N$ ($0.33$ against $0.11$ at $N = 16$, $0.23$ against $0.070$ at $N = 24$): it is the same deterministic one-loop coefficient correction analyzed in Section~\ref{sec:corrections}, now acting on each of the two vertices, together with a connected piece linking them. Two independently corrected vertices would give a factor of two; the observed factor of ${\approx}3$ is therefore consistent with a cross-vertex rung of size comparable to a single-vertex correction, but we do not compute that rung here. The operator-to-operator scatter likewise decreases with $N$ (Figure~\ref{fig:w8scatter}), but much more slowly than at $W = 4$: the relative scatter falls only from ${\approx}0.49$ to ${\approx}0.45$ across $N = 16$--$24$, and a linear extrapolation in $1/\sqrt{N}$ leaves a sizeable intercept at these sizes. This is as expected for a weight with many more competing subleading structures --- each vertex carries its own fluctuation corrections, plus genuinely new cross-vertex tensors --- so a single power law does not yet dominate in the accessible range of $N$. Appendix~\ref{app:w8} identifies the two structures responsible and shows that the parameter-free first correction better accounts for the observed scatter.

\subsection[Example: weight 2]{Example: weight $2$}
\label{sec:leading-w2}

We also briefly comment on $W=2$, which we recall is identically zero. For $X = \{i, j\}$ we have $\mu_X = 2i\,\chi_i \chi_j$, and the antiunitary selection rule of Section~\ref{sec:symmetries} --- which for $p = 4$ kills every $W = 2 \bmod 4$ string --- forces
\begin{equation}
    \xi_{\{i,j\}} = \tr\!\big(\mu_{\{i,j\}}\, \rho(\beta)\big) = 0
    \qquad \text{exactly, at every } N .
    \label{eq:w2zero}
\end{equation}
It is worth seeing how the diagrammatic expansion reproduces this zero, because the mechanism is different from the parity vanishing of odd weights: here the coupling polynomials are generically nonzero, and it is the \emph{kernels} that vanish.

\paragraph{One vertex: no diagram.} A single vertex offers four legs, of which two attach to the external string $\{i,j\}$; the remaining two must contract with each other. But the coupling $J_{ijkl}$ has four distinct indices, so the two leftover legs carry \emph{distinct} labels $k \neq l$ and cannot be Wick-paired ($\langle \chi_k \chi_l\rangle \propto \delta_{kl}$). The one-vertex tadpole
\begin{center}
\begin{tikzpicture}[line width=0.8pt, every node/.style={font=\small}]
  \coordinate (v) at (0,0);
  \draw (v) -- (120:1.2) node[shift={(120:0.25)}] {$i$};
  \draw (v) -- (60:1.2) node[shift={(60:0.25)}] {$j$};
  \draw (v) .. controls (-1.0,-1.25) and (1.0,-1.25) .. (v);
  \fill (v) circle (2.2pt);
  \node[right=2pt] at (v) {$J,\ \tau$};
  \node at (0,-1.15) {$k\,l$};
\end{tikzpicture}
\end{center}
therefore vanishes identically, independent of any kernel.

\paragraph{Two vertices: kernel cancellation.} The first structure with a nonzero coupling polynomial uses two vertices that share three internal labels $\{l, m, n\}$, one carrying the external index $i$ and the other $j$:
\begin{center}
\begin{tikzpicture}[line width=0.8pt, every node/.style={font=\small}]
  \coordinate (va) at (-1.4,0);
  \coordinate (vb) at (1.4,0);
  \draw (va) -- ++(150:1.15) node[shift={(150:0.25)}] {$i$};
  \draw (vb) -- ++(30:1.15) node[shift={(30:0.25)}] {$j$};
  \draw (va) .. controls (-0.7,0.75) and (0.7,0.75) .. (vb);
  \draw (va) -- (vb);
  \draw (va) .. controls (-0.7,-0.75) and (0.7,-0.75) .. (vb);
  \fill (va) circle (2.2pt);
  \fill (vb) circle (2.2pt);
  \node at (0,1.02) {$l\,m\,n$};
  \node[left=6pt,below=2pt] at (va) {$J_A,\ \tau_1$};
  \node[right=6pt,below=2pt] at (vb) {$J_B,\ \tau_2$};
\end{tikzpicture}
\end{center}
with $A = \{i, l, m, n\}$ and $B = \{j, l, m, n\}$. Its coupling polynomial is the off-diagonal ``self-energy'' between the two external legs,
\begin{equation}
    B_{ij} = \sum_{l < m < n} J_{ilmn}\, J_{jlmn} ,
\end{equation}
where the coupling is extended by total antisymmetry to arbitrary index order, with repeated-index components set to zero. This polynomial is generically nonzero (it is not diagonal in $i,j$, so it is not absorbed into $G$). The accompanying equal-time kernel, with both external legs pinned at $\tau = 0$ and the two vertex times integrated, is
\begin{equation}
    K(\beta) = \int_0^\beta d\tau_1\, d\tau_2\; G(\tau_1)\, G(\tau_2)\, G(\tau_1 - \tau_2)^3 .
    \label{eq:w2kernel}
\end{equation}
This integral vanishes identically. The melonic propagator is symmetric about the midpoint, $G(\beta - \tau) = G(\tau)$, so under the reflection $\tau_1 \to \beta - \tau_1$, $\tau_2 \to \beta - \tau_2$ the two single-leg factors are invariant while the relative factor flips, $G(\tau_1 - \tau_2)^3 \to G(\tau_2 - \tau_1)^3 = -G(\tau_1 - \tau_2)^3$. The integrand is thus odd under a symmetry of the square domain, so $K(\beta) = 0$. The same counting proves the result at every order: a nonvanishing $p=4$, weight-$2$ diagram with $r$ interaction vertices has $(4r-2)/2=2r-1$ propagators joining vertex times, an odd number, so global time reflection reverses its sign. Thus every higher-order kernel vanishes separately, in accordance with \eqref{eq:w2zero}.

Note that the zero is a statement about \emph{coincident} times only, since the reflection argument uses that both external legs sit at the same point $\tau = 0$. Separating the two Majoranas in imaginary time leads to an observable that is no longer required to vanish. The weight-2 observable is therefore a natural example of a \emph{time-dependent} insertion, where the polynomial $B_{ij}$ is measured through a nonzero bilocal kernel; we take this up in Section~\ref{sec:time-dep-w2imag}, recovering \eqref{eq:w2zero} as the equal-time endpoint $K(\beta) = 0$.

\section{Corrections to the leading diagrams}
\label{sec:corrections}

The leading diagrams of Section~\ref{sec:leading} are the first term of a systematic expansion, and the numerical tests already show both how well that first term does and the manner in which it fails. At weights $4$ and $8$ alike, the leading prediction is essentially exact at high temperature even at the small sizes reachable by exact diagonalization, while two distinct deviations open up as $\beta J$ grows: a through-origin \emph{slope} that drifts away from unity, and an operator-to-operator \emph{scatter} of the individual points about that slope. Both shrink as $N$ increases (Figures~\ref{fig:w4slope}, \ref{fig:w4scatter}, \ref{fig:w8slope}, \ref{fig:w8scatter}), and accounting for them is the subject of this section. We use weight $4$ and $p=4$ as the worked example throughout.

\subsection{Two types of corrections}
\label{sec:corrections-types}

Although both are finite-$N$ effects, the slope and the scatter probe orthogonal parts of the correction and, as the data already indicate, have different origins and different $N$-scalings. The ensemble projection onto the leading tensor $J_X$ controls the slope: only a contribution aligned with $J_X$ can move its disorder-averaged value, so the systematic slope shift is a deterministic renormalization of that coefficient. The scatter, by contrast, is fed by \emph{new} index structures, tensors statistically orthogonal to $J_X$ under the disorder average; being mean-zero fluctuations they displace individual operators without changing the ensemble slope, and they are the larger, relative-$1/\sqrt{N}$ effect.

It is worth stressing that although the slope is diagnosed by an ensemble fit, the exact coefficients are deterministic functions of $N$, $\beta$, and $J$ (Appendix~\ref{sec:rules-chaos}); the sample dependence of $\xi_X$ resides in the tensors. A correction that looked like a fluctuating coefficient --- the per-instance value of the melonic sum $\sum_B J_{iB}^2$, for example --- is instead a structure of higher degree in the couplings, with its own place in the graded hierarchy of Table~\ref{tab:grading}. The truncated large-$N$ evaluation below is applied instance by instance. Its accuracy, however, is established only in the ensemble or typical-realization sense tested numerically here; rare realizations are not separately controlled.

\paragraph{Fluctuation tensors.} The first kind of correction adds interaction vertices whose couplings are all \emph{distinct} from one another and from the external string: new, ``erratic'' contractions of the random couplings. Taking weight $4$ as an example, the first correction connects the four external legs to \emph{two} vertices, splitting them into two pairs $A = \{i_1,i_2\}$, $B = \{i_3,i_4\}$, with the vertices joined by two internal lines:
\begin{center}
\begin{tikzpicture}[baseline={(v.base)}, line width=0.8pt, every node/.style={font=\small}]
  \node[inner sep=0pt] (v) at (-1.0,0) {};
  \coordinate (w) at (1.0,0);
  \fill (v) circle (2.2pt);
  \fill (w) circle (2.2pt);
  \coordinate (i1) at (-2.1,0.8);
  \coordinate (i2) at (-2.1,-0.8);
  \coordinate (i3) at (2.1,0.8);
  \coordinate (i4) at (2.1,-0.8);
  \draw (v) -- (i1);
  \draw (v) -- (i2);
  \draw (w) -- (i3);
  \draw (w) -- (i4);
  \draw (v) .. controls (-0.35,0.35) and (0.35,0.35) .. (w);
  \draw (v) .. controls (-0.35,-0.35) and (0.35,-0.35) .. (w);
  \node[left] at (i1) {$i_1$};
  \node[left] at (i2) {$i_2$};
  \node[right] at (i3) {$i_3$};
  \node[right] at (i4) {$i_4$};
  \node[above] at (0,0.38) {$u$};
  \node[below] at (0,-0.38) {$v$};
  \node[above=2pt] at (v) {$J$};
  \node[above=2pt] at (w) {$J$};
\end{tikzpicture}
\end{center}
Its coupling polynomial is a new rank-four tensor,
\begin{equation}
    T^{(2)}_X = \sum_{A|B} \eta(A,B) \sum_{u<v} J_{Auv}\, J_{uvB} ,
    \qquad
    W^{(2)}(\beta) = \int_0^\beta d\tau_1\, d\tau_2\; G(\tau_1)^2\, G(\tau_2)^2\, G(\tau_1 - \tau_2)^2 ,
    \label{eq:corr-t2}
\end{equation}
with the sum over the pair-splittings of $X$ and the coefficient rule giving $\xi_X^{(2)} = +4\, W^{(2)}(\beta)\, T^{(2)}_X$. The external pairs $A$ and $B$ are disjoint, so the two displayed couplings are necessarily distinct and $T^{(2)}_X$ is a sum of products of \emph{different} Gaussian couplings, with no coincident pair anywhere in the sum. It is therefore purely quadratic in the couplings in the graded sense of Appendix~\ref{sec:rules-chaos}: it has no piece of lower degree hiding inside it, and in particular no component along $J_X$. Thus $T^{(2)}$ is statistically \emph{orthogonal} to the leading tensor, $\mathbb{E}[T^{(2)}_X J_X] = 0$. It cannot shift the ensemble slope and is precisely the structure that collapses the scatter when added to the prediction. Being one vertex beyond the leading diagram, it is of relative size $1/\sqrt{N}$, matching the observed scatter. New independent fluctuation tensors of this kind proliferate at each higher order.

\paragraph{Corrections to the melonic structure.} The second kind of correction does not add a new tensor at all; it corrects the coefficient of an existing one. The leading kernel $I_4 = \int G^4$ treats the four Majoranas of $\mu_X$ as propagating through independently dressed lines, and the melonic dressing of Section~\ref{sec:leading-melonic} already resummed the leading such correction into $G$. The next correction couples two of those lines to each other by an exchanged \emph{ladder} rung:
\begin{center}
\begin{tikzpicture}[line width=0.8pt, every node/.style={font=\small}]
  \draw (-0.9,-1.6) -- (-0.9,1.6);
  \draw (0.9,-1.6) -- (0.9,1.6);
  \draw (-0.9,0) .. controls (-0.35,0.24) and (0.35,0.24) .. (0.9,0);
  \draw (-0.9,0) .. controls (-0.35,-0.24) and (0.35,-0.24) .. (0.9,0);
  \fill (-0.9,0) circle (2.2pt); \fill (0.9,0) circle (2.2pt);
  \node at (0,0.46) {\scriptsize $u$};
  \node at (0,-0.46) {\scriptsize $v$};
  \node[above] at (-0.9,1.6) {$x_a$};
  \node[above] at (0.9,1.6) {$x_a$};
  \node[below] at (-0.9,-1.6) {$x_b$};
  \node[below] at (0.9,-1.6) {$x_b$};
\end{tikzpicture}
\end{center}
Both lines enter with a common index $x_a$ and leave with a common index $x_b$; the rung itself is the coupling $J_{x_a x_b uv}$, appearing squared under the disorder average, with its two internal lines $u, v$ summed freely over all $\binom{N-2}{2}$ values. That free internal sum is what makes the rung \emph{coherent}: $\sum_{u<v} J_{x_a x_b uv}^2$ has a nonzero disorder mean $\binom{N-2}{2}\,\sigma_J^2$, where $\sigma_J^2 = \operatorname{var}(J_I)$, so --- unlike a fluctuation tensor, whose two \emph{distinct} couplings average to zero --- the rung carries a deterministic mean and renormalizes the coefficient of the leading structure. This is what shifts the slope. Its mean is of relative order $1/N$ (a coincident coupling $\sigma_J^2 \sim N^{-3}$ against the ${\sim}N^2$ internal pairs), the smaller correction, consistent with the observed slope excess. Iterating the rung with the intermediate rail index now \emph{summed} rather than pinned builds the four-point ladder that governs the Gaussian fluctuation about the melonic saddle; we compute the leading rung quantitatively for $W = 4$ in the case study of Section~\ref{sec:corrections-w4slope}, where it accounts for the residual slope.

\paragraph{Why the two are entangled.} The clean split above --- distinct couplings giving fluctuation tensors, coincident couplings giving coefficient renormalizations --- is not always valid at the level of diagrams, and this is the central complication in the expansion. A single higher-order topology carries both: summed over generic distinct indices it is a new fluctuation tensor, but the same sum contains coincidence configurations in which a pair of couplings is Wick-paired, and those pieces collapse onto structures of lower degree in the couplings (including the self-energy melon already resummed into $G$). For later use, we refer to the degree of a coupling polynomial as its \emph{chaos order}. Chaos here refers to a representation of a random variable as a polynomial (or other expansion) composed of simple random variables, such as Gaussians; the relevance of this notion is discussed in more detail in the context of the orthogonal decomposition developed in Appendix~\ref{sec:rules}. In this language, we can say that extracting the genuinely new fluctuation part therefore requires projecting out the lower-chaos pieces already accounted for. Organizing this projection order by order is the bookkeeping that the rules of Appendix~\ref{sec:rules} are designed to handle.

\paragraph{A two-component address.} The two mechanisms are graded independently. The chaos order $r$ just introduced --- the degree of the coupling polynomial --- counts the instance couplings a structure carries, and at $p=4$ each additional one costs a relative $N^{-1/2}$. The \emph{collective order} $\ell$ counts loops in the coherent kernel corrections --- the melonic dressing already resummed into $G$ is $\ell=0$, and the ladder and self-energy corrections computed below are $\ell=1$ --- with each additional order costing a relative $1/N$. Every contribution therefore carries an address $(r,\ell)$, generic distinct-coupling structures having root-mean-square size $N^{-[(p-2)r+W]/4}N^{-\ell}$; the power counting behind these estimates is derived in Appendix~\ref{sec:rules-power}. Table~\ref{tab:grading} lists the leading addresses for the weight-four one-point function and previews the calculations of the rest of this section. Mean-zero structures compete in the scatter while coherent kernel corrections move the slope, so the degree-three tensors can be masked by the larger $T^{(2)}$ scatter even though they carry the same nominal power as the $\ell=1$ slope correction.

\begin{table}[t]
\centering
\begin{tabular}{@{}c p{0.55\linewidth} c l@{}}
\toprule
$(r,\ell)$ & contribution & size & effect \\
\midrule
$(1,0)$ & $-4 I_4\, J_X$ & $N^{-3/2}$ & the leading line \\
$(2,0)$ & $+4 W^{(2)}\, T^{(2)}_X$ & $N^{-2}$ & scatter \\
$(1,1)$ & order-$\ell=1$ kernel: the resummed rung ladder and the shift of the averaged propagator (Section~\ref{sec:corrections-w4slope}) & $N^{-5/2}$ & slope \\
$(3,0)$ & degree-three tensors (ladder, triangle, chain skeletons) & $N^{-5/2}$ & scatter \\
$(3,0)_{\rm diag}$ & diagonal $\sum_B \no{J_X J_{x_1 B}^2}$ & $N^{-3}$ & scatter \\
\bottomrule
\end{tabular}
\caption{The graded catalogue for the weight-four one-point function at $p=4$. The chaos order $r$ labels the coupling structure and the collective order $\ell$ labels the kernel correction. Generic distinct-coupling entries have size $N^{-[(p-2)r+W]/4}N^{-\ell}$. Mean-zero structures produce operator-to-operator scatter, while coherent kernel corrections move the slope. The diagonal structure denoted $(3,0)_{\rm diag}$ carries a repeated coupling and is more suppressed than its generic degree-three counterparts. The power counting is derived in Appendix~\ref{sec:rules-power}.}
\label{tab:grading}
\end{table}

\subsection[The scatter: fluctuation tensors at weight 4]{The scatter: fluctuation tensors at weight $4$}
\label{sec:corrections-w4scatter}

We now demonstrate the corrections described in Section \ref{sec:corrections-types} in the weight $4$ case, computing the leading correction to the prediction \eqref{eq:w4pred} from first principles by direct diagram analysis. As we show, the computations here suffice to bring $\beta J = 2$ in line with the quality of the uncorrected $\beta J=0.5$ comparison. Appendix \ref{app:w4-temperature} contains additional data that illustrate how the high temperature agreement at $\beta J = 0.5$ is further improved by including corrections and how the corrections continue to strongly improve the result at even lower temperature, $\beta J = 4$. One can also go to even higher order, in which case the formulation presented in Appendix~\ref{sec:rules} provides an efficient framework to deal with the complications that arise. The two finite-$N$ deviations catalogued in Section~\ref{sec:corrections-types}, the operator-to-operator \emph{scatter} and the \emph{slope} excess, are treated in turn in this subsection and the next.

\begin{figure}[t]
\centering
\includegraphics[width=0.98\textwidth]{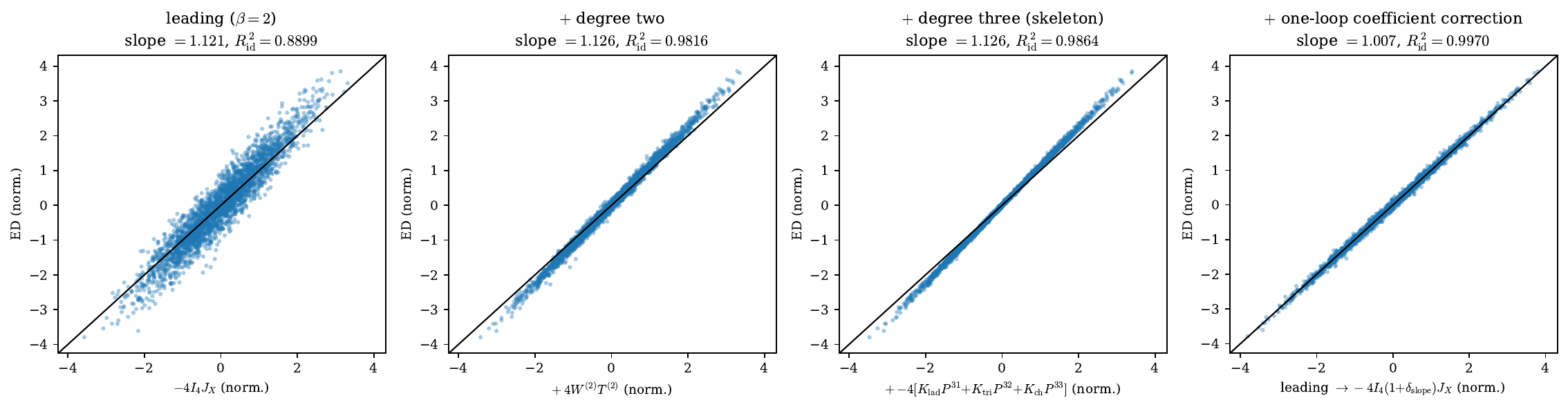}
\caption{The $W=4$ one-point function against the parameter-free prediction, order by order, at $\beta J = 2$ ($N = 10$--$20$ pooled, $20$ seeds $\times$ $24$ weight-4 strings each, axes normalized per $N$). Left to right: leading $-4 I_4 J_X$; adding the degree-two tensor $+4 W^{(2)} T^{(2)}$; adding the exactly Wick-ordered degree-three skeletons; and replacing the leading coefficient by its one-loop-corrected value $-4I_4(1+\delta_{\rm slope})J_X$ with $\delta_{\rm slope} = (\kappa_L + \kappa_S)\,\delta_{\rm rung}$ from Table~\ref{tab:w4pred}. Each panel quotes the through-origin slope as a diagnostic and the parameter-free $R^2_{\rm id}$ computed about the identity line, with no fitted coefficient. The fluctuation tensors raise $R^2_{\rm id}$ from $0.890$ to $0.986$ while barely moving the finite-sample slope ($1.121 \to 1.126$); the one-loop coefficient correction then removes the slope excess ($1.126 \to 1.007$) and raises $R^2_{\rm id}$ to $0.997$.}
\label{fig:w4order3}
\end{figure}

We start by summarizing what is achieved. The left panel of Figure \ref{fig:w4order3} shows the leading prediction at $\beta J=2$, while the remaining three panels show, in turn, the effects of adding additional fluctuation tensors at order $2$ (panel $2$, relative size $N^{-1/2}$) and $3$ (panel $3$, relative size $N^{-1}$) --- the content of this subsection --- and the effect of the resummed ladder (panel $4$, relative size $N^{-1}$) --- the content of the next subsection.

The first correction adds a second interaction vertex whose coupling is distinct from the external string. Connectedness forces the four external indices into two pairs joined by two internal lines, as drawn in Section~\ref{sec:corrections-types}. The tensor and kernel are those of \eqref{eq:corr-t2}, and their contribution is $\xi_X^{(2)} = +4\, W^{(2)}(\beta)\, T^{(2)}_X$, with the coefficient $+4$ fixed by the coefficient rule \eqref{eq:coeffrule} at $(W,r)=(4,2)$. Because $T^{(2)}$ is Wick-orthogonal to $J_X$, it moves individual operators off the leading line without shifting the ensemble slope. By the power counting it is of relative size $N^{-1/2}$, matching the scatter observed in Figure~\ref{fig:w4scatter}. Its kernel $W^{(2)}(\beta)$ is tabulated in Table~\ref{tab:w4pred}.

\paragraph{The scatter budget.} The size of this effect is fixed with no free parameter. The three pair-splittings of $X$ involve disjoint sets of couplings and are mutually uncorrelated; within one splitting, two terms with different internal pairs $(u,v)$ involve four distinct couplings and so do not correlate either; and the internal pair must avoid $X$ altogether, since a coupling with a repeated index vanishes. Hence
\beq
    \operatorname{var}\big(T^{(2)}\big) = 3 \binom{N-4}{2}\, \sigma_J^4,
    \label{eq:w4predscatter}
\eeq
and, since $T^{(2)}$ is orthogonal to $J_X$, the relative residual scatter about the leading line plotted in Figure~\ref{fig:w4scatter} is predicted to be
\beq
    r_{\rm pred}(N,\beta)
    = \frac{W^{(2)}}{I_4}\, \sqrt{3 \tbinom{N-4}{2}}\; \sigma_J
    \;\xrightarrow[N \to \infty]{}\; \frac{3J\, W^{(2)}}{I_4\, \sqrt{N}} .
\eeq
At $\beta J = 0.5$ this parameter-free amplitude agrees with the measured scatter to within a few percent over $N = 10$--$20$ (for example $0.082$ predicted against $0.082$ measured at $N = 16$). At $\beta J = 2$ it predicts $0.290$ at $N = 16$ falling to $0.249$ at $N = 24$, against the measured $0.349$ and $0.279$ of Figure~\ref{fig:w4scatter}: the shape and the $N$ dependence are right, with a residual amplitude excess of $1.21$ at $N = 16$ decreasing to $1.12$ at $N = 24$. That excess is the same finite-$N$ kernel dressing that appears as the fitted degree-two amplitude $c_2/(+4) = 1.24$ in Table~\ref{tab:w4fit}, and it shrinks with $N$ as it must. The corresponding budget at weight $8$, where two structures compete and their covariance cannot be neglected, is worked out in Appendix~\ref{app:w8}. (\texttt{numerics-v3/case\_w4/check\_scatter\_budget.py})

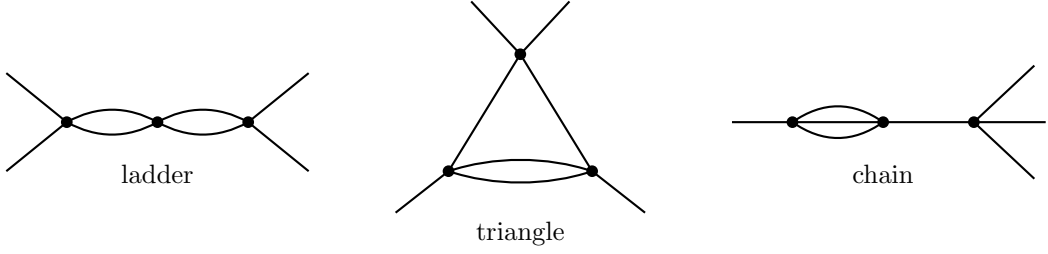
\begin{figure}[t]
\centering
\begin{tikzpicture}[line width=0.8pt, every node/.style={font=\small}]
  \begin{scope}[xshift=-4.8cm]
    \coordinate (l) at (-1.2,0); \coordinate (m) at (0,0); \coordinate (r) at (1.2,0);
    \fill (l) circle (2.2pt); \fill (m) circle (2.2pt); \fill (r) circle (2.2pt);
    \draw (l) .. controls (-0.8,0.22) and (-0.4,0.22) .. (m);
    \draw (l) .. controls (-0.8,-0.22) and (-0.4,-0.22) .. (m);
    \draw (m) .. controls (0.4,0.22) and (0.8,0.22) .. (r);
    \draw (m) .. controls (0.4,-0.22) and (0.8,-0.22) .. (r);
    \draw (l) -- (-2.0,0.65); \draw (l) -- (-2.0,-0.65);
    \draw (r) -- (2.0,0.65); \draw (r) -- (2.0,-0.65);
    \node[below=12pt] at (m) {ladder};
  \end{scope}
  \begin{scope}
    \coordinate (t) at (0,0.9); \coordinate (l) at (-0.95,-0.65); \coordinate (r) at (0.95,-0.65);
    \fill (t) circle (2.2pt); \fill (l) circle (2.2pt); \fill (r) circle (2.2pt);
    \draw (t) -- (l); \draw (t) -- (r);
    \draw (l) .. controls (-0.35,-0.45) and (0.35,-0.45) .. (r);
    \draw (l) .. controls (-0.35,-0.85) and (0.35,-0.85) .. (r);
    \draw (t) -- (-0.65,1.6); \draw (t) -- (0.65,1.6);
    \draw (l) -- (-1.65,-1.2); \draw (r) -- (1.65,-1.2);
    \node[below=15pt] at (0,-0.65) {triangle};
  \end{scope}
  \begin{scope}[xshift=4.8cm]
    \coordinate (l) at (-1.2,0); \coordinate (m) at (0,0); \coordinate (r) at (1.2,0);
    \fill (l) circle (2.2pt); \fill (m) circle (2.2pt); \fill (r) circle (2.2pt);
    \draw (l) .. controls (-0.8,0.28) and (-0.4,0.28) .. (m);
    \draw (l) -- (m);
    \draw (l) .. controls (-0.8,-0.28) and (-0.4,-0.28) .. (m);
    \draw (m) -- (r);
    \draw (l) -- (-2.0,0);
    \draw (r) -- (2.0,0.75); \draw (r) -- (2.15,0); \draw (r) -- (2.0,-0.75);
    \node[below=12pt] at (m) {chain};
  \end{scope}
\end{tikzpicture}
\caption{The three connected degree-three skeletons for a weight-four external string. From left to right the external-leg allocations among the three interaction vertices are $2$--$0$--$2$ (ladder), $2$--$1$--$1$ (triangle), and $1$--$0$--$3$ (chain); parallel internal lines are distinct summed flavors. Their pure third-chaos tensors are obtained by the exact Wick subtraction described in the text.}
\label{fig:w4degree3}
\end{figure}

At the next order the three connected degree-three skeletons shown in Figure~\ref{fig:w4degree3} enter. Their index contractions can be stated without reference to the diagrams. For each unordered pair-splitting $p|q$ of $X$, each two-element subset $p\subset X$ with ordered complement $q=(q_0,q_1)$, and each $s\in X$ with ordered complement $t=(t_0,t_1,t_2)$, define
\begin{align}
 P_{{\rm lad},X}
 &= \sum_{p|q}\eta(p,q)\,\frac14\sum_{abcd}
       J_{p_0p_1ab}J_{abcd}J_{q_0q_1cd}, \nonumber\\
 P_{{\rm tri},X}
 &= \sum_{\substack{p\subset X\\ |p|=2}}\eta(p,q_0,q_1)\,\frac12\sum_{abcd}
       J_{p_0p_1ab}J_{q_0acd}J_{q_1bcd}, \nonumber\\
 P_{{\rm ch},X}
 &= \sum_{s\in X}\eta(s,t)\,\frac16\sum_{abcd}
       J_{sabc}J_{abcd}J_{dt_0t_1t_2}.
 \label{eq:w4degree3polys}
\end{align}
Here every $J$ is the totally antisymmetric tensor of Appendix~\ref{app:conventions}; repeated indices therefore vanish. The ladder splits $X$ into two even blocks, for which $\eta$ is the well-defined sign of an unordered partition. The triangle and the chain do not: their blocks $(p, q_0, q_1)$ and $(s, t)$ are odd, the identity \eqref{eq:etaproduct} no longer holds block by block, and $\eta$ becomes order dependent, so the arguments displayed above are the ordered ones (Appendix~\ref{app:conventions-eta}). Having fixed that ordering, each odd-block class still carries one further sign --- the parity of the full ordered Majorana string, internal Wick pairings included, which for odd blocks is no longer the external-block parity alone (Appendix~\ref{app:conventions-eta}) --- and this sign is constant within the class. We absorb it into the definition of the kernel:
\begin{align}
 K_{\rm lad}&=\int d^3\tau\,
 G(\tau_1)^2G(\tau_1-\tau_2)^2G(\tau_2-\tau_3)^2G(\tau_3)^2,\nonumber\\
 K_{\rm tri}&=-\int d^3\tau\,
 G(\tau_1)^2G(\tau_2)G(\tau_3)G(\tau_1-\tau_2)G(\tau_1-\tau_3)
 G(\tau_2-\tau_3)^2,\nonumber\\
 K_{\rm ch}&=-\int d^3\tau\,
 G(\tau_1)G(\tau_1-\tau_2)^3G(\tau_2-\tau_3)G(\tau_3)^3,
 \label{eq:w4degree3kernels}
\end{align}
where each integral is over $[0,\beta]^3$ and differences use the antiperiodic continuation. Thus the complete degree-three contribution is
\beq
 \xi_X^{(3)}=-4\left(K_{\rm lad}\no{P_{{\rm lad},X}}
 +K_{\rm tri}\no{P_{{\rm tri},X}}+K_{\rm ch}\no{P_{{\rm ch},X}}\right).
 \label{eq:w4degree3contribution}
\eeq
The high-temperature limits are $K_{\rm lad}\to\beta^3/256$,
$K_{\rm tri}\to-\beta^3/768$, and $K_{\rm ch}\to+\beta^3/768$. The explicit minus signs in the definitions of $K_{\rm tri}$ and $K_{\rm ch}$ are the additional internal-pairing parities described in Appendix~\ref{app:conventions-eta}; the plain antiperiodically continued integrals, with only the ordered external-block sign retained in the coupling polynomial, would have the opposite signs.

The two signs are fixed by the exact third-order trace,
which also confirms the rest of \eqref{eq:w4degree3contribution}. Since the
$\beta^3$ term of the numerator of $\xi_X$ is $-\langle \mu_X H^3\rangle_0/6$,
the coefficient of $P_{a,X}$ in the exact cubic moment must be $24 k_a$, where
$K_a(\beta \to 0) = k_a \beta^3$. Regressing the exact moment on the three
structures --- together with $J_X \sum_A J_A^2$, the one coincident-coupling
structure at this order, which the division by $Z$ and the melonic dressing
remove --- gives
\beq
 \big\langle \mu_X H^3 \big\rangle_0
 = \tfrac{3}{32} P_{{\rm lad},X} - \tfrac{1}{32} P_{{\rm tri},X}
 + \tfrac{1}{32} P_{{\rm ch},X} + \tfrac{3}{64}\, J_X \sum_A J_A^2 ,
 \label{eq:w4cubictrace}
\eeq
with coefficient of determination $1$ to machine precision (dense traces at
$N = 8$ and $N = 10$, all quartets, several realizations;
\texttt{numerics-v3/case\_w4/check\_w4\_cubic.py}). The first three
coefficients are exactly $24 k_a$, so all three connected coefficients of
\eqref{eq:w4degree3contribution} --- including both odd-block signs --- are
determined with no reference to the melonic saddle or to any fit.

Each raw cubic polynomial contains a first-chaos component proportional to $J_X$. For a homogeneous cubic polynomial the projection is finite and exact,
\beq
    \no{P^{(3)}_a}
    = P^{(3)}_a - \frac{\sigma_J^2}{2}\,\Delta_J P^{(3)}_a
    = P^{(3)}_a - \sigma_J^2 c_a(N) J_X ,
    \qquad a\in\{\mathrm{lad},\mathrm{tri},\mathrm{ch}\},
\eeq
where $\Delta_J = \sum_A \partial^2/\partial J_A^2$ is the Laplacian in coupling space, so that $\tfrac{\sigma_J^2}{2}\Delta_J$ contracts one Wick pair; for a single coupling it reproduces the elementary identity $\no{J^3} = J^3 - 3\sigma_J^2 J$. Ensemble covariance permits no other linear rank-four tensor. Direct contraction gives the closed forms
\beq
 c_{\rm lad}=3N^2-15N+21,\qquad
 c_{\rm tri}=3N^2-3N-18,\qquad
 c_{\rm ch}=\frac{-2N^3+12N^2-34N+36}{3}.
 \label{eq:w4degree3projection}
\eeq
No ED data enter this projection. The numerical implementation evaluates
\eqref{eq:w4degree3polys} and subtracts \eqref{eq:w4degree3projection}, with
the closed forms independently checked in the test suite against explicit
coupling-space Laplacians at several sizes, and against different choices of
the external quartet. The resulting Wick-ordered tensors are exactly
orthogonal to $J_X$ under the disorder average. They are a further factor
$N^{-1/2}$ below $T^{(2)}$, so at the sizes reachable by exact diagonalization
they sit at or below the scatter floor left by the degree-two term.

The ED comparison bears this out order by order (Figure~\ref{fig:w4order3}, first three panels, in the representative $\beta J = 2$ dataset): adding the degree-two tensor raises the parameter-free identity-line statistic from $R^2_{\rm id}=0.890$ to $0.982$, and the degree-three skeletons tighten it further to $0.986$, while the through-origin slope remains at its finite-$N$ excess ($1.12$). Thus the fluctuation tensors and the systematic slope are corrected by different objects, exactly as the orthogonality argument requires. Removing the slope is the business of the next subsection (fourth panel). As a robustness check on these parameter-free amplitudes, Appendix~\ref{sec:corrections-fit} asks how much is gained by fitting one global coefficient per structure instead; the answer is little, so most of the explanatory power lies in identifying the structures rather than tuning them.

\subsection[The slope: the resummed rung at weight 4]{The slope: the resummed rung at weight $4$}
\label{sec:corrections-w4slope}

The slope excess cannot come from any fluctuation tensor: only a contribution aligned with $J_X$ can move the coefficient. By the Stein identity of the chaos decomposition (Appendix~\ref{sec:rules-chaos}, Eq.~\eqref{eq:projection}), the ensemble linear-chaos coefficient is the imaginary-time-integrated operator autocorrelation,
\beq
\alpha_X = \frac{\mathbb{E}[J_X\xi_X]}{\sigma_J^2}
= \mathbb{E}\!\left[\frac{\partial \xi_X}{\partial J_X}\right]
= -\frac14 \int_0^\beta d\tau\; \mathbb{E}\big[\langle \mu_X(\tau)\, \mu_X(0)\rangle_c\big] .
\label{eq:w4slopesusc}
\eeq
Its factorized part, in which the four Majoranas propagate independently, is $16\,G(\tau)^4$, and $-\tfrac14\int 16\,G^4 = -4 I_4$ reproduces the leading coefficient exactly. The slope \emph{excess} is therefore controlled by
\beq
    \Delta C(\tau)
    = \mathbb{E}\big[\langle \mu_X(\tau)\mu_X(0)\rangle_c\big]
      - 16\,G(\tau)^4 :
\eeq
the four fermions remain more jointly correlated than independent propagation predicts, and the integral of that enhancement gives the excess.

\begin{figure}[t]
\centering
\begin{tikzpicture}[line width=0.8pt, every node/.style={font=\small}]
  \draw (0,0.6) -- (2.2,0.6);
  \draw (0,-0.6) -- (2.2,-0.6);
  \fill (1.1,0.6) circle (2.2pt);
  \fill (1.1,-0.6) circle (2.2pt);
  \draw (1.1,0.6) .. controls (0.9,0.2) and (0.9,-0.2) .. (1.1,-0.6);
  \draw (1.1,0.6) .. controls (1.3,0.2) and (1.3,-0.2) .. (1.1,-0.6);
  \node[above] at (0.45,0.6) {$i$};
  \node[above] at (1.75,0.6) {$j$};
  \node[below] at (0.45,-0.6) {$i$};
  \node[below] at (1.75,-0.6) {$j$};
  \node[left] at (0.93,0) {\scriptsize $u$};
  \node[right] at (1.27,0) {\scriptsize $v$};
  \node at (2.85,0) {$+$};
  \draw (3.5,0.6) -- (6.9,0.6);
  \draw (3.5,-0.6) -- (6.9,-0.6);
  \foreach \x in {4.5,5.9}{
    \fill (\x,0.6) circle (2.2pt);
    \fill (\x,-0.6) circle (2.2pt);
    \draw (\x,0.6) .. controls (\x-0.2,0.2) and (\x-0.2,-0.2) .. (\x,-0.6);
    \draw (\x,0.6) .. controls (\x+0.2,0.2) and (\x+0.2,-0.2) .. (\x,-0.6);
  }
  \node[above] at (3.95,0.6) {$i$};
  \node[above] at (5.2,0.6) {$k$};
  \node[above] at (6.45,0.6) {$j$};
  \node[below] at (3.95,-0.6) {$i$};
  \node[below] at (5.2,-0.6) {$k$};
  \node[below] at (6.45,-0.6) {$j$};
  \node at (7.75,0) {$+\;\cdots\;=$};
  \draw (8.5,0.6) -- (9.25,0.28);
  \draw (8.5,-0.6) -- (9.25,-0.28);
  \draw (10.25,0.28) -- (11.0,0.6);
  \draw (10.25,-0.28) -- (11.0,-0.6);
  \draw[fill=gray!18,rounded corners=2pt] (9.25,-0.38) rectangle (10.25,0.38);
  \node at (9.75,0) {$\mathcal R$};
  \node[above] at (8.45,0.6) {$i$};
  \node[below] at (8.45,-0.6) {$i$};
  \node[above] at (11.05,0.6) {$j$};
  \node[below] at (11.05,-0.6) {$j$};
\end{tikzpicture}
\caption{The rung and its resummation. Left: the elementary rung --- a single coupling $J_{ijuv}$ shared between two vertices (dots), its internal pair $u,v$ summed over $\binom{N-2}{2}$ values; the pair of lines enters with common flavor $i$ and leaves with common flavor $j$. Middle: iterating the rung with the intermediate pair flavor $k$ summed. Each added rung costs $\sigma_J^2 \sim N^{-3}$, repaid exactly by its internal pair (${\sim}\,N^2$) and the summed flavor $k$ (${\sim}\,N$), so every term of the ``necklace'' tower is the same relative order $1/N$. Right: the geometric sum defines the effective four-leg rung $\mathcal R=K(1-K)^{-1}$ (shaded box), whose integrated amplitude relative to the single rung defines $\kappa_L(\beta)$. The box acts on an antisymmetric pair but does not specify a linkage of its external fermion legs; contracting it with the two terms of $F_0$ produces the direct and exchanged routings. In the collective language of Appendix~\ref{sec:rules}, $\mathcal R$ is the external-leg-dressed $\sigma$ propagator.}
\label{fig:rungresum}
\end{figure}
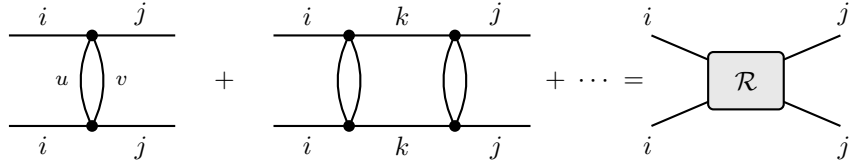

\paragraph{The ladder sum.} At leading order the connected correlator is produced by a single coupling shared between two interaction vertices, a \emph{rung} joining two of the four Majorana lines (the diagram of Section~\ref{sec:corrections-types}). The single rung, however, does not act alone at this order. Repeating it along the pair of lines, with the intermediate pair flavor \emph{summed}, costs nothing (Figure~\ref{fig:rungresum}): each added rung brings one more factor of $\sigma_J^2 \sim 1/N^3$, which is repaid by its two new internal lines (${\sim}\,N^2$) and the freely summed intermediate flavor (${\sim}\,N$). Every term of this ladder tower is therefore of the \emph{same} relative order $1/N$ and the whole geometric series must be summed at once. Doing so on the melonic saddle produces the effective four-leg rung $\mathcal R \equiv K(1-K)^{-1}$, shown as the shaded box in Figure~\ref{fig:rungresum}, and multiplies the integrated single-rung amplitude by a temperature-dependent factor $\kappa_L(\beta)$. The box acts on an antisymmetric fermion pair but does not prescribe how its four external legs are linked. Terminating it on the free pair propagator $F_0(12;34)=-G(\tau_{13})G(\tau_{24})+G(\tau_{14})G(\tau_{23})$ gives the connected pair kernel $\mathcal L=\mathcal R F_0$; the two terms of $F_0$ are the direct and exchanged routings. (In the collective language of Appendix~\ref{sec:rules}, $\mathcal R$ is the external-leg-dressed $\sigma$ propagator, and the geometric sum uses the Bethe--Salpeter kernel $K = -(p-1)J^2\, G(\tau_{13})G(\tau_{24})G(\tau_{34})^{p-2}$.) The complete derivation and its direct translation into the numerical implementation are given in Appendix~\ref{app:w4-oneloop}.

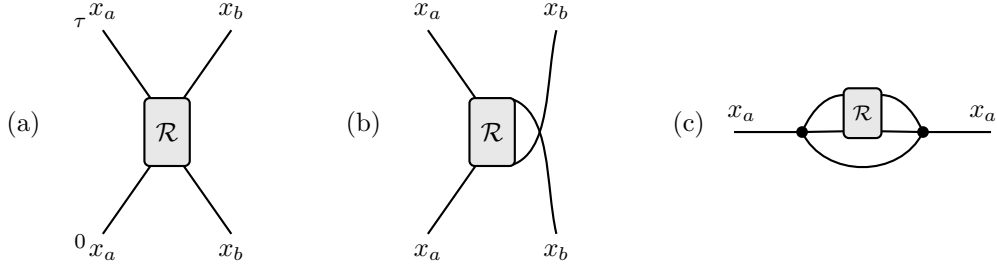
\begin{figure}[t]
\centering
\begin{tikzpicture}[line width=0.8pt, every node/.style={font=\small}]
  \node at (-1.9,0.1) {(a)};
  \node at (-1.15,1.45) {\scriptsize $\tau$};
  \node at (-1.15,-1.45) {\scriptsize $0$};
  \draw (-0.85,1.35) -- (-0.22,0.45);
  \draw (0.85,1.35) -- (0.22,0.45);
  \draw (-0.85,-1.35) -- (-0.22,-0.45);
  \draw (0.85,-1.35) -- (0.22,-0.45);
  \draw[fill=gray!18,rounded corners=2pt] (-0.30,-0.45) rectangle (0.30,0.45);
  \node at (0,0) {$\mathcal R$};
  \node[above] at (-0.85,1.35) {$x_a$};
  \node[above] at (0.85,1.35) {$x_b$};
  \node[below] at (-0.85,-1.35) {$x_a$};
  \node[below] at (0.85,-1.35) {$x_b$};
  \node at (2.6,0.1) {(b)};
  \draw (3.45,1.35) -- (4.08,0.45);
  \draw (3.45,-1.35) -- (4.08,-0.45);
  \draw (4.52,-0.45) .. controls (5.1,-0.3) and (5.0,0.7) .. (5.15,1.35);
  \draw (4.52,0.45) .. controls (5.1,0.3) and (5.0,-0.7) .. (5.15,-1.35);
  \draw[fill=gray!18,rounded corners=2pt] (4.0,-0.45) rectangle (4.6,0.45);
  \node at (4.3,0) {$\mathcal R$};
  \node[above] at (3.45,1.35) {$x_a$};
  \node[below] at (3.45,-1.35) {$x_a$};
  \node[above] at (5.15,1.35) {$x_b$};
  \node[below] at (5.15,-1.35) {$x_b$};
  \node at (6.9,0.1) {(c)};
  \draw (7.5,0) -- (8.4,0);
  \draw (8.4,0) .. controls (8.55,0.34) and (8.75,0.49) .. (8.95,0.49);
  \draw (9.45,0.49) .. controls (9.65,0.49) and (9.85,0.34) .. (10.0,0);
  \draw (8.4,0) -- (8.95,0.01);
  \draw (9.45,0.01) -- (10.0,0);
  \draw (8.4,0) .. controls (8.7,-0.62) and (9.7,-0.62) .. (10.0,0);
  \fill (8.4,0) circle (2.2pt);
  \fill (10.0,0) circle (2.2pt);
  \draw[fill=gray!18,rounded corners=2pt] (8.95,-0.08) rectangle (9.45,0.58);
  \node at (9.2,0.25) {\scriptsize $\mathcal R$};
  \draw (10.0,0) -- (10.9,0);
  \node[above] at (7.6,0) {$x_a$};
  \node[above] at (10.8,0) {$x_a$};
\end{tikzpicture}
\caption{Contractions of the effective four-leg rung $\mathcal R$ in the autocorrelation $\langle \mu_X(\tau)\mu_X(0)\rangle$, $X = \{x_a x_b x_c x_d\}$; top legs sit at time $\tau$, bottom legs at $0$, and the two spectator lines, contributing $G(\tau)^2$, are not drawn. The shaded box acts on an antisymmetric fermion pair and does not itself select a linkage. (a) The direct term of $F_0$, giving the kernel $D$ in \eqref{eq:w4rung}. (b) The exchanged term, giving $-\mathcal{E}$; the difference $D-\mathcal{E}$ vanishes at $\tau=0,\beta$, as required by $\mu_X^2=1$. (c) Insertion of $\mathcal R$ into the melonic self-energy. Both direct and exchanged fermion routings are implicit; the panel denotes the complete one-loop propagator shift $\mathbb{E}[G]=G_*+\delta G/N$, which gives the channel-(S) factor $\kappa_S$, rather than one selected bare linkage.}
\label{fig:runglandings}
\end{figure}

\paragraph{Where the rung lands.} The effective rung can be contracted with the surrounding fermion lines through the two terms of $F_0$, shown in Figure~\ref{fig:runglandings}(a,b) as the direct and exchanged routings. When it is inserted into a self-energy subdiagram as in panel (c), both routings are implicit; the panel denotes the resulting one-loop propagator shift, not one selected bare linkage.

Across a pair of lines the two spectator legs contribute $G(\tau)^2$, and the pair-channel contribution is
\beq
\Delta C_L(\tau) = \frac{16\binom{4}{2}}{N-1}\,G(\tau)^2\,
\mathcal L(\tau,0;\tau,0),
\qquad \mathcal L = K(1-K)^{-1}F_0 ,
\label{eq:w4rung}
\eeq
This is the complete pair channel at collective order $\ell=1$: every term of the necklace tower is the same order, so nothing is truncated. The prefactor uses $\binom{N-2}{2}\sigma_J^2=3J^2/(N-1)$, with $\binom{4}{2}$ counting the choice of pair.
The single rung has $[KF_0](\tau,0;\tau,0)=3J^2[D(\tau)-\mathcal E(\tau)]$, with direct and exchange kernels
\beq
D = G^2 * G^2 * G^2, \qquad \mathcal{E}(\tau) = \int_0^\beta d\sigma_1 d\sigma_2\; G(\tau{-}\sigma_1)G(\sigma_1)\,G(\tau{-}\sigma_2)G(\sigma_2)\,G(\sigma_1{-}\sigma_2)^2 .
\eeq
At single-rung order, the direct attachment (panel a) gives $D$; the exchanged attachment (panel b) gives $-\mathcal{E}$. Their difference vanishes at $\tau = 0, \beta$, enforcing $\Delta C_L(0) = 0$ from $\mu_X^2 = 1$, and turns the direct kernel's edge-peaked profile into the $\beta/2$-peaked dome that is observed. This is exactly the pair-channel object that was cancelled by parity in the one-point analysis but survives here, where the autocorrelation carries no $J_X$ factor. Resummation changes the time-dependent shape; $\kappa_L$ is the ratio only after integration against $G^2$, as defined in Appendix~\ref{app:w4-oneloop}. At single-rung order the induced fractional correction to the leading coefficient is
\beq
\delta_{\rm rung}(N,\beta) = \frac{1}{16\, I_4}\int_0^\beta \Delta C_L(\tau)\, d\tau \;\Big|_{\mathcal L=KF_0} = \binom{4}{2}\binom{N-2}{2}\sigma_J^2 \cdot \frac{\int_0^\beta G^2 (D-\mathcal{E})}{I_4} \;=\; O(1/N) .
\eeq

On a single line the story starts one step later. A rung with both ends on the same line is forced, at a single Wick pair, to be the melon itself --- the coupling shares one index with the line, and its remaining three legs close into the three internal lines of the sunset --- and the melon is already resummed into $G$; with the exact-count variance \eqref{eq:leadingsigma} it is exact, leaving nothing behind. The first genuine single-line effect therefore appears at two coupling pairs: the effective rung decorating the \emph{interior} of the melon (panel c), with both fermion routings implicit, which shifts the disorder-averaged propagator, $\mathbb{E}[G] = G_* + \delta G/N$. Feeding $\delta G$ back through the leading kernel --- any of the four lines can be decorated, counted automatically by $\delta(G^4) = 4 G^3\, \delta G$ --- renormalizes the coefficient by a further factor $\kappa_S(\beta)$. It starts at $O((\beta J)^2)$ and is negative, partly cancelling the pair channel.

\paragraph{Why the list is complete.} Every correction at this order is built from one \emph{elementary} Wick-paired coupling added to the leading diagram --- the necklace continuations of Figure~\ref{fig:rungresum} iterate that elementary object without changing its order, and have already been summed into $\mathcal R$ --- and it is classified by how many of the four indices of that coupling lie in $X$. One shared index is the melon: already inside $G$, and exact. Two shared indices is the rung: the direct and exchanged attachments of Figure~\ref{fig:runglandings} together with the necklace tower of Figure~\ref{fig:rungresum}. Three shared indices tie three external lines to a single summed internal flavor, of coherent size $N\sigma_J^2 \sim N^{-2}$ --- a full order below the rung --- and four shared indices is smaller still. On a single line the one-pair topology is the melon, so single-line effects begin at two pairs, panel (c). Finally, no tadpole diagram appears anywhere in this list: a coupling pair cannot hang a disconnected closed loop off an external line, because the second vertex carries the same four flavors as the first and is dragged onto the same line, reproducing the melon (or else decouples into a vacuum bubble that the normalization removes). In the collective formalism of Appendix~\ref{sec:rules} the same statement reappears as the saddle-point cancellation of tadpoles. The one-loop slope correction is therefore exactly the sum of two channels at the stated order: the pair channel (L) --- the resummed rung, factor $\kappa_L$ --- and the single-line channel (S) --- the propagator shift, factor $\kappa_S$.

Both factors are computed once on the numerical saddle (Appendix~\ref{app:numerics}) and are collected in Table~\ref{tab:w4pred}. The full parameter-free slope correction is
\beq
\xi_X^{\rm slope} = -4\, I_4\,\big(1 + \delta_{\rm slope}\big)\, J_X, \qquad \delta_{\rm slope}(N,\beta) = \big[\kappa_L(\beta) + \kappa_S(\beta)\big]\, \delta_{\rm rung}(N,\beta) .
\label{eq:w4slopecorr}
\eeq
It is worth noting that the two factors in \eqref{eq:w4slopecorr} are of different status. The bare amplitude $\delta_{\rm rung}$ is exact at finite $N$: its index count is $\binom{4}{2}\binom{N-2}{2}\sigma_J^2 = 18J^2/(N-1)$, with no expansion involved. The resummation factors $\kappa_L$ and $\kappa_S$, by contrast, are the $N \to \infty$ one-loop values, evaluated on the melonic saddle; their own finite-$N$ corrections belong to the next collective order and are not retained. The product is therefore accurate to relative order $1/N$ within the $\ell = 1$ kernel, which is the accuracy claimed, and the residuals seen below at small $N$ and low temperature are of the expected two-loop size.

\begin{table}[t]
\centering
\begin{tabular}{c cc c cc c c}
\toprule
$\beta J$ & $I_4$ & $W^{(2)}$ & $\delta_{\rm rung}$ & $\kappa_L$ & $\kappa_S$ & $\kappa_L+\kappa_S$ & $\delta_{\rm slope}$ \\
\midrule
$0.5$ & $0.0306$ & $0.0038$ & $0.0096$ & $\phantom{-}0.991$ & $-0.021$ & $0.970$ & $0.0094$ \\
$1.0$ & $0.0579$ & $0.0139$ & $0.0363$ & $\phantom{-}0.969$ & $-0.071$ & $0.898$ & $0.0326$ \\
$1.5$ & $0.0802$ & $0.0278$ & $0.0746$ & $\phantom{-}0.944$ & $-0.131$ & $0.813$ & $0.0606$ \\
$2.0$ & $0.0975$ & $0.0428$ & $0.1193$ & $\phantom{-}0.921$ & $-0.187$ & $0.735$ & $0.0877$ \\
$3.0$ & $0.1205$ & $0.0713$ & $0.2140$ & $\phantom{-}0.892$ & $-0.272$ & $0.620$ & $0.1326$ \\
\bottomrule
\end{tabular}
\caption{Parameter-free $W=4$ predictions of the current rules, evaluated on the melonic saddle. Kernels $I_4$, $W^{(2)}$ are temperature-only; the bare rung amplitude $\delta_{\rm rung}$ and the assembled slope correction $\delta_{\rm slope}$ are quoted at $N = 20$ (both scale as $1/(N-1)$). The resummed-ladder factor $\kappa_L$ (channel L) and the self-energy factor $\kappa_S$ (channel S) are the $N\to\infty$ one-loop factors derived in Appendix~\ref{app:w4-oneloop}; the predicted slope excess is $\delta_{\rm slope} = (\kappa_L+\kappa_S)\,\delta_{\rm rung}$. The self-energy channel grows from a $2\%$ trim at high temperature to a $30\%$ reduction of the ladder by $\beta J = 3$.}
\label{tab:w4pred}
\end{table}

The two channels together determine how much of the bare rung survives: from Table~\ref{tab:w4pred}, $\kappa_L+\kappa_S$ falls from $0.97$ at $\beta J = 0.5$ to $0.62$ at $\beta J = 3$. At weak coupling the ladder is essentially the whole story and the bare rung is nearly exact; at strong coupling the negative self-energy channel removes a third of it. This is the sharpest content of the current rules relative to the single-rung treatment: the slope correction is not $\delta_{\rm rung}$ but $(\kappa_L+\kappa_S)\,\delta_{\rm rung}$, and both factors are computed from first principles.

\begin{figure}[t]
\centering
\includegraphics[width=0.76\textwidth]{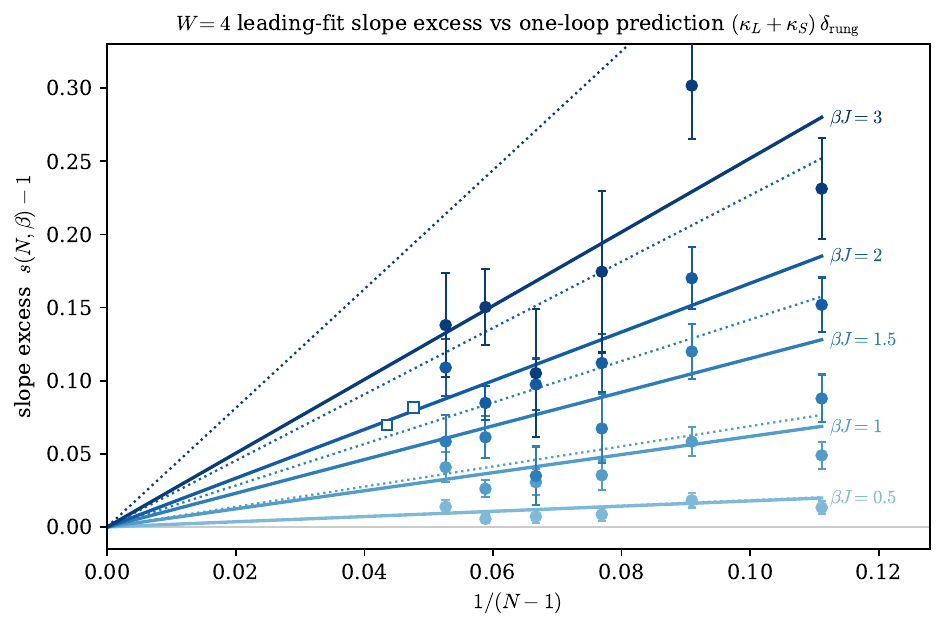}
\caption{The $W=4$ leading-fit slope excess $s(N,\beta) - 1$ against $1/(N-1)$, compared with the parameter-free one-loop prediction $\delta_{\rm slope} = (\kappa_L + \kappa_S)\,\delta_{\rm rung}$ (solid lines) and the bare-rung normalization $\delta_{\rm rung}$ (dotted); shading orders the temperatures ($\beta J = 0.5, 1, 1.5, 2, 3$). Filled circles: pooled through-origin slopes from the dense one-point dataset ($N = 10$--$20$; $20$ seeds at $\beta J = 0.5, 1, 2$ and $10$ at $1.5, 3$; bootstrap errors over seeds). Open squares: the large-$N$ push at $\beta J = 2$ ($N = 22, 24$, $160$ seeds). The prediction is linear in $1/(N-1)$ by construction; within its larger operator-sampling errors this estimator is consistent with the resummed prediction, while the spectrum-only estimator of Figure~\ref{fig:w4anchor} provides the decisive comparison.}
\label{fig:w4slopecheck}
\end{figure}

Figure~\ref{fig:w4slopecheck} makes the comparison against exact diagonalization. Because $\delta_{\rm rung} \propto 1/(N-1)$ exactly, the prediction at each temperature is a straight line through the origin in $1/(N-1)$, with slope fixed by the kernels --- nothing is fitted. Despite the sampling noise of the operator-resolved estimator, the measured through-origin slope excess is consistent with the resummed line across $\beta J = 0.5$--$3$ and $N = 10$--$24$, and the large-$N$ $\beta J=2$ data favor it over the bare-rung normalization. The sharper spectrum-only test below resolves the factor $\kappa_L + \kappa_S$ without operator-sampling noise. Applied operator by operator, the same correction gives the fourth panel of Figure~\ref{fig:w4order3}: at $\beta J = 2$ it moves the pooled slope from $1.126$ to $1.007$ while leaving the collapsed scatter nearly unchanged.

\subsection{The energy anchor: an exact test of the one-loop coefficient}
\label{sec:corrections-energy}

The aggregate slope obeys an exact identity that fixes the same correction thermodynamically. Since $H = \sum_A (J_A/4)\mu_A$, one has $\sum_{|X|=4} J_X \mu_X = 4H$, so summing the one-point functions against their own couplings gives the thermal energy,
\beq
\sum_{|X|=4} J_X\, \xi_X = 4\,\langle H\rangle_\beta .
\eeq
The through-origin slope is therefore identically $\langle H\rangle_\beta / \langle H\rangle_{\rm mel}$, with the melonic value $\langle H\rangle_{\rm mel} = -I_4 \sum_X J_X^2$. The slope excess is thus the finite-$N$ correction to the thermal energy, or equivalently the $\beta$ derivative of the corresponding correction to $-\ln Z$. Its leading $1/N$ piece is the Gaussian fluctuation about the melonic $(G,\Sigma)$ saddle, $+\partial_\beta F_1$ with $F_1 = \tfrac12 \operatorname{Tr}\big[\log(1-K) + K\big]$. This is the same ladder-plus-self-energy object as \eqref{eq:w4slopecorr}, now summed over all operators: an aggregate check of $\kappa_L + \kappa_S$ that does not require resolving individual operators.

The $+K$ inside $F_1$ deserves a comment, since it is what makes the ring sum start at \emph{two} rungs. One has $\tfrac12\operatorname{Tr} K = -\tfrac34 J^2 \beta I_4$, so the one-rung term of the raw determinant is $-\tfrac12\operatorname{Tr}K = +\tfrac34J^2\beta I_4$; this is cancelled exactly by the counterterm $-N(\hat{J}^2 - J^2)\!\iint\!G_*^4/8 \to -\tfrac34J^2\beta I_4$, where $\hat{J}^2 = \sigma_J^2 N^3/6$ is the collective-normalization coupling of \eqref{eq:jhat}, tracking the difference between the exact-count and more standard large-$N$ variance conventions (we verify the discretized identity to the expected $O(1/M)$ in the grid size $M$ of Appendix~\ref{app:w4-oneloop-check}). Equivalently, the one-rung ring is the coherent single-Wick-pair vacuum term, already spent in the normal ordering of the instance vertices; counting it again in the collective sector would double count it.

\begin{figure}[t]
\centering
\includegraphics[width=0.76\textwidth]{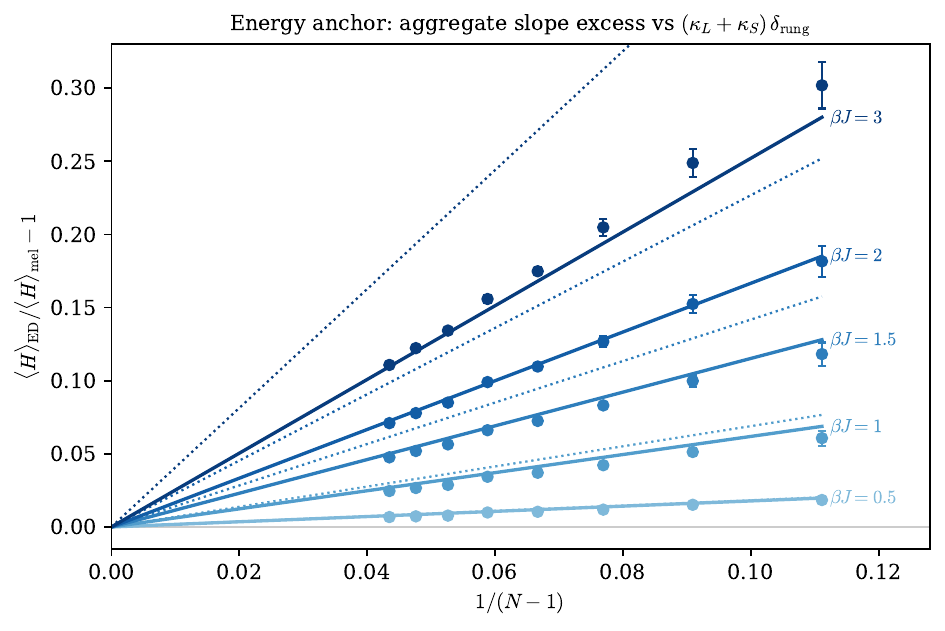}
\caption{The energy anchor: the aggregate slope excess $\langle H\rangle_{\rm ED}/\langle H\rangle_{\rm mel} - 1$, which by the exact identity $\sum_X J_X \xi_X = 4\langle H\rangle_\beta$ equals the $J_X$-weighted through-origin slope over \emph{all} weight-four operators, measured from the ED spectrum alone ($N = 10$--$24$, $20$ seeds, errors are seed-to-seed SE). Solid lines: the parameter-free one-loop prediction $(\kappa_L+\kappa_S)\,\delta_{\rm rung}$; dotted: bare-rung normalization. The spectrum-only estimator removes all operator-sampling noise, so the data cleanly resolve the resummation factor: at $\beta J = 2$, $N = 24$ the measurement is $0.0710(9)$ vs.\ $0.0724$ predicted ($0.0986$ bare).}
\label{fig:w4anchor}
\end{figure}

It is also the sharpest numerical test available, because $\langle H\rangle_\beta$ needs only the spectrum and is operator-scatter free. Figure~\ref{fig:w4anchor} shows $\langle H\rangle_{\rm ED}/\langle H\rangle_{\rm mel} - 1$ across the full range: the data follow the parameter-free line $(\kappa_L + \kappa_S)\,\delta_{\rm rung}$ at every temperature, with statistical errors at the $10^{-3}$ level by $N = 24$. At $\beta J = 2$, $N = 24$ the measured excess is $0.0710(9)$ against the predicted $0.0724$, while the bare-rung value $0.0986$ lies far outside the statistical error bars, thus showing that the thermodynamic anchor resolves not just the rung but the resummation factor on top of it. The percent-level residuals that remain (visible mostly at $\beta J = 3$ and small $N$) are the expected size of the two-loop terms. The identification of the curve is also confirmed on the theory side: evaluating $+\partial_\beta F_1$ directly from the eigenvalues of the Bethe--Salpeter kernel on the saddle (\texttt{numerics-v3/case\_w4/check\_trlog.py}, Richardson-extrapolated in the grid) reproduces $(\kappa_L + \kappa_S)\,\delta_{\rm rung}\,\langle H\rangle_{\rm mel}$ to better than one percent at every temperature, while the ladder-only normalization $\kappa_L\,\delta_{\rm rung}$ is off by $30\%$ at $\beta J = 3$. This means the two routes to the one-loop energy agree, verifying the normalization bookkeeping used in the channel split.

\section{Generalization to time-dependent correlations}
\label{sec:time-dep}

So far every Majorana of the external string sat at the same imaginary time. The rules extend with almost no change to \emph{time-dependent} insertions, in which the Majoranas are placed at distinct points on the thermal circle. The object of interest is the generalized one-point function $\xi_X(\tau_1, \ldots, \tau_W)$ defined in \eqref{eq:bilocaldef}, with insertion times $\tau_a \in [0, \beta)$, the same normalization phase as the equal-time string \eqref{eq:mu} (so $\xi_X(0, \ldots, 0) = \xi_X$), and $T_\tau$ ordering the Majoranas around the circle with the usual fermionic signs.

\subsection{Rules for general times}
\label{sec:time-dep-rules}

The diagram rules are unchanged but for one modification. Previously every external leg ran from a vertex at integrated time $\tau_v$ to the common point $\tau = 0$, contributing a propagator $G(\tau_v)$; now external leg $a$ runs from its own \emph{fixed} insertion time $\tau_a$ to the vertex it attaches to, contributing $G(\tau_a - \tau_v)$, and the $\tau_a$ are held fixed rather than integrated. Everything else is identical --- the coupling polynomials, the coefficient, the fermion signs, the linked-cluster cancellation, and the melonic dressing. Only the kernels change: each becomes a function of the insertion times, with every $\tau_a$ entering as the fixed argument of one propagator factor. For instance, the leading $W = 4$ diagram with legs at $\tau_1, \ldots, \tau_4$ carries the kernel
\begin{equation}
    \int_0^\beta d\tau\; G(\tau_1 - \tau)\, G(\tau_2 - \tau)\, G(\tau_3 - \tau)\, G(\tau_4 - \tau) ,
    \label{eq:w4timedep}
\end{equation}
one integrated vertex time and one propagator per external leg; setting all $\tau_a = 0$ recovers $I_4(\beta) = \int_0^\beta G^4$.

Two features make time-dependent insertions especially useful. First, unequal times weaken some of the equal-time selection rules of Section~\ref{sec:symmetries}. Fermion parity continues to kill every odd-weight correlator, but the antiunitary rule that forces an equal-time weight-2 string to vanish instead constrains the reality properties of its bilocal continuation, which can be nonzero at separated times; the equal-time zero survives as a property of the kernel. Second, the kernels admit analytic continuation toward real time, so the same fixed coupling polynomials can be confronted with early real-time data. We illustrate both points with the weight-2 bilocal. Our analysis will consider only the leading non-vanishing diagrams, but one could mirror the weight four discussion in Section~\ref{sec:corrections} to improve the prediction at larger $\beta J$.

\subsection[Example: weight 2, imaginary time]{Example: weight $2$, imaginary time}
\label{sec:time-dep-w2imag}

Return to $X = \{i, j\}$, now with the two Majoranas separated in imaginary time. It is convenient to use unit-normalized Majoranas
\begin{equation}
    \gamma_i = \sqrt{2}\,\chi_i,
    \qquad \{\gamma_i,\gamma_j\}=2\delta_{ij},
    \qquad
    \mu_{ij}(\tau_a, \tau_b) = i\,\gamma_i(\tau_a)\gamma_j(\tau_b),
    \qquad i \ne j ,
\end{equation}
normalized to reduce to the weight-2 string at $\tau_a = \tau_b$. The relevant diagram is exactly the bridge of Section~\ref{sec:leading-w2} --- two vertices sharing three internal labels $\{l, m, n\}$, one carrying the external index $i$ and the other $j$ --- but now with its external legs anchored at the separated times $\tau_a, \tau_b$:
\begin{center}
\begin{tikzpicture}[line width=0.8pt, every node/.style={font=\small}]
  \coordinate (ei) at (-4.2,0);
  \coordinate (v1) at (-1.5,0);
  \coordinate (v2) at (1.5,0);
  \coordinate (ej) at (4.2,0);
  \draw (ei) -- (v1);
  \draw (v2) -- (ej);
  \draw (v1) .. controls (-0.5,0.75) and (0.5,0.75) .. (v2);
  \draw (v1) -- (v2);
  \draw (v1) .. controls (-0.5,-0.75) and (0.5,-0.75) .. (v2);
  \fill (v1) circle (2.2pt);
  \fill (v2) circle (2.2pt);
  \node[left=6pt,above=3pt] at (-1.5,0.1) {$J_{ilmn},\ \sigma_1$};
  \node[right=6pt,above=3pt] at (1.5,0.1) {$J_{jlmn},\ \sigma_2$};
  \node[above] at (0,0.55) {$l$};
  \node[above] at (0,0.02) {$m$};
  \node[below] at (0,-0.58) {$n$};
  \node[left] at (ei) {$\chi_i(\tau_a)$};
  \node[right] at (ej) {$\chi_j(\tau_b)$};
\end{tikzpicture}
\end{center}
The coupling polynomial is unchanged, the off-diagonal bridge $B_{ij} = \sum_{l<m<n} J_{ilmn} J_{jlmn}$, but the kernel no longer vanishes. Writing the real-valued correlator and its prediction,
\begin{equation}
\begin{aligned}
    R_{ij}(\tau)
    &= \big\langle \gamma_i(\tau)\gamma_j(0)\big\rangle_\beta
     =-i\big\langle\mu_{ij}(\tau,0)\big\rangle_\beta
     \approx 2\, K_\beta(\tau)\, B_{ij}, \\
    K_\beta(\tau)
    &= \int_0^\beta d\sigma_1\, d\sigma_2\;
       G(\tau - \sigma_1)\, G(\sigma_1 - \sigma_2)^3\, G(\sigma_2) .
\end{aligned}
    \label{eq:w2bridge}
\end{equation}
where $G$ is continued antiperiodically. (The bilocal $\langle\mu_{ij}(\tau,0)\rangle_\beta$ is purely imaginary, so $R_{ij}(\tau)$ is real; there is no residual $1/2!$, as the two assignments of the distinct vertices $A = \{i,l,m,n\}$, $B = \{j,l,m,n\}$ to the two times cancel it.) The kernel is a double convolution on the thermal circle, $K_\beta = G * F * G$ with $F = G^3$, so it is cleanest to evaluate in Matsubara frequency as $K_\beta(i\omega_n) = G(i\omega_n)^2\, F(i\omega_n)$. It vanishes at coincidence, $K_\beta(0) = 0$ --- the equal-time zero \eqref{eq:w2zero} of the reflection argument \eqref{eq:w2kernel} --- and is nonzero away from $\tau = 0$, where the polynomial $B_{ij}$ becomes measurable.

\begin{figure}[t]
\begin{center}
\includegraphics[width=0.98\textwidth]{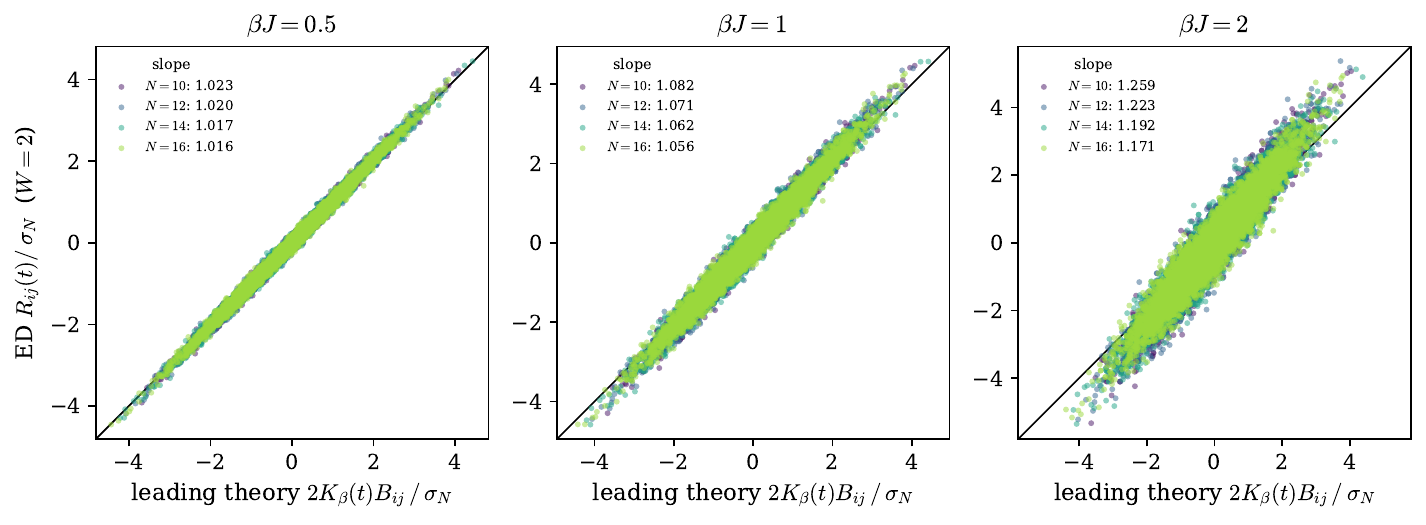}
\end{center}
\caption{ED weight-2 bilocal $R_{ij}(\tau)=\langle\gamma_i(\tau)\gamma_j(0)\rangle_\beta$ versus the leading bridge prediction $2 K_\beta(\tau) B_{ij}$ of \eqref{eq:w2bridge}, at $\beta J = 0.5, 1, 2$ (left to right), for $N = 10$--$16$ (color), pooling separations $\tau/\beta \in \{0.25, 0.5, 0.75\}$ and normalized per $N$ by the standard deviation of the prediction. Legends give the per-$N$ through-origin slope. As for the equal-time strings, the prediction is nearly exact at high temperature and degrades with $\beta J$, with the slope drifting toward unity as $N$ grows.}
\label{fig:w2bridge}
\end{figure}

\paragraph{Numerical test.} Figure~\ref{fig:w2bridge} compares the dense-ED bilocal $R_{ij}(\tau)$ against the prediction \eqref{eq:w2bridge} for $N = 10$--$16$ at $\beta J = 0.5, 1, 2$, pooling the separations $\tau/\beta \in \{0.25, 0.5, 0.75\}$ over all $\binom{N}{2}$ pairs and $40$ disorder realizations. The story mirrors the equal-time examples: the prediction is nearly exact at high temperature (through-origin slope ${\approx}1.02$ at $\beta J = 0.5$), and both the slope excess and the operator-to-operator scatter grow with $\beta J$, reaching slopes ${\approx}1.17$--$1.26$ at $\beta J = 2$. The slope again decreases toward unity with $N$ (from $1.26$ at $N = 10$ to $1.17$ at $N = 16$ at $\beta J = 2$), consistent with the leading bilocal formula becoming exact as $N \to \infty$; the underlying $\langle\mu_{ij}(\tau,0)\rangle_\beta$ is purely imaginary, $R_{ij}(\tau)$ is real, and $K_\beta(0) = 0$ holds to numerical precision, as the selection rule requires.

\begin{figure}[t]
\begin{center}
\includegraphics[width=0.72\textwidth]{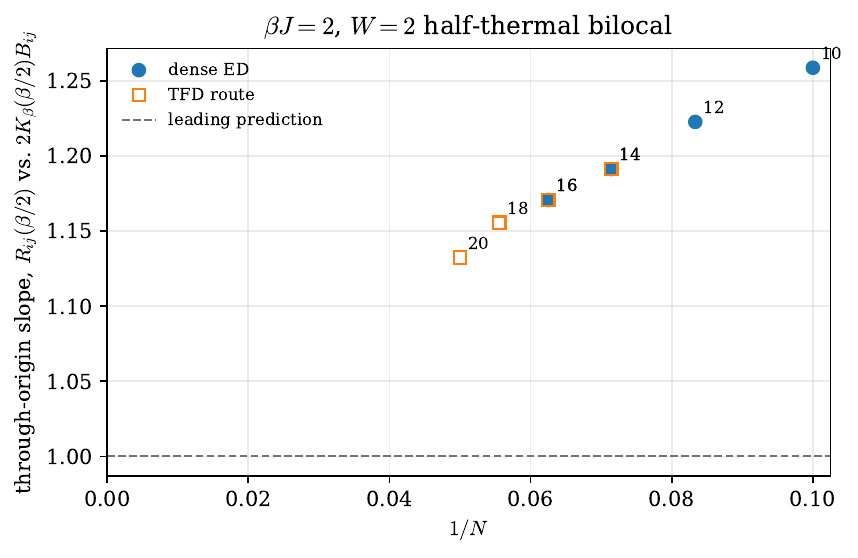}
\end{center}
\caption{Through-origin slope of the half-thermal weight-2 bilocal $R_{ij}(\beta/2)$ against the bridge prediction $2 K_\beta(\beta/2) B_{ij}$ at $\beta J = 2$, versus $1/N$. Filled circles: dense ED ($N = 10$--$16$, $40$ seeds, all pairs). Open squares: the thermofield-double route ($N = 14$--$20$, $40$ seeds, up to $120$ pairs each), which prepares $e^{-\beta H/2}$ acting on the infinite-temperature double by Krylov imaginary-time evolution and needs no eigensystem. The two routes coincide at the overlap sizes, and the slope falls monotonically toward the ideal value $1$ (dashed).}
\label{fig:w2tfd}
\end{figure}

\paragraph{Larger $N$ via the thermofield double.} The half-thermal separation $\tau = \beta/2$ admits a route to larger sizes that avoids the dense eigensystem. Writing
\beq
    R_{ij}(\beta/2) = \frac{1}{Z}\tr\!\big( e^{-\beta H/2}\, \gamma_i\, e^{-\beta H/2}\, \gamma_j \big)
\eeq
and regarding $M = e^{-\beta H/2}$ as a state on two copies of the system --- the imaginary-time-evolved infinite-temperature thermofield double --- the bilocal becomes a single inner product, $R_{ij}(\beta/2) = \langle \gamma_i M, M \gamma_j\rangle / \langle M, M\rangle$. Preparing $M$ costs one Krylov application of $e^{-\beta H/2}$ and no diagonalization, so $N = 20$ is comfortable. Figure~\ref{fig:w2tfd} shows the through-origin slope of $R_{ij}(\beta/2)$ against the prediction $2 K_\beta(\beta/2) B_{ij}$ at $\beta J = 2$ as a function of $1/N$: the dense-ED and thermofield-double routes coincide at the overlap sizes $N = 14, 16$ (same realizations, independent algorithms), and the slope continues its monotone descent toward unity, from ${\approx}1.26$ at $N = 10$ to ${\approx}1.13$ at $N = 20$. The equal-time value $R_{ij}(0)$, which the selection rule forces to vanish, is computed alongside and is zero to numerical precision.

\subsection[Example: weight 2, real time]{Example: weight $2$, real time}
\label{sec:time-dep-w2real}

Nothing in the diagram rules ties the insertion times to the imaginary axis, so we now continue the weight-2 bilocal toward real time. A natural object is the thermally regulated correlator at complex separation $\tau = \beta/2 + it$,
\beq
    R_{ij}(\beta/2 + it) = \frac{1}{Z} \tr\!\big( e^{-\beta H/2}\, \gamma_i(t)\, e^{-\beta H/2}\, \gamma_j \big),
    \qquad
    \gamma_i(t) = e^{iHt}\, \gamma_i\, e^{-iHt} ,
    \label{eq:w2realtimedef}
\eeq
the symmetric Wightman function, which reduces at $t = 0$ to the half-thermal point of Figure~\ref{fig:w2tfd} and splits the thermal weight evenly around the two insertions. This regulator is the analytically best-behaved route into real time. The function is real for any Hermitian $H$: both $e^{-\beta H/2}\gamma_i(t)e^{-\beta H/2}$ and $\gamma_j$ are Hermitian, and the trace of their product is real. For $p = 0 \bmod 4$, the antiunitary symmetry of Section~\ref{sec:symmetries} additionally makes it even in $t$. The exact-diagonalization data below confirm these constraints to machine precision.

The prediction continues along with the observable. The coupling polynomial is untouched and only the kernel must be continued, $R_{ij} \approx 2\, K(\beta/2 + it)\, B_{ij}$. The continuation cannot be performed term by term in the Matsubara sum, which diverges off the imaginary axis; instead it goes through the spectral representation. Because the kernel is a convolution in imaginary time, in frequency space it is the product $K(i\omega_n) = G(i\omega_n)^2 F(i\omega_n)$ with $F(\tau) = G(\tau)^3$. Retarded analyticity and the large-frequency falloff select the continuation $K(z) = G(z)^2 F(z)$. Both factors follow from the spectral function of the melonic propagator, obtained by solving the SD equations directly in real frequency. In particular, $F^R$ is reconstructed from the pointwise Wightman products $F^{>/<}(t)=[G^{>/<}(t)]^3$; it is not simply $[G^R]^3$. With the convention
\beq
K(z)=\int d\omega\,\frac{\rho_K(\omega)}{\omega-z},
\qquad
\rho_K(\omega)=\frac{1}{\pi}\operatorname{Im}K^R(\omega),
\eeq
the symmetric line is a single damped integral,
\beq
    K(\beta/2 + it) = \int d\omega\; \frac{\rho_K(\omega)\, e^{-i \omega t}}{2\cosh(\beta\omega/2)},
    \qquad
    \rho_K = \frac{1}{\pi} \mathrm{Im}\big[ G^R(\omega)^2 F^R(\omega) \big],
    \label{eq:w2realtimekernel}
\eeq
with the $1/\cosh$ providing exponential control; numerical details and cross-checks against the imaginary-time solver are collected in Appendix~\ref{app:numerics-spectral}.

Figure~\ref{fig:w2realtime} confronts this parameter-free prediction with exact diagonalization at $\beta J = 1$ and $2$, fitting the measured $R_{ij}(\beta/2+it)$ across all pairs to the single amplitude $c_N(t)$ multiplying $B_{ij}$. Two features stand out. First, the kernel is strongly time-dependent --- $|K(\beta/2+it)|$ \emph{grows} over $0 \le tJ \le 2$ by a factor ${\approx} 3.6$ at $\beta J = 2$ and ${\approx}12$ at $\beta J = 1$ --- and the fitted bridge component tracks this growth in full. The spectral weight $\rho_K$ has positive and negative parts and obeys the sum rule $\int d\omega\,\rho_K(\omega)=0$, so the value at $t=0$ is strongly cancelled and real-time evolution progressively relaxes that cancellation. Second, the ratio of measurement to prediction (insets) is nearly independent of $t$: it is the familiar finite-$N$ coefficient excess, ${\approx}1.20, 1.18, 1.15, 1.13$ at $N = 14, 16, 18, 20$ for $\beta J = 2$, and an already much smaller ${\approx}1.07$--$1.04$ at $\beta J = 1$, matching its imaginary-time value at $t = 0$ and decreasing with $N$ exactly as elsewhere in the paper. Thus the analytic continuation captures the time dependence of the projection onto the leading bridge tensor over the tested window.

\begin{figure}[t]
\begin{center}
\includegraphics[width=0.78\textwidth]{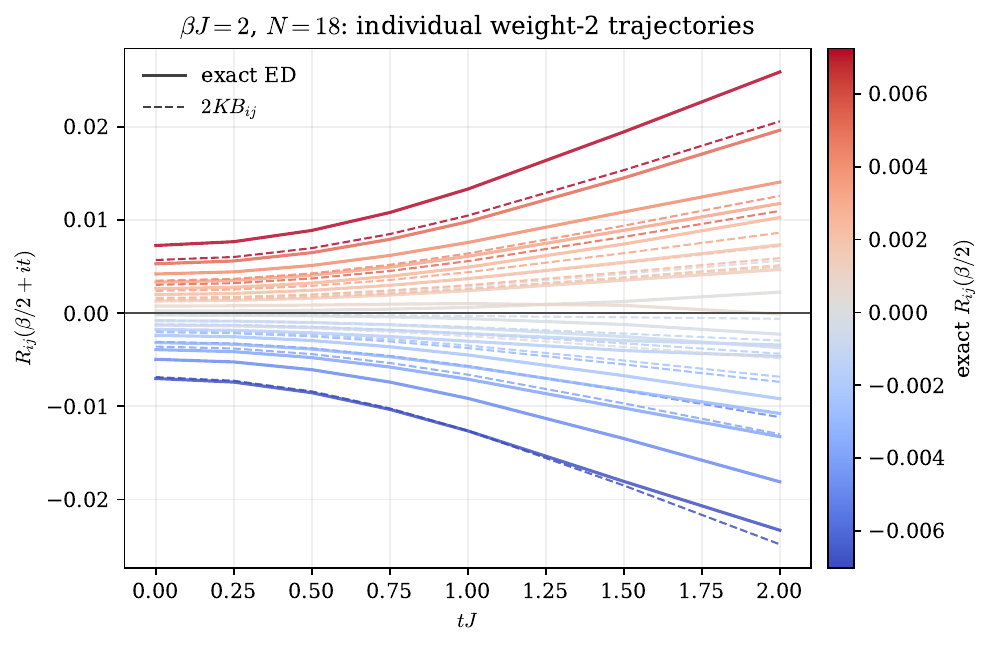}
\end{center}
\caption{Selected individual weight-2 correlator trajectories at $\beta J=2$ and $N=18$. Solid curves are exact-diagonalization values of $R_{ij}(\beta/2+it)$; dashed curves are the parameter-free leading predictions $2K(\beta/2+it)B_{ij}$ for the same disorder realization and pair. Colors encode the exact value at $t=0$, with trajectories chosen deterministically across its quantiles. The common growth is captured by the continued bridge kernel, while the systematic coefficient excess and subleading coupling structures produce the visible residuals.}
\label{fig:w2realtimetraj}
\end{figure}

This successful coefficient test does not mean that $B_{ij}$ exhausts the operator-resolved correlator. At $\beta J=2$ the bridge alone explains $R^2\simeq0.88$--$0.93$ of the variation over pairs. A representative cubic replacement tensor,
\beq
P_{ij}=\sum_{a,b,c,d,e}J_{iabc}J_{jade}J_{bcde},
\eeq
raises this to $R^2\simeq0.98$--$0.997$ when given an independently fitted coefficient (one could also compute this coefficient from the analytic continuation of a more complicated integral of propagators); its relative importance grows with $t$. Figure~\ref{fig:w2realtimetraj} makes the same point at the level of individual fermion correlators: the parameter-free bridge prediction captures the common leading motion, while finite-$N$ residuals remain visible along individual trajectories.

The spectral representation \eqref{eq:w2realtimekernel} is numerically well conditioned by the thermal damping factor. Exact diagonalization validates the continued leading coefficient over the tested window $0\le tJ\le2$. At later times, neglected coupling structures and finite-$N$ corrections may become comparable to the leading term; determining the breakdown scale requires a separate analysis as discussed in the Outlook of Section \ref{sec:outlook}.

\begin{figure}[t]
\begin{center}
\includegraphics[width=0.98\textwidth]{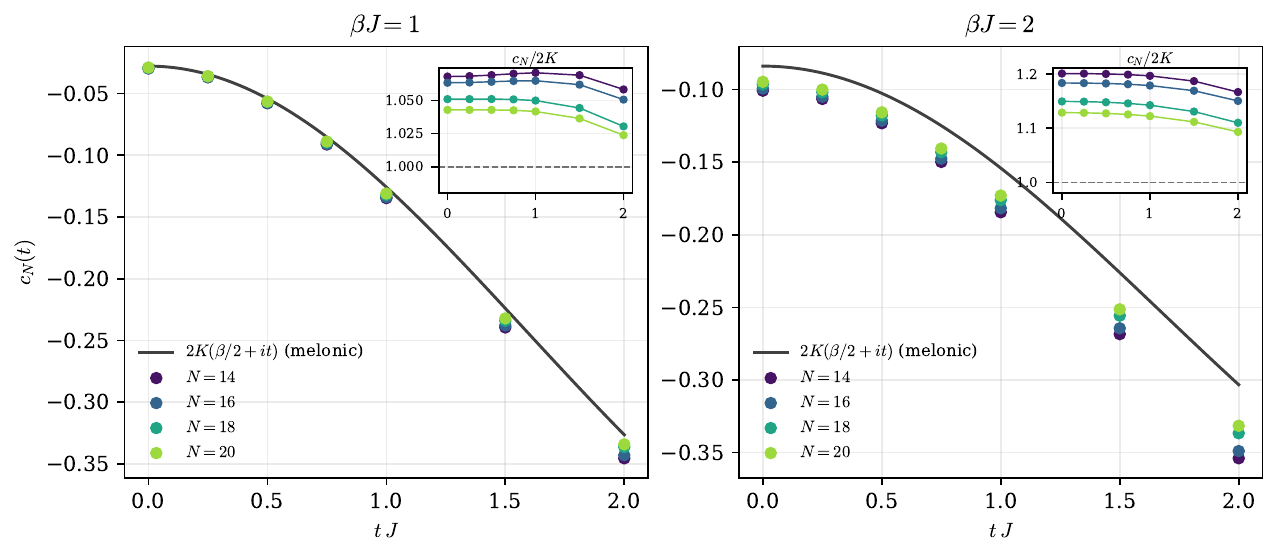}
\end{center}
\caption{The weight-2 bridge amplitude at complex separation $\tau = \beta/2 + it$, at $\beta J = 1$ (left) and $\beta J = 2$ (right) ($N = 14$--$20$, $10$ seeds, all pairs). Each panel shows the fitted amplitude $c_N(t)$ of the bridge polynomial $B_{ij}$ (points; real to machine precision, as Hermiticity requires) against the parameter-free continued kernel $2 K(\beta/2 + it)$ from the real-frequency melonic solution (curve). Insets: the ratio $c_N(t) / 2K(\beta/2+it)$: the continuation captures the full time dependence, leaving the familiar near-$t$-independent finite-$N$ coefficient excess, which matches the imaginary-time slope at $t = 0$ and decreases with $N$.}
\label{fig:w2realtime}
\end{figure}

\section{Application to Maldacena--Qi}
\label{sec:mq}

The Maldacena--Qi (MQ) model~\cite{MaldacenaQi2018} was introduced to describe a simple eternal wormhole and has a number of interesting dynamical properties~\cite{MaldacenaMilekhin2021,GarciaGarciaNosaka2019,PluggeLantagneHurtubiseFranz2020}. In particular, its ground state is known to resemble a thermofield double (TFD) of the SYK model from which it is built~\cite{MaldacenaQi2018,Cottrell2019}. However, the overlap between the MQ ground state and the closest TFD state still decays exponentially with system size~\cite{MaldacenaQi2018,schuster2025coolingsachdevyekitaevmodelusing,khor2026preparinghighfidelitythermofielddouble}. As an application of the fixed-instance rules developed in this paper, we identify a leading mismatch in low-weight observables and use it to construct a nearby Hamiltonian whose TFD has substantially higher fidelity at the sizes studied. This low-weight matching does not establish an asymptotic fidelity exponent or a bound on the global overlap. The key point is that the MQ ground state can be understood as a TFD state for a ``nearby'' Hamiltonian~\cite{khor2026preparinghighfidelitythermofielddouble}, and one can use the single-instance formulas to read off the couplings in this Hamiltonian order by order. We first describe the MQ model and the modified diagrammatic rules for it before turning to the ground state application.

\subsection{MQ rules}
\label{sec:mq-rules}

The model is defined from two copies of the same fixed SYK instance,
\beq
    H_{\rm MQ}
    =
    H_L + H_R + i\mu \sum_{j=1}^N \chi_j^L \chi_j^R ,
    \qquad
    H_s =
    -\sum_{i<j<k<l} J_{ijkl}\,
    \chi_i^s \chi_j^s \chi_k^s \chi_l^s ,
    \label{eq:HMQ}
\eeq
with $s=L,R$ and $\{\chi_i^s,\chi_j^{s'}\}=\delta_{ij}\delta_{ss'}$. The $J_I$ couplings are the same on the two sides and carry the same normalization as in \eqref{eq:H4}. The rules are structurally similar to the case of a single SYK instance, except that there are two flavors of fermions to keep track of.

In analytic formulas, $\beta$ and $\mu$ retain their dimensional meanings. All numerical temperatures and MQ couplings below, including table headings and figure axes, are quoted as the dimensionless combinations $\beta J$ and $\mu/J$, respectively. Note that the MQ model has $2N$ fermions, so the $N$ values used in the numerics below are correspondingly smaller than those quoted in Sections \ref{sec:leading} and \ref{sec:corrections}.

The large-$N$ propagator is the $2\times2$ matrix
\beq
    G_{ss'}(\tau)
    =
    \langle T_\tau\, \chi_i^s(\tau)\chi_i^{s'}(0)\rangle ,
\eeq
with $G_{RR}=G_{LL}$ and $G_{RL}=-G_{LR}$. On a thermal circle of length $\beta_{\rm MQ}$ it obeys
\beq
    \Sigma_{ss'}(\tau) = J^2 G_{ss'}(\tau)^3,
    \qquad
    \big[G^{-1}\big](i\omega_n)
    =
    \begin{pmatrix}
        -i\omega_n-\Sigma_{LL} & i\mu-\Sigma_{LR}\\
        -i\mu+\Sigma_{LR} & -i\omega_n-\Sigma_{LL}
    \end{pmatrix}.
    \label{eq:mqsd}
\eeq
Diagrammatically, each quartic interaction vertex carries an independent side label $s_v\in\{L,R\}$, summed over, and a line from side $s$ to side $s'$ carries $G_{ss'}$. Although the rules make sense for any $\beta_{\rm MQ}$, we are interested here in the ground state. We compute its kernels by taking $\beta_{\rm MQ}$ large compared with the inverse MQ gap, so the thermal circle may be replaced by the line up to exponentially small thermal corrections.

\paragraph{A two-point check.}
The simplest probes of the MQ ground state $|G\rangle$ are the cross-side two-point functions $\langle G|\chi_i^L \chi_j^R|G\rangle$. The diagonal correlator requires no instance-specific coupling polynomial: at leading order it is the off-diagonal matrix propagator itself,
\beq
    \langle G|\chi_i^L \chi_i^R|G\rangle = G_{LR}(0^+) + \cdots,
\eeq
purely imaginary in our conventions. Exact diagonalization at $N=8$--$12$ reproduces the matrix-SD value at the percent level for $\mu/J \ge 1$ (within $0.3\%$ at $\mu/J=2$), with a larger ${\sim}4\%$ drift at $\mu/J = 0.5$.

For $i \ne j$ the leading contribution is instance-specific, and the diagram is the weight-2 bridge of Section~\ref{sec:time-dep-w2imag} --- two vertices $J_{ilmn}$ and $J_{jlmn}$ sharing three internal lines --- with each vertex summed over its side label. The coupling polynomial is the same bridge $B_{ij} = \sum_{l<m<n} J_{ilmn}J_{jlmn}$; only the kernel changes:
\beq
\begin{aligned}
    \langle G|\chi_i^L \chi_j^R|G\rangle
    &\approx i\,\widehat K(\mu)\, B_{ij}, \\
    \widehat K(\mu)
    &= \mathrm{Im} \sum_{s_1, s_2=L,R} \int d\tau_1\, d\tau_2\,
    G_{L s_1}(-\tau_1)\, G_{s_1 s_2}(\tau_1 - \tau_2)^3\,
    G_{s_2 R}(\tau_2) .
\end{aligned}
    \label{eq:mqlrbridge}
\eeq
where the external legs run from the probe time $0$ into the two vertices ($G_{LL}$ is odd in $\tau$ and $G_{LR}$ even, so the reversed legs carry signs). Every coloring contains an odd number of $LR$ propagators, so the sum is purely imaginary, as the observable requires. In the free-dimer limit the integrals are elementary, $\widehat K \to -1/(64\mu^2)$. Exact diagonalization at $N = 8$--$12$ (four realizations, all pairs) confirms the prediction with the familiar finite-$N$ pattern: the through-origin slope of $\mathrm{Im}\,\langle \chi_i^L\chi_j^R\rangle$ against $\widehat K(\mu) B_{ij}$ is $1.02$--$1.05$ at $\mu/J=2$, $1.13$--$1.17$ at $\mu/J=1$, and $1.4$--$1.5$ at $\mu/J=0.5$, growing with the effective $\beta_* J$ like every coefficient excess in the one-sided thermal tests.

This computation is the thermofield-double bilocal of Section~\ref{sec:time-dep-w2imag} in a two-sided representation. In the TFD of a single copy, the defining identity of the infinite-temperature double, $\chi_j^R|I\rangle = i\, \chi_j^L |I\rangle$, folds the cross-side correlator onto the half-thermal bilocal:
\beq
    \langle {\rm TFD}(\beta)|\, \chi_i^L \chi_j^R\, |{\rm TFD}(\beta)\rangle
    = \frac{i}{2}\, R_{ij}(\beta/2)
    = i\,\delta_{ij}\, G(\beta/2) + i\,K_\beta(\beta/2)\, B_{ij} + \cdots ,
    \label{eq:tfdlr}
\eeq
the observable tested at large $N$ in Figure~\ref{fig:w2tfd}. The MQ ground state realizes the same structure --- diagonal propagator plus bridge --- with the explicit half-circle evolution traded for the off-diagonal components of the matrix propagator: $G_{LR}(0^+)$ stands in for $i\,G(\beta/2)$, and $\widehat K(\mu)$ for $K_{\beta}(\beta/2)$. The correspondence is quantitative once $\beta$ is chosen as the effective temperature $\beta_*(\mu)$ constructed in the next subsection: the two kernels then agree exactly in the free-dimer limit, and at $J=1$ they agree to $0.5\%$ at $\mu/J = 2$, $1.7\%$ at $\mu/J = 1$, and $5\%$ at $\mu/J = 0.5$. Thus the two-point correlator gives a first parameter-free comparison between the MQ ground state and the TFD selected below. Its residual kernel mismatch also previews the four-point mismatch that will determine the corrected Hamiltonian.

\subsection[Leading weight 4 diagrams and the MQ--TFD map]{Leading weight $4$ diagrams and the MQ--TFD map}
\label{sec:mq-w4-leading}

Let $X=\{i<j<k<l\}$ and probe the MQ ground state $|G\rangle$ with the left string $\mu_X^L$. At leading order there is a single quartic vertex, but it may sit on either side. The two colorings give
\beq
    \langle G|\mu_X^L|G\rangle
    =
    -4\,\widehat I_4(\mu)\, J_X + \cdots ,
    \qquad
    \widehat I_4(\mu)
    =
    \int d\tau\,
    \big[ G_{LL}(\tau)^4 + G_{LR}(\tau)^4 \big] .
    \label{eq:mqw4lead}
\eeq
Here and below the integral is over the long MQ thermal circle, or equivalently over the line in the ground-state limit. Compare this with the single-SYK thermal formula \eqref{eq:w4pred} with couplings $\tilde J_X$,
\beq
    \langle \mu_X\rangle_{\beta,\tilde J}
    =
    -4\, I_4(\beta)\, \tilde J_X + \cdots .
\eeq
The leading $W=4$ results match provided
\beq
    \tilde J_X = J_X ,
    \qquad
    I_4(\beta_*) = \widehat I_4(\mu).
    \label{eq:mqbetastar}
\eeq
This implicitly defines the effective TFD temperature $\beta_*(\mu)$ at which the weight-four expectation values match to leading order. It is a large-$N$ kernel prediction depending only on $\mu/J$, not a temperature fitted to finite-$N$ expectation values or fidelity data.

\begin{figure}[t]
\begin{center}
\includegraphics[width=0.90\textwidth]{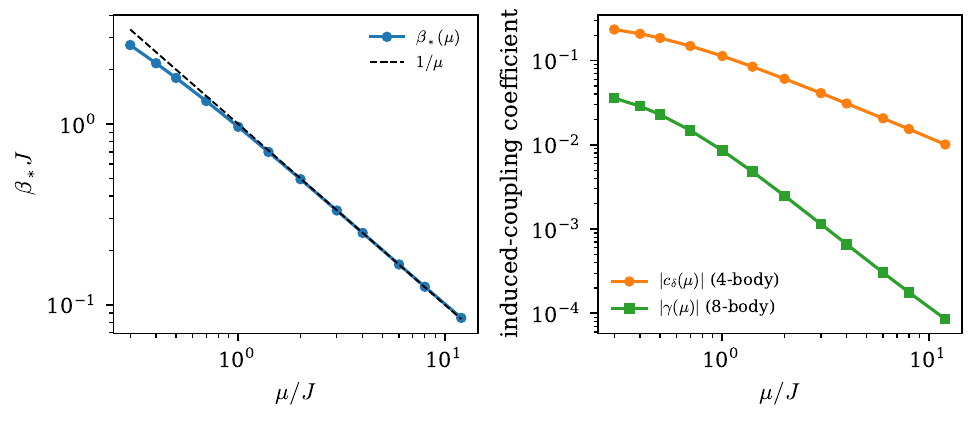}
\end{center}
\caption{Left: the dimensionless temperature map $\beta_*J$ versus $\mu/J$, obtained by matching the leading $W=4$ expectation value, $I_4(\beta_*)=\widehat I_4(\mu)$. The dashed line is the large-$\mu/J$ free-dimer asymptote $\beta_*J=J/\mu$. Right: magnitudes of the induced-coupling coefficients versus $\mu/J$. The four-body coefficient $c_\delta$ is the one used in the simplified nearby-Hamiltonian story; the induced eight-body coefficient $\gamma$ (Section~\ref{sec:mq-higherorder}) is much smaller and has little effect on the state-overlap comparison.}
\label{fig:mqmap}
\end{figure}

For example, in the large-$\mu$ limit,
\beq
    G_{LL}(\tau) = \frac12 \sgn(\tau)e^{-\mu|\tau|},
    \qquad
    G_{LR}(\tau) = \frac{i}{2}e^{-\mu|\tau|},
\eeq
so $\widehat I_4 = 1/(16\mu)$ and hence $\beta_*\to 1/\mu$. This is the strong-$LR$-coupling limit in which the MQ ground state is almost a collection of Bell pairs, just as the TFD state is at high temperature. The interacting numerical solution follows this asymptote closely at large $\mu/J$ and bends upward when $\mu/J$ becomes order one; see Figure~\ref{fig:mqmap}.

\subsection{First fluctuation order and the nearby four-body Hamiltonian}
\label{sec:mq-fourbody}

Now keep the first fluctuation tensor for $W=4$. In the one-sided thermal problem this was the tensor $T_X^{(2)}$ in \eqref{eq:corr-t2}; for clarity in this section we also write
\beq
    P^{(4)}_X \equiv T_X^{(2)} .
\eeq
The single-SYK thermal result through this order is
\beq
    \langle \mu_X\rangle_{\beta,\tilde J}
    =
    -4\, I_4(\beta)\, \tilde J_X
    +4\, W^{(2)}(\beta)\, P^{(4)}_X(\tilde J)
    +\cdots ,
    \label{eq:thermalw4fluc}
\eeq
where $W^{(2)}$ is defined in \eqref{eq:corr-t2}. The MQ side has the same coupling polynomial but a side-summed kernel,
\beq
    \langle G|\mu_X^L|G\rangle
    =
    -4\, \widehat I_4(\mu)\, J_X
    +4\, \widehat W^{(2)}(\mu)\, P^{(4)}_X(J)
    +\cdots ,
    \label{eq:mqw4fluc}
\eeq
with
\beq
    \widehat W^{(2)}(\mu)
    =
    \sum_{s_1,s_2=L,R}
    \int d\tau_1 d\tau_2\,
    G_{L s_1}(\tau_1)^2\,
    G_{s_1 s_2}(\tau_1-\tau_2)^2\,
    G_{L s_2}(\tau_2)^2 .
    \label{eq:mqWhat}
\eeq
After imposing $I_4(\beta_*)=\widehat I_4(\mu)$, the leading terms agree for $\tilde J=J$, but the first-fluctuation terms differ by
\beq
    4\big[\widehat W^{(2)}(\mu)-W^{(2)}(\beta_*)\big]\,P^{(4)}_X(J).
\eeq
To this order, substituting $\tilde J=J+\delta J$ into $P_X^{(4)}(\tilde J)$ only changes terms beyond the retained accuracy. The mismatch can therefore be absorbed into the leading thermal term by the four-body shift
\beq
    \tilde J_X = J_X + \delta J_X,
    \qquad
    \boxed{\;
    \delta J_X
    =
    c_\delta(\mu)\, P^{(4)}_X(J),
    \qquad
    c_\delta(\mu)
    =
    -\frac{\widehat W^{(2)}(\mu)-W^{(2)}(\beta_*)}{I_4(\beta_*)}.
    \;}
    \label{eq:mqdK}
\eeq
Thus, through the first fluctuation order, the MQ ground-state expectation values match the thermal expectation values of the nearby four-body Hamiltonian
\beq
    H_{\tilde J}
    =
    -\sum_{i<j<k<l}
    \big(J_{ijkl}+\delta J_{ijkl}\big)
    \chi_i\chi_j\chi_k\chi_l .
    \label{eq:Hnear4}
\eeq
This correction is not primarily a rescaling of the SYK variance. At $N=10$ in the numerical ensemble below, $\operatorname{var}(\delta J)$ is only about $1.9\%$, $0.71\%$, and $0.21\%$ of $\operatorname{var}(J)$ at $\mu/J=0.5,1,2$, and $\delta J$ is nearly uncorrelated with $J$ (the fitted slope of $\delta J$ against $J$ is at most $5\times10^{-3}$ in magnitude). It is instead a structured, instance-dependent direction in four-body coupling space built from the same first fluctuation tensor that improves the one-sided thermal prediction.

\subsection{Numerical tests of the four-body story}
\label{sec:mq-numerics}

\begin{figure}[t]
\begin{center}
\includegraphics[width=0.92\textwidth]{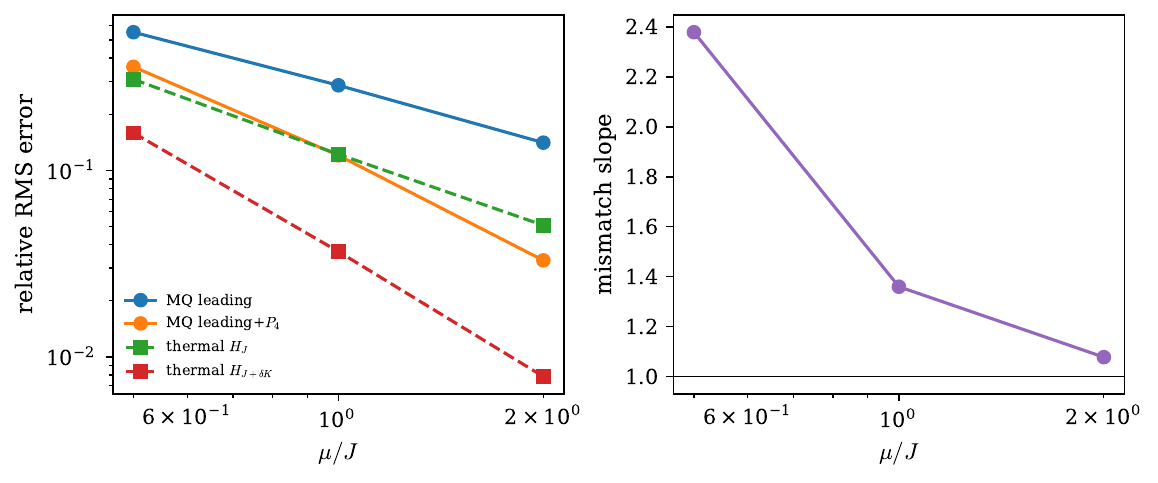}
\end{center}
\caption{Direct $W=4$ test of the simplified MQ matching at $N=10$ (eight disorder realizations, all $210$ quartets per realization), shown versus $\mu/J$. Left: relative RMS error against the exact MQ ground-state $W=4$ expectation values. Adding the first fluctuation tensor improves the diagrammatic MQ formula, and replacing $H_J$ by $H_{\tilde J}$ improves the exact thermal expectation values. Right: slope of the exact MQ--thermal mismatch against the predicted first-fluctuation mismatch $4[\widehat W^{(2)}-W^{(2)}]P^{(4)}_X$. The drift above one at smaller $\mu/J$ is a finite-size coefficient effect, analogous to the slope excess in the one-sided thermal tests.}
\label{fig:mqw4match}
\end{figure}

Figure~\ref{fig:mqw4match} tests the $W=4$ matching directly at $N=10$ Majoranas per side. For each of eight disorder realizations and each $\mu/J\in\{0.5,1,2\}$, we compute all $\binom{10}{4}=210$ left-side $W=4$ expectation values in the exact MQ ground state and compare them with two kinds of predictions. First, we compare the diagrammatic formulae \eqref{eq:mqw4lead} and \eqref{eq:mqw4fluc}. Adding the first fluctuation tensor substantially improves the MQ result: the relative RMS error drops from $0.55$ to $0.36$ at $\mu/J=0.5$, from $0.29$ to $0.12$ at $\mu/J=1$, and from $0.14$ to $0.033$ at $\mu/J=2$.

Second, we compare exact thermal expectation values at $\beta_*(\mu)$ for the bare Hamiltonian $H_J$ and the corrected Hamiltonian $H_{\tilde J}$. This is the closer analogue of the TFD statement, since no truncation of the thermal side is made. The relative RMS error against the MQ result drops
\beq
\begin{aligned}
    0.308 &\to 0.159 && (\mu/J=0.5),\\
    0.122 &\to 0.0367 && (\mu/J=1),\\
    0.0508 &\to 0.00784 && (\mu/J=2).
\end{aligned}
\eeq
The fitted slope of the exact MQ--thermal mismatch against the predicted tensor $4[\widehat W^{(2)}-W^{(2)}]P^{(4)}_X$ is close to one at $\mu/J=2$ ($1.08$), but is larger at smaller $\mu/J$ ($1.36$ at $\mu/J=1$ and $2.38$ at $\mu/J=0.5$). We interpret this as the same finite-size physics already visible in the one-sided thermal story: deterministic ladder corrections shift coefficients, while higher fluctuation tensors improve the operator-to-operator scatter. The important point for the MQ application is that the simple four-body correction already captures the dominant direction.

\begin{figure}[t]
\begin{center}
\includegraphics[width=0.94\textwidth]{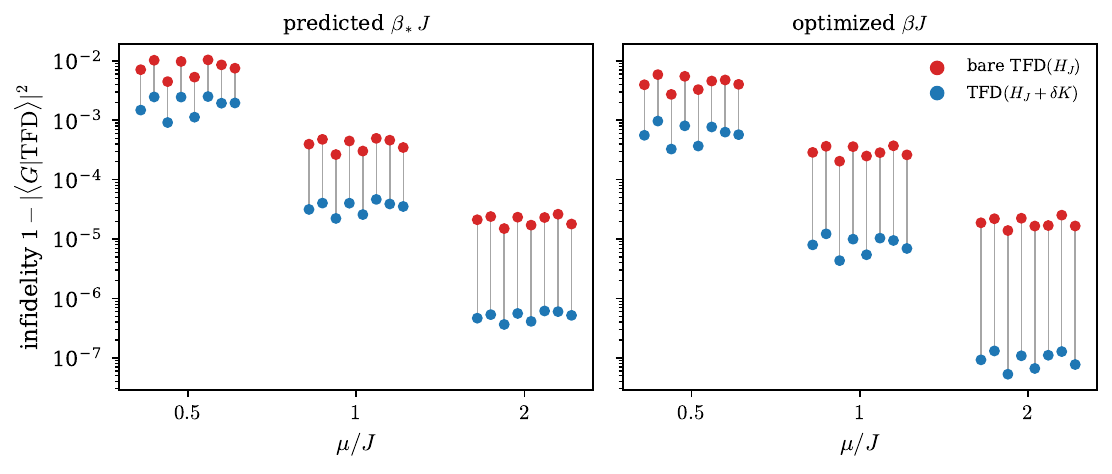}
\end{center}
\caption{Infidelity of the exact MQ ground state against the bare TFD of $H_J$ (red) and the corrected TFD of $H_{\tilde J}$ (blue), at $N=10$ Majoranas per side, shown versus $\mu/J$. The left panel uses the parameter-free large-$N$ prediction $\beta_*J$ from \eqref{eq:mqbetastar}; the right panel optimizes $\beta J$ separately for each sample and Hamiltonian. Each vertical segment connects the bare and corrected infidelity for one disorder realization.}
\label{fig:mqfidelity}
\end{figure}

\paragraph{State overlap.} The same correction improves the state-level comparison. We construct the infinite-temperature TFD $|I\rangle$ as the ground state of the bilinear term in \eqref{eq:HMQ}, and for a single-copy Hamiltonian $H$ define
\beq
    |{\rm TFD}(\beta;H)\rangle
    =
    \frac{e^{-\beta H_L/2}|I\rangle}
    {\|e^{-\beta H_L/2}|I\rangle\|}.
\eeq
There are two distinct tests, shown in Figure~\ref{fig:mqfidelity}. The stricter, parameter-free comparison fixes $\beta=\beta_*(\mu)$ from \eqref{eq:mqbetastar}, with no finite-$N$ temperature adjustment. At $N=10$, replacing $H_J$ by $H_{\tilde J}$ lowers the mean infidelity from $7.93\times10^{-3}$ to $1.86\times10^{-3}$ at $\mu/J=0.5$, from $3.99\times10^{-4}$ to $3.51\times10^{-5}$ at $\mu/J=1$, and from $2.09\times10^{-5}$ to $5.08\times10^{-7}$ at $\mu/J=2$.

The second test asks how well each Hamiltonian can match the MQ ground state after optimizing its temperature separately for each sample. Define
\beq
    \mathcal F_{\rm opt}(H)
    =
    \max_\beta |\langle G|{\rm TFD}(\beta;H)\rangle|^2 .
\eeq
The optimized mean infidelity improves from $4.34\times10^{-3}$ to $6.25\times10^{-4}$ at $\mu/J=0.5$, from $2.98\times10^{-4}$ to $8.33\times10^{-6}$ at $\mu/J=1$, and from $1.90\times10^{-5}$ to $9.56\times10^{-8}$ at $\mu/J=2$: factors of about $7$, $36$, and $200$, respectively. The fidelity-optimal $\beta J$ for the bare TFD exceeds $\beta_*J$ by $20\%$, $5.5\%$, and $1.3\%$ on average at $\mu/J=0.5,1,2$, and the four-body correction cuts the shift roughly in half, to $10\%$, $2.7\%$, and $0.7\%$. The excess grows with $\beta_*J$ and shrinks when the matching improves, echoing the finite-$N$ coefficient excesses in the one-sided thermal tests.

\paragraph{Size dependence.} Because the ground state and the TFD are prepared by sparse Krylov methods, it is straightforward to extend the comparison to $N=12$ and $N=14$ Majoranas per side ($16$ disorder realizations each, with the $N=10$ ensemble extended to $32$). The corrected comparison, and its advantage over the bare TFD, improve with $N$. At the parameter-free temperature $\beta_*(\mu)$, the ratio of bare to corrected mean infidelity grows from $4.3$ to $5.8$ at $\mu/J=0.5$, from $12$ to $15$ at $\mu/J=1$, and from $43$ to $57$ at $\mu/J=2$ as $N$ goes from $10$ to $14$, with the per-sample optimized ratios behaving similarly ($6.7 \to 8.4$, $34 \to 39$, ${\approx}190 \to {\approx}200$). The mechanism is one-sided: the bare infidelity is flat or mildly increasing over this range, while the corrected infidelity decreases (at $\mu/J = 0.5$, from $2.0\times10^{-3}$ to $1.3\times10^{-3}$ at fixed $\beta_*J$). The temperature shift shows the same trend, the bare excess falling from $21\%$ to $16\%$ at $\mu/J = 0.5$ and the corrected one from $10\%$ to $7\%$.

\subsection{Beyond the four-body correction}
\label{sec:mq-higherorder}

The nearby-Hamiltonian construction continues in two directions: higher-order corrections to the four-body coupling and induced eight- and higher-body interactions. Here we give the first diagnostic of that tower.

The structure of the higher-body tower follows the same logic as $\delta J$. Take the case of weight $8$. The leading MQ diagrams factorize into two copies of the $W=4$ vertex, so once $I_4(\beta_*)=\widehat I_4(\mu)$ the leading $W=8$ expectation value agrees automatically with the single-SYK thermal formula \eqref{eq:w8pred}: no new coupling is needed at this order. At the first fluctuation order the comparison produces two kinds of mismatch. One multiplies the same fluctuation tensor that entered the $W=4$ analysis; it is absorbed automatically by the four-body shift already in hand, through the cross term $J\,\delta J$ in the factorized leading formula. The other is genuinely new: an induced eight-body coupling proportional to the chain polynomial $P_{{\rm chain},X}$ appearing in \eqref{eq:w8corrected}, cubic in the underlying couplings, whose coefficient $\gamma(\mu)$ is again a ratio of a side-summed kernel mismatch to a thermal normalization, computed from the same matrix-SD data as $c_\delta$. Numerically $\gamma = -0.023$, $-0.0086$, $-0.0025$ at $\mu/J = 0.5, 1, 2$ (right panel of Figure~\ref{fig:mqmap}), an order of magnitude below the four-body coefficient throughout.

Both statements are visible in the exact data. We regress the differences in exact weight-eight expectation values between the MQ ground state and the thermal state of $H_{\tilde J}$ at $\beta_*$ on the predicted structures, using $N=14$, all $\binom{14}{8}=3003$ octets, and four disorder realizations. The coefficient of the fluctuation-tensor structure is suppressed by two to four orders of magnitude relative to its value without the four-body shift ($2\times10^{-3}$ versus $0.33$ at $\mu/J=0.5$, and $4\times10^{-6}$ versus $0.030$ at $\mu/J=2$), confirming its absorption into $\delta J$. The coefficient of the new eight-body structure is reproduced in sign and in order of magnitude, but not quantitatively at smaller $\mu/J$: the mean fitted coefficients are $-0.1283$, $-0.01774$, and $-0.00306$ at $\mu/J=0.5,1,2$, against predictions $-0.02282$, $-0.00857$, and $-0.00249$, ratios of $5.6$, $2.1$, and $1.2$. At $\mu/J=2$ this is the familiar finite-$N$ excess. At $\mu/J=0.5$ it is not: the matched temperature is only $\beta_*J=1.8$, where the kernel-dressing excesses seen elsewhere in this paper are $1.2$--$1.4$, and the fitted coefficient is within a factor $1.4$ of the four-body coefficient $c_\delta=-0.185$ rather than an order of magnitude below it. The origin of this discrepancy is open; a plausible candidate is that the regression at $N=14$ absorbs higher-order structures correlated with $P_{\rm chain}$. At the state level, however, the induced eight-body term lies below the residual floor: adding it to $H_{\tilde J}$ changes the mean infidelity by at most a few percent of itself at these sizes. Thus the four-body correction carries the observable improvement in the state overlap, while the eight-body structure is a detectable next term in the formal nearby-Hamiltonian expansion, not yet quantitatively understood at small $\mu/J$.

\section{Outlook}
\label{sec:outlook}

We have developed a fixed-instance diagrammatic expansion for SYK observables in which explicit coupling tensors carry the sample dependence and melonic kernels carry the temperature and time dependence. In the $p=4$ example, the weight-four one-point function provides the most detailed test: fluctuation tensors account for the operator-resolved scatter, while collective loops account for the coherent slope correction. Weight eight, unequal imaginary and real times, and the MQ model show that the same organization applies more broadly. In particular, while we focused on observables with vanishing disorder average, the same formalism applies to observables with a nonzero average, such as the fermion autocorrelation function, where it yields the instance-dependent corrections to that average.

The mathematical status of this expansion has two distinct parts. At fixed finite $N$, the Wick--Hermite chaos decomposition of Appendix~\ref{sec:rules} is complete and converges in $L^2$ of the disorder. What is not established is a uniform large-$N$ truncation bound: one must control both the melonic and collective evaluation of each kernel and the remainder from omitted chaos and collective orders. Every fixed truncation can be evaluated with polynomial classical effort, and the observed $N^{-1/2}$ hierarchy suggests that accuracy $N^{-a}$ may require an order proportional to $a$, but we have not proved the error bound needed to turn that suggestion into a general polynomial-time approximation theorem. Establishing such bounds, including their dependence on $\beta J$ and their behavior for rare disorder realizations, is a central theoretical open problem.

There are a variety of forms such bounds might take. For example, considering the leading term for $W=4$, $\xi_X^{(r=1)} = - 4 I_4 J_X$, a reasonable conjecture is that for any fixed $\beta J$ there is an $N_*(\beta J)$ and $C(\beta J)$ such that
\begin{equation}
\mathbb{E}_J\left[\left(\xi_X - \xi_X^{(r=1)}\right)^2 \right] \leq \frac{C}{N^4}
\end{equation}
for $N \geq N_*$. The $N^{-4}$ power appears because it is the variance of the $r=2$ term and the other corrections are further suppressed in $N$. More generally, it is natural to conjecture successively stronger inverse-power bounds when all chaos and collective contributions through a specified power-counting order are retained. It is less clear what uniform statement remains valid when $\beta J$ is allowed to grow with $N$; the melonic organization used here plausibly requires restricting to $\beta J\ll N$.

Real time raises a separate question. The analytically continued bridge kernel agrees with exact evolution over the time window tested in Section~\ref{sec:time-dep-w2real}, but the spectral representation used for the continuation does not by itself determine how long a fixed diagrammatic truncation remains accurate. A possible breakdown scale is $t \sim \Gamma^{-1} \log N$, when an exponentially decaying leading kernel (decay rate $\Gamma$) becomes comparable to power-suppressed corrections, but this remains a heuristic. Computing the first real-time corrections and extending the rules to out-of-time-order correlators would provide sharper tests and might help extract large-$N$ chaotic behavior from finite-$N$ data.

More efficient organization will also be important to go beyond the orders considered here. New effects include sample-to-sample fluctuations of the melonic sums $\sum_B J_{iB}^2$ as well as a rapidly growing set of skeletons. The fixed-disorder collective-field formulation of Appendix~\ref{sec:rules} exhibits these per-instance melon fluctuations and gives an all-orders rule for the normalized observable, so it may be a useful starting point for automation or resummation.

Similar methods may apply to mean-field quantum spin glasses and to systems perturbed away from the all-to-all limit. Such extensions would help isolate how much classical leverage comes specifically from statistical symmetry and dense connectivity, despite the entanglement and magic of the underlying thermal state and the existence of a rigorous quasi-polynomial algorithm in a complementary regime~\cite{Zlokapa2026}.

The MQ application suggests a further direction. The parameter-free map $\beta_*(\mu)$ and the induced four-body coupling $\delta J$ identify a nearby Hamiltonian whose TFD improves both the fixed-temperature expectation values and the optimized state fidelity. Extending the nearby-Hamiltonian expansion to higher orders, and determining whether it remains controlled at larger $N$ and smaller $\mu/J$, would clarify the structure of the MQ ground state and could inform protocols for preparing high-fidelity SYK TFD states~\cite{Bentsen_2024,schuster2025coolingsachdevyekitaevmodelusing,khor2026preparinghighfidelitythermofielddouble}.

Finally, the high-weight regime that motivated part of this work~\cite{BettaqueSwingle2026} requires a different scaling limit. For a weight $W=4m$ string in the $p=4$ model, the leading diagrams give $\xi_X = (-4 I_4)^m \sum_P \eta(P) \prod_{A \in P} J_A$ and hence
\beq
    \operatorname{var}(\xi_X)
    = \frac{(4m)!}{(4!)^m m!}\,(4I_4)^{2m}\,\sigma_J^{2m} .
\eeq
When $m\propto N$, this has the form $e^{-N\phi}$ with $\phi=\phi_0 (W/N)\ln(N/W)+\cdots$. The leading diagrams predict $\phi_0 = 3/4$, whereas the path-integral analysis gives $\phi_0 \simeq0.94$ at $\beta J = 20$ from a fit to SD data~\cite{BettaqueSwingle2026}. This mismatch is expected to probe diagrams that are subleading at fixed weight but become competitive when $W\sim N$. Resumming that class would connect the present low-weight expansion to the extensive-weight regime.

\section*{Acknowledgments}

We thank Alexander Zlokapa and Amir Raz for helpful conversations, Val Bettaque for collaboration on related work, and Brian Khor, Martin Sasieta, and Nadie Li for collaboration on the investigation of thermofield double states. This paper was produced with AI tools, partly as an ongoing experiment in agentic AI for physics. A brief first-person account follows.

\paragraph{The use of AI tools.} The core idea in this paper had been on my mind for several years, and I decided it would provide an interesting test of the physics applications of AI models, by which I mean agentic systems built upon large language models. I focused on an approach in which the AI agents operated in a relatively tight loop with me, in contrast to a more autonomous approach. The models used include GPT 5.5 and 5.6 Sol via Codex and Opus 4.8, 5, and Fable 5 via Claude Code. I largely let the models do the initial writing and I significantly edited the resulting text, similar to how I approach writing a paper with a beginning graduate student. All references were supplied with URL links and individually checked. The models provided a large number of cross-checks of the results and each other. The diagram figures required the most iteration during this process, especially when using GPT 5.5 and Opus 4.8. I independently hand-coded numerical checks of the first two orders using different random seeds, reviewed the analytic and numerical outputs, and take responsibility for the results. In the final stages of preparation, Fable 5.1 and GPT 6 Astra became available, so I also used them to run checks of all the files.

\paragraph{Code availability.} The numerical code (SD solver, parity-block and dense ED, kernels, and all data and figure scripts, with a test suite) is available in the \texttt{numerics-v3} directory of the repository accompanying this paper, \url{https://github.com/physicsmonkey/syk-diagrams}.

\appendix

\section{Conventions and the orientation sign}
\label{app:conventions}

This appendix collects the conventions used throughout the paper and then defines the orientation sign $\eta$ carried by the coupling polynomials.

\subsection{Collected conventions}

\begin{itemize}
    \item \emph{Majoranas.} $\{\chi_i, \chi_j\} = \delta_{ij}$, $\chi_i^2 = \tfrac12$ \eqref{eq:chialg}. The infinite-temperature state is the normalized trace, $\langle \cdot \rangle_0 = 2^{-N/2} \tr(\cdot)$, with free Euclidean propagator $G_0(\tau) = \tfrac12 \sgn(\tau)$.
    \item \emph{Strings.} For a sorted subset $X = \{x_1 < \cdots < x_W\}$ of weight $W = |X|$, $\mu_X = i^{W(W-1)/2}\, 2^{W/2}\, \chi_{x_1} \cdots \chi_{x_W}$ \eqref{eq:mu}; these are Hermitian, square to one, and are orthonormal under $\langle \cdot \rangle_0$.
    \item \emph{Couplings.} Independent couplings are labeled by sorted index sets, $J_I = J_{i_1 \cdots i_p}$ with $i_1 < \cdots < i_p$. We extend this tensor to arbitrary ordered indices by total antisymmetry, $J_{i_{\pi(1)}\cdots i_{\pi(p)}}=\sgn(\pi)J_{i_1\cdots i_p}$, and set it to zero whenever indices repeat. When a coupling appears with a composite label, as in $J_{Auv}$, this antisymmetric extension and the displayed index order are understood. The variance of each independent sorted component is the exact-count normalization \eqref{eq:H}, for which we write
    \beq
        \sigma_J^2 \equiv \operatorname{var}(J_I) = \frac{J^2}{\binom{N-1}{p-1}} \sim (p-1)!\; J^2\, N^{-(p-1)} .
    \eeq
    The unsubscripted $\sigma$ is reserved for the fluctuation of the collective self-energy field in Appendix~\ref{sec:rules} (and, subscripted by numbers, for imaginary-time integration variables).
    \item \emph{Propagators.} Matsubara frequencies are fermionic, $\omega_n = 2\pi(n + \tfrac12)/\beta$, with antiperiodic continuation $G(\tau + \beta) = -G(\tau)$. The melonic propagator $G$ solves the SD equations \eqref{eq:SD}; in Appendix~\ref{sec:rules} the same object is written $G_*$, to distinguish it from the fluctuating collective field.
    \item \emph{Kernels.} Every kernel is built from the melonic $G$, e.g.\ $I_4(\beta) = \int_0^\beta G(\tau)^4\, d\tau$ and $W^{(2)}(\beta)$ of \eqref{eq:corr-t2}.
    \item \emph{Wick ordering.} $\no{J^\alpha} = \prod_A \sigma_J^{\alpha_A}\, \He_{\alpha_A}(J_A/\sigma_J)$, with $\He_m$ the probabilists' Hermite polynomials (Appendix~\ref{sec:rules-chaos}); Wick monomials are mean zero for $|\alpha| \geq 1$ and orthogonal under the disorder average.
    \item \emph{Coefficient rule.} A connected diagram with $r$ vertices contributing to a weight-$W$ observable carries the prefactor $(-1)^r\, 2^{W/2}$ \eqref{eq:coeffrule}.
\end{itemize}

\subsection{The orientation sign}
\label{app:conventions-eta}

A diagram assigns the external Majoranas of $\mu_X$ to its interaction vertices in groups, and writing every coupling with sorted indices leaves behind a combinatorial sign from reordering the Majoranas into the sorted order of $X$. The bookkeeping is captured by a single identity for products of disjoint even strings.

Let $X$ be a sorted index set of even weight $W$ and let $P = \{B_1, \ldots, B_K\}$ be a partition of $X$ into blocks of even sizes $W_1, \ldots, W_K$. Define $\eta(P) = \pm 1$ to be the parity of the permutation carrying the concatenation $(B_1 B_2 \cdots B_K)$ --- each block sorted internally, the blocks taken in any order --- to sorted $X$. Then:
\begin{enumerate}
    \item $\eta(P)$ is well defined on \emph{unordered} partitions: exchanging two adjacent blocks reorders the concatenation by $W_i W_j$ transpositions, an even number when the block sizes are even.
    \item The product of the corresponding strings reproduces $\mu_X$ up to exactly this sign:
    \beq
        \mu_{B_1}\, \mu_{B_2} \cdots \mu_{B_K} = \eta(P)\, \mu_X .
        \label{eq:etaproduct}
    \eeq
\end{enumerate}
To verify \eqref{eq:etaproduct}, note that both sides contain the same Majoranas up to ordering and phases. Reordering the concatenated product into sorted order costs the sign $\eta(P)$, while the phase conventions of \eqref{eq:mu} combine as
\beq
    \prod_k i^{W_k(W_k-1)/2} = i^{W(W-1)/2}\; i^{-\sum_{k<l} W_k W_l} ,
\eeq
and the excess exponent $\sum_{k<l} W_k W_l$ is a multiple of four when every $W_k$ is even, so no phase survives. (Consistently with item 1, even strings on disjoint index sets mutually commute.) Taking the normalized trace against $\mu_X$ gives $\langle \mu_X\, \mu_{B_1} \cdots \mu_{B_K} \rangle_0 = \eta(P)$.

Two even-block instances appear in the main text:
\begin{itemize}
    \item \emph{Leading weight-$4m$ diagrams} \eqref{eq:w8pred}: each vertex absorbs a quartet of external Majoranas, $P$ is a partition of $X$ into quartets, and the diagram's coupling polynomial is $\eta(P) \prod_{A \in P} J_A$.
    \item \emph{Pair splittings} \eqref{eq:corr-t2}: the two vertices of the fluctuation tensor $T^{(2)}$ absorb the external Majoranas in pairs, and $\eta(A,B)$ is the sign of the two-block partition $X = A \cup B$. For example, at $X = \{1,2,3,4\}$: $\eta(\{1,2\}|\{3,4\}) = +1$, $\eta(\{1,3\}|\{2,4\}) = -1$, and $\eta(\{1,4\}|\{2,3\}) = +1$.
\end{itemize}

This unordered-partition identity does not cover a general diagram: a vertex
may absorb an odd number of external Majoranas, as in the triangle and chain
of Figure~\ref{fig:w4degree3}. Exchanging two odd blocks changes the
concatenation parity, so their order is part of the data. In that case we fix
the displayed vertex and leg ordering, write each coupling using the totally
antisymmetric convention, and take the sign from the parity of the
\emph{full} ordered Majorana string, including the internal Wick pairings.
The result need not be the external-block parity alone. With the convention
of \eqref{eq:w4degree3polys}, this full calculation supplies one additional
minus sign for each of the weight-four triangle and chain; those signs are
included explicitly in \eqref{eq:w4degree3kernels}. The weight-eight chain is
fixed in the same way by the exact cubic trace of
Appendix~\ref{app:w8-cubic}. Thus $\eta(P)$ by itself is used only for
even-block partitions; odd-block topologies use their explicitly ordered
contractions and signed kernels. These assignments are also checked against
the exact high-temperature traces in the numerical test suite.

\section{Systematic Organization of Higher Orders}
\label{sec:rules}

The dressed-diagram expansion of the main text is economical at the orders computed there: one enumerates a small set of skeletons, dresses every internal line by the melonic propagator, and performs the necessary Wick subtractions explicitly. At higher orders the bookkeeping becomes cumbersome. A single topology can contain both a new mean-zero coupling tensor and, through coincidences among its couplings, renormalizations of lower structures already carried by the dressed propagator or by a collective kernel. The degree-three tensors and ladder corrections of Sections~\ref{sec:corrections-w4scatter} and~\ref{sec:corrections-w4slope} are the first place where this distinction matters in practice.

This appendix gives an optional reorganization designed for that higher-order problem; none of the main-text results depends on adopting it. The organization has two ingredients. First, the Wick--Hermite chaos decomposition separates the coupling dependence into orthogonal, mean-zero structures. Second, an exact fixed-disorder collective path integral generates the same structures with their melonic dressing and coherent kernel corrections already organized. The normalization by $Z$ adds one further class of diagrams, those connected only through the subtraction vertex, identified in Appendix~\ref{sec:rules-pathint} and included in the rules of Appendix~\ref{sec:rules-diagrams}. Throughout the appendix $G_*$ denotes the melonic saddle propagator called $G$ in the main text, reserving $G$ for the fluctuating collective field.

\subsection{Chaos organization of the coupling structures}
\label{sec:rules-chaos}

The exact starting point is furnished by the Gaussian statistics of the couplings. For a multi-index $\alpha = (\alpha_A)$ with total degree $r = |\alpha| = \sum_A \alpha_A$, define the \emph{Wick monomial} --- the normal-ordered product with respect to the Gaussian coupling measure ---
\beq
    \no{J^\alpha} \;=\; \prod_A \sigma_J^{\alpha_A}\, \He_{\alpha_A}\!\big(J_A/\sigma_J\big),
    \qquad \text{e.g.} \quad
    \no{J_A J_B} = J_A J_B \;\; (A \neq B), \qquad
    \no{J_A^2} = J_A^2 - \sigma_J^2,
    \label{eq:wick}
\eeq
with $\He_m$ the probabilists' Hermite polynomials and $\sigma_J^2 = \operatorname{var}(J_I)$. Wick monomials are orthogonal under the disorder average, $\mathbb{E}\big[\no{J^\alpha}\, \no{J^\beta}\big] = \delta_{\alpha\beta}\, \alpha!\, \sigma_J^{2r}$ (with $\alpha! = \prod_A \alpha_A!$), and complete: every observable with finite disorder variance has a unique expansion
\beq
    \xi_X(J) = \sum_\alpha c_{\alpha,N}(\beta)\, \no{J^\alpha},
    \qquad
    c_{\alpha,N} = \frac{\mathbb{E}\big[\xi_X\, \no{J^\alpha}\big]}{\alpha!\, \sigma_J^{2r}}
    = \frac{1}{\alpha!}\, \mathbb{E}\big[\partial_J^\alpha\, \xi_X\big],
    \label{eq:projection}
\eeq
the second equality following from Gaussian integration by parts. These coefficients are exact deterministic functions of $N$, $\beta$, and $J$.

The derivatives in \eqref{eq:projection} are explicit thermal correlators. For example, with $p=4$ and $\partial H / \partial J_A = \mu_A / 4$, first-order perturbation theory in imaginary time (Duhamel's formula) gives
\beq
    \frac{\partial \xi_X}{\partial J_A}
    = -\frac{1}{4} \int_0^\beta ds\, \Big[ \big\langle \mu_A(s)\, \mu_X \big\rangle_\beta - \big\langle \mu_A \big\rangle_\beta \big\langle \mu_X \big\rangle_\beta \Big],
    \label{eq:duhamel}
\eeq
and iterating produces $r$-fold Duhamel integrals of fully connected correlators: derivatives acting on the $1/Z$ normalization supply the disconnected subtractions. Thus the exact chaos coefficients reduce the fixed-instance problem to disorder-averaged connected correlators, which can then be evaluated in the large-$N$ expansion.

Statistical $O(N)$ symmetry packages the individual Wick monomials into a much smaller set of covariant tensors. For $p=4$, regard $J$ as an element of $\Lambda^4 V$, $V=\mathbb{R}^N$, and the weight-$W$ observable as an antisymmetric tensor-valued equivariant function, $\xi(gJ)=g\cdot\xi(J)$. At each chaos order $r$, choose a basis $T_{r,a}(J;X)$ of covariant Wick-ordered contractions with the external slot. The exact expansion and the subsequent kernel expansion may then be displayed together as
\begin{equation}
    \xi_X(J)
    = \sum_{r,a} \mathcal{K}_{r,a}(N,\beta)\,T_{r,a}(J;X),
    \qquad
    \mathcal{K}_{r,a}(N,\beta)
    \sim \sum_{\ell\geq0} N^{-\ell}\,\mathcal{K}^{(\ell)}_{r,a}(\beta) .
    \label{eq:basisexpansion}
\end{equation}
The first equality is the exact finite-$N$ chaos expansion expressed in a covariant basis; the second is its formal large-$N$ evaluation by melonic and collective diagrams. At each fixed $r$ the covariant space is finite, but a basis inside a multiplicity space is not unique: different contraction bases are related by finite linear transformations, and an orthogonal presentation may be obtained by diagonalizing $\mathbb{E}[T_{r,a}T_{r,b}]$. What is canonical is the projection onto each chaos sector and the coefficient of every Wick monomial, not a particular choice of covariant basis within that sector.

Four consequences deserve emphasis:
\begin{enumerate}
    \item \emph{Different chaos orders are canonically separated.} No contribution can belong simultaneously to a lower-order coefficient renormalization and a higher-order mean-zero structure.
    \item \emph{The exact kernels $\mathcal K_{r,a}(N,\beta)$ are deterministic by construction}, not by a self-averaging assumption. A purported fluctuating kernel is instead a higher-chaos structure.
    \item \emph{The basis contains diagonal monomials.} Structures built from repeated couplings, such as $\sum_B \no{J_B^2}$, are invisible to any enumeration of distinct-coupling products, but they are genuine mean-zero structures with a definite (suppressed) size; the worked example below produces one.
    \item \emph{The chaos grading refines the slope/scatter split of Section~\ref{sec:corrections-types}.} Fully Wick-paired configurations renormalize lower kernels, while unpaired Wick monomials carry instance-to-instance fluctuations.
\end{enumerate}
Because the decomposition is orthogonal, the chaos series converges in $L^2$ of the disorder at fixed $N$. This exact statement does not by itself bound the error from truncating the chaos or collective expansions at large $N$; that separate question is discussed in the power counting below and in the Outlook.

The practical classification is now short: enumerate the covariant Wick-monomial structures, including their diagonal cousins; evaluate their deterministic coefficients from the connected correlators in \eqref{eq:projection}; and, where useful, choose an orthogonal basis by diagonalizing the finite Gram matrix. When the coefficients are evaluated from the normalized observable in \eqref{eq:projection}, the chaos projection is exact. A calculation starting from unnormalized diagrams must first include the normalization and then re-express the result in Wick monomials.

\paragraph{Worked example: the melon on a leg.} The simplest nontrivial example is the self-energy melon dressing one external leg of the leading vertex. Its coupling polynomial is
\beq
P_X(J) = J_X \sum_{u<v<w} J_{x_1 u v w}^2 ,
\eeq
the leading coupling $J_X$ times the sum over all couplings sharing the external index $x_1$ (the two vertices of the melon carry the same coupling, hence the square). Wick-order each term: the single coincident triple $(u,v,w) = (x_2,x_3,x_4)$ gives $J_X^3 = \no{J_X^3} + 3\sigma_J^2 J_X$, while every other triple $B$ gives $J_X J_{x_1 B}^2 = \no{J_X J_{x_1 B}^2} + \sigma_J^2 J_X$. Collecting, and using the exact count $\binom{N-1}{3}\sigma_J^2 = J^2$,
\beq
P_X(J) = \big( J^2 + 2\sigma_J^2 \big)\, J_X
\;+\; J_X \!\!\sum_{B \neq (x_2 x_3 x_4)} \!\!\! \no{J_{x_1 B}^2}
\;+\; \no{J_X^3}\, .
\eeq
The decomposition does everything at once. The deterministic piece $J^2 J_X$ is precisely the exact-count self-energy \eqref{eq:leadingsigma}: it renormalizes the coefficient of $J_X$ coherently, and it is exactly what the dressed propagator $G_*$ already resums, so the systematic expansion must not generate it again (the remaining $2\sigma_J^2 = O(N^{-3})$ is the finite-$N$ residue of the coincident triple). The rest is not discarded: $\sum_B \no{J_{x_1 B}^2}$ is a sum of ${\sim}\binom{N-1}{3}$ independent mean-zero terms --- the \emph{per-instance fluctuation of the melon} --- a genuine diagonal structure of relative size $N^{-(p-1)/2} = N^{-3/2}$ that contributes scatter at high order. Nothing is projected out ``on pain of double counting'': the Wick ordering has already routed the coherent part to the propagator and the fluctuation to its own graded order. The next subsection builds this Wick ordering into the action and identifies the one further class of diagrams introduced by the normalization.

\subsection{Fixed-disorder collective generating functional}
\label{sec:rules-pathint}

The chaos decomposition defines the higher-order structures but does not by itself provide the most efficient way to evaluate their kernels. In the disorder-averaged theory, the bilocal action and its fluctuations can be developed through collective-field and two-particle-irreducible methods~\cite{JevickiSuzuki2016,BenedettiGurau2018,ArefevaKhramtsovTikhanovskaya2018}. Here a useful fixed-instance generating device is obtained by integrating in the collective fields without integrating out the fermions. The resulting formal path integral retains the fixed couplings explicitly while routing their coherent contractions into the melonic and collective sectors. Its subtraction term implements Wick ordering of the interaction exponential at the integrand level.

Consider the path integral for a single instance of the partition function,
\begin{equation}
    Z = \int \mathcal{D}\psi \exp\left( - \sum_i \frac{1}{2} \int d\tau\, \psi_i \partial_\tau \psi_i \;-\; I_{\mathrm{inst}} \right),
    \qquad
    I_{\mathrm{inst}} = \sum_A \int d\tau\, i^{p/2} J_A \psi_A(\tau),
\end{equation}
where the $\psi_i(\tau)$ are Grassmann variables, $J_A$ is a single instance coupling, and $\psi_A$ denotes the product of the corresponding Grassmann variables. The standard move is to next average $Z$ and introduce the collective field variables $G(\tau,\tau')$ and $\Sigma(\tau,\tau')$, but this removes the physics of a single instance. Instead, define the subtraction term to be the second cumulant of the instance vertex,
\beq
    I_{\mathrm{sub}} \;\equiv\; \frac{1}{2}\, \mathbb{E}_J\big[ I_{\mathrm{inst}}^2 \big]
    \;=\; \frac{\sigma_J^2}{2} \sum_A \int d\tau\, d\tau' \prod_{a \in A} \psi_a(\tau) \psi_a(\tau'),
    \label{eq:isub}
\eeq
where the average is over the coupling distribution at fixed $\psi$.\footnote{The phases work out for even $p$: reordering $\psi_A(\tau)\psi_A(\tau')$ into $\prod_{a \in A} \psi_a(\tau)\psi_a(\tau')$ costs $(-1)^{p(p-1)/2} = i^{p(p-1)}$, which combines with the vertex factors $(i^{p/2})^2 = i^p$ to give $i^{p^2} = 1$.} Because $I_{\mathrm{inst}}$ is linear in the Gaussian couplings, with commuting (Grassmann-even) coefficients, adding $I_{\mathrm{sub}}$ to the action is nothing but the Wick ordering of Appendix~\ref{sec:rules-chaos} applied to the interaction:
\beq
    e^{-I_{\mathrm{inst}} - I_{\mathrm{sub}}} \;=\; \frac{e^{-I_{\mathrm{inst}}}}{\mathbb{E}_J\big[ e^{-I_{\mathrm{inst}}} \big]} \;=\; \no{e^{-I_{\mathrm{inst}}}}\, .
    \label{eq:normalorder}
\eeq
Expanding the right-hand side in powers of the interaction generates Wick monomials directly: a cluster of $k$ vertices carrying the same coupling contributes $\no{J_A^k} = \sigma_J^k \He_k(J_A/\sigma_J)$, never a bare power. This identity accounts for the subtraction diagrams within the interaction exponential. It does not commute with forming a normalized expectation value.

The compensating factor $e^{+I_{\mathrm{sub}}}$ that \eqref{eq:normalorder} strips out of $e^{-I_{\mathrm{inst}}}$ is the disorder mean $\mathbb{E}_J[e^{-I_{\mathrm{inst}}}]$, and it is the seed of the collective action. A Grassmann identity makes this explicit: the bilinears $b_i = \psi_i(\tau) \psi_i(\tau')$ are commuting and nilpotent ($b_i^2 = 0$), so all power sums $\sum_i b_i^k$ with $k \geq 2$ vanish and Newton's identities collapse the elementary symmetric polynomial to a pure power,
\beq
    \sum_A \prod_{a \in A} \psi_a(\tau) \psi_a(\tau') \;=\; e_p(b_1, \ldots, b_N) \;=\; \frac{1}{p!} \left[ \sum_i \psi_i(\tau) \psi_i(\tau') \right]^p .
    \label{eq:grassmannid}
\eeq
Therefore, identically,
\begin{equation}
\begin{aligned}
    I_{\mathrm{sub}} &= \frac{N \hat{J}^2}{2p} \int d\tau\, d\tau' \left[ \frac{1}{N} \sum_i \psi_i(\tau) \psi_i(\tau') \right]^p ,\\
    \hat{J}^2 &\equiv \frac{\sigma_J^2 N^{p-1}}{(p-1)!} = J^2 \left[ 1 + \binom{p}{2} \frac{1}{N} + O(N^{-2}) \right] :
\end{aligned}
    \label{eq:jhat}
\end{equation}
the subtraction term is written through the flavor-averaged bilinear with a slightly shifted coupling $\hat{J}$. Its $1/N$ excess over $J$ tracks the difference between the restricted distinct-index count in the Hamiltonian and the unrestricted collective power.

Now integrate in the collective fields by inserting the identity
\begin{equation}
    1 = \int \mathcal{D}G \mathcal{D} \Sigma \exp\left( - \frac{N}{2} \int d\tau d\tau'  \Sigma(\tau,\tau') \left[G(\tau ,\tau') - \frac{1}{N} \sum_i \psi_i(\tau) \psi_i(\tau')\right] \right),
\end{equation}
an integral representation of a $\delta$-function. On its support the bilinear form \eqref{eq:jhat} of the compensator may be transcribed onto the collective field, $[N^{-1} \sum_i \psi_i \psi_i]^p \to G^p$, which turns it into the familiar collective potential --- but at the shifted coupling $\hat{J}^2$, not at $J^2$.

This is the place to fix the bookkeeping of the two normalizations, because two different variances are genuinely in play and they should not be conflated. The \emph{instance} couplings carry their true variance $\sigma_J^2 = J^2/\binom{N-1}{p-1}$, and nothing about that may be adjusted: the Wick ordering \eqref{eq:normalorder} is exact only at the true variance, and normal-ordering against any other value would leave coincident clusters with nonzero means, breaking the chaos grading at relative order $1/N$. The \emph{collective} sector, by contrast, is ours to organize, and the convenient choice is the standard large-$N$ normalization: write the collective potential at $J^2$ --- the coefficient it would have had if the couplings carried the usual large-$N$ variance $(p-1)!\, J^2/N^{p-1}$ --- so that the saddle, the $(g,\sigma)$ propagator, and every kernel in this paper are the familiar $N$-independent objects. The entire mismatch between the two normalizations is then isolated in a single deterministic counterterm vertex,
\beq
    \Delta I \;\equiv\; \frac{N (\hat{J}^2 - J^2)}{2p} \int d\tau\, d\tau'\; G(\tau,\tau')^p ,
    \qquad
    \hat{J}^2 - J^2 = \binom{p}{2}\, \frac{J^2}{N} + O(N^{-2}) ,
    \label{eq:deltaI}
\eeq
a purely collective term of absolute order $N^0$ in the action, i.e.\ relative order $1/N$ against the $O(N)$ collective potential. Assembling the pieces, the fixed-disorder collective representation is
\begin{equation}
    Z = \int \mathcal{D}\psi \mathcal{D} G \mathcal{D} \Sigma\, e^{- I},
    \qquad
    I = I_{\mathrm{avg}} + I_{\mathrm{inst}} + I_{\mathrm{sub}} - \Delta I,
\end{equation}
with $I_{\mathrm{inst}}$, $I_{\mathrm{sub}}$, and $\Delta I$ as above and
\begin{equation}
    I_{\mathrm{avg}} = \frac{1}{2} \int d\tau d\tau'\, \sum_i \psi_i(\tau)\left[\delta(\tau-\tau')\partial_{\tau'} - \Sigma\right] \psi_i(\tau') + \frac{N}{2} \int d\tau d\tau' \left[  G \Sigma - \frac{J^2}{p}  G^p \right].
\end{equation}
If the representation is summed without truncation, it is simply an exact rewriting of the original fixed-instance path integral. Its usefulness is computational: the usual collective action coexists with the single-instance data, and the latter enters only through the normal-ordered combination \eqref{eq:normalorder}. Expanding the collective sector about its saddle gives formal large-$N$ expansions of the numerator and of $Z$; the rule for their ratio is derived below.

The observable is inserted directly: for $\xi_X$ one adds the Grassmann image of the string \eqref{eq:bilocaldef}, $i^{W(W-1)/2}\, 2^{W/2}\, \psi_{x_1}(\tau_1) \cdots \psi_{x_W}(\tau_W)$, to the numerator and divides by $Z$ (all insertion times coincide for the equal-time one-point function). In the diagrams below the insertion appears as $W$ external fermion legs with pinned --- not summed --- flavors and pinned times, and the division by $Z$ implements the fixed-disorder normalization of Section~\ref{sec:leading-taylor}. In a Wick-ordered expansion its effect includes contractions between numerator and denominator monomials, not just the deletion of vacuum components.

\paragraph{Normalization and chaos projection.}
Write the inserted path integral as $\mathcal N_X(J)$, so that $\xi_X=\mathcal N_X/Z$. Although both $\mathcal N_X$ and $Z$ can be expanded in Wick monomials, expanding their ratio produces ordinary products of these monomials, which regenerate lower chaos orders. Already for one Gaussian coupling,
\beq
\no{J}\,\no{J^2}=\no{J^3}+2\sigma_J^2 J.
\eeq
For example, in the scalar model $H=J\mu/4$, $\mu^2=1$, the ratio of the Wick-ordered numerator and denominator, with $a=\beta/4$, begins
\beq
\frac{-a\no{J}-a^3\no{J^3}/6+\cdots}
     {1+a^2\no{J^2}/2+\cdots}
=-aJ+\frac{a^3}{3}\no{J^3}+a^3\sigma_J^2J+O(a^5).
\eeq
The linked-cluster theorem locates these terms in the formal expansion. The full action $I=I_{\rm avg}+I_{\rm inst}+I_{\rm sub}-\Delta I$ contains ordinary interactions, the subtraction term included: by \eqref{eq:isub}, $I_{\rm sub}$ is a two-time ``dumbbell'' vertex with the $2p$ legs of one coupling $A$. Its sign is fixed by its appearance in $e^{-I}$. Hence $\xi_X=\mathcal N_X/Z$ is the sum of diagrams connected to the observable insertion, with the usual bare vertex weights and symmetry factors, where a dumbbell counts as a link. Treat the entire observable insertion as a root. Define the underlying \emph{wiring} by deleting the links between the two halves of each dumbbell while retaining all fermion lines, collective connections, vertex times, and the root. Group diagrams that have this same wiring and differ only in which pairs of equal-$A$ slots are joined into dumbbells.

If the wiring is connected, every dumbbell assignment is present. The generating identity behind \eqref{eq:normalorder}, $e^{-Jv-\sigma_J^2v^2/2}=\sum_k\no{J^k}(-v)^k/k!$, applied separately to each coupling label, sums the family to the global Wick monomial $\no{J^\alpha}$ times its wiring kernel. If the wiring is disconnected, only dumbbell assignments that connect every component to the root survive the division by $Z$.

The signs of this second class follow from the same generating identity. For one coupling and two commuting vertex sources $v,w$,
\beq
\no{e^{-J(v+w)}}
=\no{e^{-Jv}}\,\no{e^{-Jw}}\,e^{-\sigma_J^2vw}.
\label{eq:wickbridgeexp}
\eeq
After summing dumbbells internal to each component, the family with at least one bridge between the two is therefore generated by
\beq
\no{e^{-Jv}}\,\no{e^{-Jw}}\bigl(e^{-\sigma_J^2vw}-1\bigr).
\eeq
For components with $m,n$ slots of that coupling, the coefficient of $(-v)^m(-w)^n/(m!n!)$ is
\begin{align}
\mathcal B_{mn}(J)
&=\no{J^{m+n}}-\no{J^m}\,\no{J^n} \nonumber\\
&=-\sum_{j=1}^{\min(m,n)}j!\binom{m}{j}\binom{n}{j}
  \sigma_J^{2j}\,\no{J^{m+n-2j}}.
\label{eq:wickproduct}
\end{align}
Thus the bridged family is the negative of the positive cross-contraction sum in the ordinary Wick-product identity. Individual bridge assignments retain the signs from the subtraction interaction. With several wiring components, only assignments whose bridge graph connects all components to the root are kept; the two-component formula alone does not impose that condition. Any remaining bare coupling products must also be decomposed into Wick monomials before a definite chaos order is assigned.

In the scalar example, $\mathcal B_{12}=-2\sigma_J^2J$ multiplies the vertex-expansion coefficient $-a^3/2$, giving precisely the $+a^3\sigma_J^2J$ term above. In the SYK path integral, the simplest bridge attaches a vacuum component consisting of one further $J_A$ vertex to a slot whose label $A$ is already fixed by the original wiring. There is no new independent sum over $A$, so this decoration costs $\sigma_J^2\sim N^{-3}$ at $p=4$, relative to the wiring it decorates at fixed $\beta J$. It has the same parametric suppression as the coincidence residue in the worked example of Appendix~\ref{sec:rules-chaos}; this comparison does not identify their coefficients or kernels. The attached component is a subdiagram like any other and obeys the power counting of Appendix~\ref{sec:rules-power}: each further vertex in it costs the usual factor $N^{-(p-2)/4}$, and its own coincidences are Wick-ordered by the same argument. The bridged class therefore begins at relative order $\sigma_J^2$ and decreases from there; we do not attempt a proof of a uniform bound. The explicit cubic projections and the separately checked one-loop channels of the main text are independent of this bookkeeping.

The perturbative expansion begins by expanding $G = G_* + g$ and $\Sigma = \Sigma_* + \sigma$, where $G_*$ and $\Sigma_*$ are the fixed solutions of the SD equations of Section~\ref{sec:leading-melonic}, i.e.\ at coupling $J$. All the terms that are quadratic in $\psi_i$, $\sigma$, and $g$ are declared to be the bare action,
\begin{align}
I_0 = \, &  \frac{1}{2} \int d\tau d\tau'\, \sum_i \psi_i(\tau)\left[\delta(\tau-\tau')\partial_{\tau'} - \Sigma_*(\tau,\tau')\right]
\psi_i(\tau') \\ \nonumber
& + \int d\tau d\tau' \left[ \frac{N}{2}  g \sigma - \frac{N(p-1)J^2}{4} G_*^{p-2} g^2 \right].
\end{align}
The terms that involve only $G_*$ and $\Sigma_*$ do not contribute to $\xi_X$ because they cancel between the numerator and denominator. The terms \emph{linear} in the fluctuations are kept, and they organize themselves into subtractions. Collecting everything linear in $\sigma$ --- the $\sigma\psi\psi$ coupling from the fermion kinetic term together with the c-number source $\tfrac{N}{2}\int G_*\,\sigma$ left behind by the shift --- gives
\beq
    -\frac{1}{2} \int d\tau\, d\tau'\; \sigma(\tau,\tau') \left[ \sum_j \psi_j(\tau)\psi_j(\tau') - N G_*(\tau,\tau') \right] :
    \label{eq:sigmasub}
\eeq
the field $\sigma$ couples to the \emph{subtracted} fermion bilinear, whose expectation value under $I_0$ vanishes identically because the $I_0$ propagator is $G_*$ itself. The terms linear in $g$ from $I_{avg}$, namely $\tfrac{N}{2}\int [\Sigma_* - J^2 G_*^{p-1}]\, g$, vanish identically as well, because the collective potential and the saddle now carry the \emph{same} coupling $J^2$. The only surviving linear term comes from the counterterm vertex. Expanding \eqref{eq:deltaI} about the saddle,
\beq
    \Delta I = \frac{N(\hat{J}^2 - J^2)}{2p}\int G_*^p
    \;+\; \frac{N(\hat{J}^2 - J^2)}{2}\int G_*^{p-1}\, g
    \;+\; O\!\big( N (\hat{J}^2 - J^2)\, g^2 \big) ,
    \label{eq:deltaIexp}
\eeq
the constant cancels between numerator and denominator, and the quadratic term is a relative-$1/N$ correction to the collective kernel, beyond the order used in this paper. The linear term is a c-number source for $g$, and its effect is computed by a single contraction: the source pulls on $\sigma$ through the off-diagonal block of $D_0$,
\beq
    \langle \sigma \rangle \;=\; \frac{2}{N} \cdot \frac{N(\hat{J}^2 - J^2)}{2}\, G_*^{p-1} \;=\; (\hat{J}^2 - J^2)\, G_*^{p-1} ,
\eeq
which shifts the fermion self-energy by the first-order response to $J^2 \to \hat{J}^2$ at fixed $G_*$. To keep the notation consistent with Section~\ref{sec:corrections-w4slope}, write the disorder-averaged propagator as
\beq
    \mathbb{E}[G] = G_* + \frac{1}{N}\,\delta G + O(N^{-2}),
    \qquad
    \delta G = \delta G_{\rm norm}+\delta G_{\rm tad} .
\eeq
Dressing the source by the linearized SD response gives the normalization contribution
\beq
    \delta G_{\rm norm}
    = N(\hat{J}^2 - J^2)\, \partial_{J^2} G_*
    = \binom{p}{2}\, J^2\, \partial_{J^2} G_* + O(N^{-1}) ,
    \label{eq:deltaGshift}
\eeq
with $\partial_{J^2}G_*$ the derivative of the SD solution. The second contribution, $\delta G_{\rm tad}$, is generated by the cubic collective vertices. Together they give the channel-(S) factor $\kappa_S$, while the pair-channel ladder is carried by the quadratic $(g,\sigma)$ sector and gives $\kappa_L$. As a check on this bookkeeping, the combined one-loop correction reproduces $+\partial_\beta \tfrac12\operatorname{Tr}[\log(1-K)+K]$ to better than one percent over every temperature in Table~\ref{tab:w4pred}. (An alternative bookkeeping expands around the saddle at $\hat{J}$, removing the normalization source but making every kernel $N$-dependent through $\beta\hat{J}(N)$; we do not use it here.) The diagrammatic meaning of the subtraction in \eqref{eq:sigmasub} is spelled out next.

The bare action is supplemented by a number of vertices. First, all the higher order terms in $I_{\mathrm{avg}}$,
\beq
I_{\mathrm{avg,vert}} = - \frac{1}{2} \int d\tau d\tau'\, \sum_i \sigma \psi_i(\tau) \psi_i(\tau') - \frac{N J^2}{2p} \int d\tau d\tau' \left[ \binom{p}{3} G_*^{p-3} g^3 + \cdots \right],
\eeq
together with the normal-ordered instance vertex, i.e.\ $I_{\mathrm{inst}}$ and $I_{\mathrm{sub}}$, and the counterterm vertex $\Delta I$, whose expansion \eqref{eq:deltaIexp} supplies the $1/N$ source and kernel counterterms. The perturbative expansion uses propagators drawn from $I_0$ and vertices drawn from $I_{\mathrm{avg,vert}}$, $I_{\mathrm{inst}}$, $I_{\mathrm{sub}}$, and $\Delta I$. We now state the diagrammatic rules.

\subsection{Diagrammatic rules}
\label{sec:rules-diagrams}

The first ingredient is the propagators. The fermion propagator is
\beq
G_{0,ij}(\tau,\tau') = G_*(\tau,\tau') \delta_{ij}.
\eeq
The appearance of $G_*$ indicates that this propagator already contains all the resummed melonic diagrams.

The $g$ and $\sigma$ propagators are coupled. The quadratic action for $(g,\sigma)$ can be written as
\beq
\frac{1}{2} \int d\tau d\tau' (g, \sigma)(\tau,\tau')\, \Upsilon(\tau,\tau')\, (g, \sigma)^T(\tau,\tau')
\eeq
where
\beq \Upsilon =
\begin{bmatrix}
   -\frac{N (p-1)J^2}{2} G_*(\tau,\tau')^{p-2} & \frac{N}{2} \\ \frac{N}{2}  & 0
\end{bmatrix}.
\eeq
The propagator is the inverse of $\Upsilon$ and is diagonal in time,
\beq
D_0(\tau_1,\tau_1',\tau_2,\tau_2') =
\begin{bmatrix}
   0 & \frac{2}{N} \\ \frac{2}{N}  & \frac{2(p-1)J^2}{N} G_*(\tau_1,\tau_1')^{p-2}
\end{bmatrix} \delta(\tau_1 - \tau_2)\delta(\tau_1'-\tau_2').
\eeq
Note that $g$ and $\sigma$ propagators go from two times to two times because these are bilocal fields.

These propagators may join into vertices. The first vertex from $I_{avg,vert}$ allows two fermions of the same flavor to convert into a $\sigma$; there are also vertices in which three or more $g$s combine. All the physics of the particular instance is contained in clusters of the $I_{inst}$ vertex, in which $p$ fermions meet with amplitude proportional to $J_A$. The linked-cluster argument of Appendix~\ref{sec:rules-pathint} gives the rule for the normalized observable:
\begin{quote}
\emph{Draw diagrams connected to the observable root using the instance vertex, subtraction dumbbell, collective vertices, and counterterm $\Delta I$, with every internal fermion line dressed ($G_*$). Group them by the wiring left after deleting dumbbell links, retaining all other connections. A connected wiring contributes the global Wick monomial $\no{J^\alpha}$ of its vertex couplings. For disconnected wirings, retain only dumbbell assignments connecting every component to the root, with the signs and symmetry factors fixed by the action, as in \eqref{eq:wickbridgeexp}. When several couplings bridge at once, the connectivity condition is imposed by inclusion--exclusion, and the resulting bare coupling products are decomposed into Wick monomials before chaos orders are assigned.}
\end{quote}
Within a connected wiring, grouping the subtraction diagrams implements global Wick ordering. For an isolated pair of instance vertices, a melonic self-energy insertion carries $\no{J_A^2}$ rather than $J_A^2$: its coherent part, which would double count the dressing already inside $G_*$, is simply absent, while its mean-zero remainder survives at its own graded order. Summed over the couplings on a line of flavor $i$, that remainder is $\sum_B \no{J_{iB}^2} = \sum_B J_{iB}^2 - J^2$, the per-instance melon fluctuation of Appendix~\ref{sec:rules-chaos}, with zero mean and relative root-mean-square size $N^{-(p-1)/2}$. Similarly, the coherent part of the naive rung of Section~\ref{sec:corrections-types} is carried by the $\sigma$ propagator $D_{\sigma\sigma}$, whose kernel is the elementary Bethe--Salpeter rung, while the mean-zero remainder $\no{J_{x_a x_b u v}^2}$ survives as a diagonal fluctuation structure. When other vertices share the same coupling, its occurrences must be Wick-ordered together, as in the worked example of Appendix~\ref{sec:rules-chaos}. The dumbbell-bridged wirings supply the additional normalization terms, beginning at relative order $\sigma_J^2$.

\paragraph{Tadpoles: how the collective sector avoids a second melon.} The normal ordering protects the instance sector, but there is a second route to a spurious melon: a fermion line of flavor $i$ can emit a $\sigma$ at the $\sigma\psi\psi$ vertex, and the $\sigma$ can land on a closed fermion loop of any flavor $j$. Since $D_{\sigma\sigma} \sim (p-1) J^2 G_*^{p-2}/N$ while the loop sum supplies a factor $N G_*$, the result is of order $J^2 G_*^{p-1}$ --- melon-sized, with no $1/N$ suppression. This contribution is cancelled, but by a different mechanism: the saddle. By \eqref{eq:sigmasub} the $\sigma$ vertex couples to the \emph{subtracted} bilinear $\sum_j \psi_j\psi_j - N G_*$, and a bare closed loop evaluates to exactly $N G_*$ because the $I_0$ propagator is $G_*$ itself: a collective line terminating on an undecorated loop carries coefficient zero. This is the usual statement that fluctuations about a true saddle have no tadpoles. Like the normal ordering, the protection leaves a graded remainder rather than nothing: the counterterm $\Delta I$ supplies the normalization response $\delta G_{\rm norm}$, while loops formed from the cubic collective vertices produce $\delta G_{\rm tad}$. Their sum is the one-loop propagator shift $\delta G$ entering channel (S) of the case study.

The relation to the direct diagrams of Section~\ref{sec:corrections-w4slope} is now precise. In channel (L), the effective rung of Figure~\ref{fig:rungresum} is the dressed $\sigma\sigma$ collective propagator and lands across pairs of external lines as in Figure~\ref{fig:runglandings}(a,b). On a single line, however, one elementary contracted pair is just the melon already contained in $G_*$ and leaves no correction. The leading nonmelonic self-energy remainder therefore begins with Figure~\ref{fig:runglandings}(c): a rung bridges two internal lines of the melonic self-energy, with its necklace continuations implicit. This direct-diagram object is the channel-(S) face of the collective result $\delta G=\delta G_{\rm norm}+\delta G_{\rm tad}$; the $O(J^2/N)$ terms in its two collective pieces cancel, so the remainder begins at $O(J^4/N)$. It must not be represented as a lone elementary $\sigma$ rung attached to an otherwise dressed external line.

The two subtractions are the same move made in two sectors. The instance vertex carries its coupling normal-ordered against the disorder average ($I_{sub}$); the $\sigma$ vertex carries its fermion bilinear normal-ordered against the saddle \eqref{eq:sigmasub}. Neither vertex may be dropped --- both are essential, and both come pre-subtracted --- and the practical rule is uniform: a collective line ending on a fermion loop contributes the loop \emph{minus} $N G_*$, an object with zero leading value, deterministic $1/N$ corrections, and mean-zero fluctuations, automatically graded rather than melon-sized.

To summarize: $I_{inst}$ carries the single-instance data, entering through the normal-ordered combination \eqref{eq:normalorder}; its vertex clusters generate the structures $T_{r,a}$ with the melonic dressing already included. The collective $(g,\sigma)$ sector carries the coherent $1/N$ corrections to their kernels. Without truncation the representation is exact, and the normalized observable is the sum of connected wirings, each carrying its Wick monomial, plus the dumbbell-bridged wirings selected by the root-connectivity condition. The latter retain their subtraction signs, begin at relative order $\sigma_J^2$, and are power-counted like any other subdiagram. Truncating the saddle and collective expansions gives the formal large-$N$ approximation used in practice.

\subsection{Power counting and the map to the main text}
\label{sec:rules-power}

The rules above generate infinitely many diagrams. The melonic dominance of the averaged SYK expansion admits a direct combinatorial proof~\cite{BonzomNadorTanasa2019}, and related methods classify leading and next-to-leading topologies in colored variants~\cite{BonzomLionniTanasa2017}. For the fixed-instance structures considered here, at fixed $r$, fixed external weight, and fixed $\beta J$, generic sizes can be estimated by random-walk counting of the free index sums. The basic input is that each coupling is a mean-zero Gaussian of typical size
\beq
\sigma_J = \sqrt{\operatorname{var}(J_I)} = \sqrt{\frac{J^2}{\binom{N-1}{p-1}}} \;\sim\; \sqrt{(p-1)!}\; J\, N^{-(p-1)/2}.
\eeq
Consider a connected diagram with $v$ instance vertices, external weight $W$, and therefore
\beq
L = \frac{pv - W}{2}
\eeq
internal fermion lines. For a generic maximal-index contraction, these lines carry $L$ independently summed flavors. The couplings contribute $\sigma_J^v \sim N^{-v(p-1)/2}$; when their labels are distinct, the Wick monomials from different index assignments add in quadrature and the sums contribute $N^{L/2}$ rather than $N^L$. Thus
\beq
\text{(generic fluctuation structure)} \;\sim\; \sigma_J^v\, N^{L/2} \;\sim\; N^{-[(p-2)v + W]/4}.
\label{eq:powercount}
\eeq

For this generic class, each additional vertex at fixed external weight costs a relative factor $N^{-(p-2)/4}$, which is $N^{-1/2}$ at $p=4$. The leading weight-$W$ diagrams saturate \eqref{eq:powercount}: $W=4$, $v=1$ gives $N^{-3/2}$, while $W=8$, $v=2$ gives $N^{-3}$. The two-vertex fluctuation tensor of Section~\ref{sec:corrections-types} has $v=2$, $W=4$, and $L=2$, hence size $N^{-2}$ --- one factor $N^{-1/2}$ below the leading weight-four structure, as observed in its scatter.

Second, the estimate \eqref{eq:powercount} assumes distinct couplings and must be applied after coherent contractions have been separated from the fluctuation structures. A bare coincident pair $J_A^2$ has a nonzero mean and can sum coherently. The melon is the extreme case: its sum over ${\sim}N^{p-1}$ couplings is $O(1)$ and must be resummed into $G_*$; the rung is the next such object. In the rules of Appendix~\ref{sec:rules-diagrams} these coherent pieces never appear in a structure diagram: they live in $G_*$ and in the collective $(g,\sigma)$ sector, and what a coincidence leaves behind at the instance vertices is its mean-zero Wick remainder, which is \emph{suppressed} relative to the distinct-coupling count, not enhanced. The dumbbell-bridged wirings are counted separately and begin at relative order $\sigma_J^2\sim N^{-3}$, as derived in Appendix~\ref{sec:rules-pathint}. For example, the melon fluctuation in Appendix~\ref{sec:rules-chaos} is of order $N^{-3}$, below the distinct-coupling degree-three weight-four tensors at $N^{-5/2}$.

The coherent physics is graded separately by the collective $1/N$ expansion. Each $(g,\sigma)$ propagator costs $1/N$ and each closed flavor sum or collective vertex restores the corresponding power, so an additional connected collective order is generically suppressed by $1/N$. We denote this order by $\ell$; for ordinary collective diagrams it agrees with loop order, while deterministic counterterm insertions such as $\Delta I$ are assigned the same $\ell$ by their explicit $N$ scaling. The grading is not by rung number: composing the Bethe--Salpeter kernel with freely summed intermediate flavors adds powers of $K$ but no further power of $1/N$, so the order-$\ell=1$ pair channel contains the full $(1-K)^{-1}$ ladder. The melon is the $\ell=0$ saddle object already contained in $G_*$.

Every contribution therefore has the two-component address introduced in Section~\ref{sec:corrections-types}: the chaos order $r$ of its structure and the collective order $\ell$ of its kernel. Generic distinct-coupling structures have root-mean-square size $N^{-[(p-2)r+W]/4}$, with a further relative factor $N^{-\ell}$ from the kernel expansion; diagonal structures may be more suppressed. Table~\ref{tab:grading} catalogues the leading addresses for the weight-four one-point function. Keeping all addresses above a target power gives a natural \emph{formal} truncation prescription, but controlling the omitted $L^2$ tail requires an additional error bound not established here.

\section[One-loop coefficient correction at weight 4]{One-loop coefficient correction at weight $4$}
\label{app:w4-oneloop}

This appendix derives the two factors $\kappa_L$ and $\kappa_S$ used in
Section~\ref{sec:corrections-w4slope}.  Every normalization entering
Table~\ref{tab:w4pred} is displayed below in the same convention used by
\texttt{numerics-v3/syk/oneloop.py}.  No finite-$N$ data or fitted coefficient
enters either factor.  We set $J=1$ in the implementation, so its argument
called \texttt{beta} is the dimensionless combination $\beta J$; the formulas
below retain $J$ where it makes the counting clearer.

\subsection{Target quantity and bare-rung normalization}
\label{app:w4-oneloop-target}

The linear-chaos coefficient is the susceptibility \eqref{eq:w4slopesusc}.
Its factorized value is $-4I_4$, and we write the coherent correction as
\beq
    \alpha_X=-4I_4\big(1+\delta_{\rm slope}\big),
    \qquad I_4=\int_0^\beta d\tau\,G(\tau)^4.
\eeq
Introduce the direct and exchanged one-rung kernels
\begin{align}
 D(\tau)&=\int d\tau_1d\tau_2\,
 G(\tau-\tau_1)^2G(\tau_1-\tau_2)^2G(\tau_2)^2,\\
 \mathcal{E}(\tau)&=\int d\tau_1d\tau_2\,
 G(\tau-\tau_1)G(\tau_1)
 G(\tau-\tau_2)G(\tau_2)G(\tau_1-\tau_2)^2,
\end{align}
and the integrated shape
\beq
    R_\beta\equiv\int_0^\beta d\tau\,
    G(\tau)^2\big[D(\tau)-\mathcal{E}(\tau)\big].
    \label{eq:oneloop-R}
\eeq
The difference $D-\mathcal{E}$ is forced by fermion exchange and vanishes at the two
equal-time endpoints.  With the exact-count variance, the bare coherent rung
of \eqref{eq:w4rung} simplifies algebraically:
\beq
 \binom{4}{2}\binom{N-2}{2}\sigma_J^2
 =6\,\frac{(N-2)(N-3)}{2}
 \frac{6J^2}{(N-1)(N-2)(N-3)}
 =\frac{18J^2}{N-1}.
\eeq
Consequently
\beq
    \boxed{\quad
    \delta_{\rm rung}(N,\beta)
    =\frac{18J^2}{N-1}\frac{R_\beta}{I_4}.
    \quad}
    \label{eq:oneloop-deltarung}
\eeq
This exact $1/(N-1)$ expression is the common normalization against which
both one-loop channels are quoted.

\subsection{Channel L: the resummed pair ladder}
\label{app:w4-oneloop-L}

For antisymmetric bilocal functions, define the free pair propagator and the
Bethe--Salpeter kernel by
\begin{align}
 F_0(12;34)&=-G(\tau_{13})G(\tau_{24})
             +G(\tau_{14})G(\tau_{23}),\\
 (Kf)(12)&=-3J^2\int d\tau_3d\tau_4\,
 G(\tau_{13})G(\tau_{24})G(\tau_{34})^2 f(34).
 \label{eq:oneloop-K}
\end{align}
Composition includes integration over both intermediate times.  The connected
pair correlator beyond free propagation is
\beq
    \mathcal L\equiv K(1-K)^{-1}F_0
    =\sum_{n\geq1}K^nF_0.
    \label{eq:oneloop-ladder}
\eeq
The first term satisfies
\beq
   [KF_0](\tau,0;\tau,0)=3J^2[D(\tau)-\mathcal{E}(\tau)].
   \label{eq:oneloop-single-rung}
\eeq
The two spectator Majoranas supply $G(\tau)^2$, so the ratio of the resummed
pair channel to the bare rung is
\beq
 \boxed{\quad
 \kappa_L(\beta)=
 \frac{\displaystyle\int_0^\beta d\tau\,G(\tau)^2
       \mathcal L(\tau,0;\tau,0)}
      {\displaystyle\int_0^\beta d\tau\,G(\tau)^2
       [KF_0](\tau,0;\tau,0)}.
 \quad}
 \label{eq:oneloop-kappaL}
\eeq
This includes every necklace continuation but no free-pair term.  In the
code, \texttt{ladder\_kappa} generates $K^nF_0$ successively.  Each
application of $K$ is two dense matrix multiplications, the sum stops only
after the last term is below $10^{-9}$ times the first in max norm, and the
tail ratio is recorded as a spectral-radius estimate.  The production number
is the linear Richardson extrapolation
$2\kappa_L(M)-\kappa_L(M/2)$ at $M=512$.

\subsection{Channel S: shift of the averaged propagator}
\label{app:w4-oneloop-S}

After the fermions are integrated out, the standard $p=4$ collective action
per flavor, organized at coupling $J$, is
\beq
 \mathcal S[G,\Sigma]
 =-\frac12\operatorname{Tr}\log(\partial_\tau-\Sigma)
 +\frac12\int d\tau_1d\tau_2
 \left[\Sigma_{12}G_{12}-\frac{J^2}{4}G_{12}^4\right].
 \label{eq:oneloop-action}
\eeq
The exact-count coupling distribution instead produces
\beq
 \widehat J^2
 =\frac{\sigma_J^2N^3}{6}
 =J^2\frac{N^3}{(N-1)(N-2)(N-3)}
 =J^2\left(1+\frac6N+O(N^{-2})\right).
 \label{eq:oneloop-Jhat}
\eeq
We keep the saddle at $J$ and represent the difference by
\beq
 \Delta I=\frac{N(\widehat J^2-J^2)}8\int d\tau_1d\tau_2\,G_{12}^4,
 \qquad I=I_J-\Delta I.
 \label{eq:oneloop-counterterm}
\eeq

Write $G=G_*+g$ and $\Sigma=\Sigma_*+s$, with $g$ and $s$ antisymmetric.
Expanding \eqref{eq:oneloop-action} gives
\begin{align}
 \mathcal S_2={}&\frac14\operatorname{Tr}(G_*sG_*s)
 +\frac12\int s_{12}g_{12}
 -\frac{3J^2}{4}\int G_{*,12}^2g_{12}^2,
 \label{eq:oneloop-S2}\\
 \mathcal S_3={}&\frac16\operatorname{Tr}(G_*s)^3
 -\frac{J^2}{2}\int G_{*,12}g_{12}^3.
 \label{eq:oneloop-S3}
\end{align}
Let $\phi=(g,s)$ denote independent antisymmetric components and define
\beq
 H_{ab}=\frac{\partial^2\mathcal S_2}{\partial\phi_a\partial\phi_b},
 \qquad
 T_{abc}=\frac{\partial^3\mathcal S_3}
 {\partial\phi_a\partial\phi_b\partial\phi_c},
 \qquad P=H^{-1}.
\eeq
Since the full action is $N\mathcal S$, the Gaussian covariance is $P/N$.
Expanding once in the cubic vertex gives
\beq
 \boxed{\quad
 N\langle\phi_a\rangle_{\rm tad}
 =-\frac12P_{ab}T_{bcd}P_{cd}.
 \quad}
 \label{eq:oneloop-tadpole}
\eeq

The counterterm supplies a separate deterministic response.  Its linear term
shifts $J^2$ by $6J^2/N$ at leading order, hence
\beq
 \delta G_{\rm norm}(\tau)
 =6J^2\partial_{J^2}G_*(\tau).
 \label{eq:oneloop-dGnorm}
\eeq
Combining it with the $g$ component of \eqref{eq:oneloop-tadpole},
\beq
 \mathbb E[G_{\rm inst}(\tau)]
 =G_*(\tau)+\frac1N\delta G(\tau)+O(N^{-2}),
 \qquad
 \delta G=\delta G_{\rm tad}+\delta G_{\rm norm}.
 \label{eq:oneloop-dG}
\eeq
Inserting this shift into $I_4$ and dividing by the large-$N$ limit of
\eqref{eq:oneloop-deltarung} gives
\beq
 \boxed{\quad
 \kappa_S(\beta)
 =\frac{2}{9J^2}\,
 \frac{\displaystyle\int_0^\beta d\tau\,G_*(\tau)^3\delta G(\tau)}
      {R_\beta}.
 \quad}
 \label{eq:oneloop-kappaS}
\eeq
Indeed, the numerator enters the fractional correction as
$4(NI_4)^{-1}\int G_*^3\delta G$, while
$\delta_{\rm rung}=18J^2R_\beta/(NI_4)+O(N^{-2})$.  Thus the factor $2/9$
in \texttt{kappa\_S\_from\_deltaG} is algebraic, not empirical.

\subsection{Discrete Hessian and cubic contraction}
\label{app:w4-oneloop-discrete}

For completeness, use an $M$-point circle with spacing
$\Delta\tau=\beta/M$, evaluate $A_{ab}=G_*(\tau_a-\tau_b)$ spectrally at
integer offsets, set $A_{aa}=0$, and retain one coordinate per pair $a<b$.
Extending $g$ and $s$ antisymmetrically to matrices, the code evaluates
\begin{align}
 S_2^{(M)}={}&\frac{\Delta\tau^4}{4}\operatorname{tr}(AsAs)
 +\frac{\Delta\tau^2}{2}\sum_{ab}s_{ab}g_{ab}
 -\frac{3J^2\Delta\tau^2}{4}\sum_{ab}A_{ab}^2g_{ab}^2,
 \label{eq:oneloop-S2M}\\
 S_3^{(M)}={}&\frac{\Delta\tau^6}{6}\operatorname{tr}(As)^3
 -\frac{J^2\Delta\tau^2}{2}\sum_{ab}A_{ab}g_{ab}^3.
 \label{eq:oneloop-S3M}
\end{align}
The derivatives of \eqref{eq:oneloop-S2M} form $H$, which is inverted
directly.  Two finite-grid identities fix all factors and signs:
\beq
 H_{ss}=-\Delta\tau^4F_0,
 \qquad
 P_{gg}=(1-K)^{-1}F_0.
 \label{eq:oneloop-discrete-identities}
\eeq
Both are enforced in \texttt{tests/test\_oneloop.py}.  The second also
computes $\kappa_L$ inside the Hessian route, independently of the Neumann
series.

For the cubic contraction, in pair coordinates $b=(i,j)$, $i<j$, the $ggg$
source is
\beq
 Y^g_b\equiv T^{ggg}_{bcd}P^{gg}_{cd}
 =-6J^2\Delta\tau^2 A_{ij}P^{gg}_{bb}.
\eeq
Extend $P^{ss}$ antisymmetrically to $\mathcal P_{ab,cd}$ and define
\beq
 Q_{ad}=\sum_{bc}\mathcal P_{ab,cd}A_{bc},
 \qquad
 R_{bc}=\sum_{ad}\mathcal P_{ab,cd}A_{da}.
\eeq
The $sss$ source is
\beq
 \mathcal Y^s=\frac{\Delta\tau^6}{2}
 \left[(AQA)^T+A^T R A^T\right],
 \qquad
 Y^s_{(ij)}=\mathcal Y^s_{ij}-\mathcal Y^s_{ji},
\eeq
and the desired $g$ tadpole is
\beq
 \delta g_{\rm tad}
 =-\frac12\left(P_{gg}Y^g+P_{gs}Y^s\right).
 \label{eq:oneloop-discrete-tadpole}
\eeq
The optimized $sss$ formula is tested against a direct third derivative of
\eqref{eq:oneloop-S3M}.  Diagonal averaging with an antiperiodic wrap then
recovers $\delta G_{\rm tad}(\tau)$.

The response \eqref{eq:oneloop-dGnorm} is evaluated by the centered difference
\beq
 6J^2\partial_{J^2}G_*
 \simeq 6J^2\frac{G_*(J^2+\epsilon)-G_*(J^2-\epsilon)}{2\epsilon},
 \qquad \epsilon=0.02.
\eeq
At $\beta J=2$, reducing $\epsilon$ to $0.01$ and $0.005$ changes
$\kappa_S$ by less than $1.5\times10^{-5}$ on every tested grid.  Production
uses $M=96,128$ and a linear $1/M$ extrapolation.  At $\beta J=2$ the raw
values at $M=32,48,64,80$ are
$-0.17521,-0.17888,-0.18075,-0.18189$; their $1/M$ trend extrapolates to
$-0.1865$, consistent with the production extrapolation quoted in
Table~\ref{tab:w4pred}.

\subsection{Assembly and determinant check}
\label{app:w4-oneloop-check}

The full coherent correction is
\beq
 \boxed{\quad
 \delta_{\rm slope}(N,\beta)
 =\big[\kappa_L(\beta)+\kappa_S(\beta)\big]
 \delta_{\rm rung}(N,\beta).
 \quad}
\eeq
There is an independent theory-side check that never uses this split.  The
one-loop contribution to $-\log Z$ is
\beq
 F_1(\beta)=\frac12\operatorname{Tr}\big[\log(1-K)+K\big].
 \label{eq:oneloop-trlog}
\eeq
The $+K$ removes the one-rung term already carried by the exact-count
normalization.  Since $\tfrac12\operatorname{Tr}K=-\tfrac34J^2\beta I_4$,
the raw determinant term $-\tfrac12\operatorname{Tr}K$ is cancelled by the
$-\Delta I$ term in \eqref{eq:oneloop-counterterm}.  Because $F_1$ contributes
to $-\log Z$, the energy correction is $+\partial_\beta F_1$.  Direct
differentiation reproduces the strict large-$N$ channel prediction
\beq
\begin{aligned}
 \Delta E_1
 &=\lim_{N\to\infty}\big(\kappa_L+\kappa_S\big)\delta_{\rm rung}(N,\beta)
 \mathbb E\big[\langle H\rangle_{\rm mel}\big] \\
 &=-\frac92 J^4\big(\kappa_L+\kappa_S\big)R_\beta
\end{aligned}
\eeq
to better than one percent at every temperature in Table~\ref{tab:w4pred}.
Here $\mathbb E[\langle H\rangle_{\rm mel}]=-NJ^2I_4/4$, so the comparison
tests the limiting order-one energy correction; the extra finite-$N$ factor
$N/(N-1)$ is not part of this determinant check.
At $\beta J=0.5,1,1.5,2,3$, the determinant-to-channel ratios are
\beq
 1.008,\qquad 1.001,\qquad 1.000,\qquad 0.999,\qquad 0.999,
\eeq
respectively.  The script \path{case_w4/check_trlog.py} reports this
comparison but never alters either $\kappa$.

The dependency is one-way: the SD saddle determines $G_*$; $G_*$ determines
$\delta_{\rm rung}$, $\kappa_L$, and $\kappa_S$; those theory-only numbers
are written to \texttt{case\_w4\_pred.csv}; and only afterward do plotting
scripts read ED data.  The fitted amplitudes of Table~\ref{tab:w4fit} are
computed in a separate script and are not imported by this calculation.

\section[Additional tests at weight 4]{Additional tests at weight $4$}
\label{app:w4-extra}

\subsection{Comparison to freely fitted coefficients of coupling tensors}
\label{sec:corrections-fit}

The weight-four predictions of Section~\ref{sec:corrections} are parameter-free: every coefficient is fixed by the rules. As a robustness check, Table~\ref{tab:w4fit} instead allows one global amplitude for each of the leading and degree-two structures, pooling $N=10$--$20$ at fixed temperature. The fitted amplitudes agree with the rule values to $1$--$2\%$ at high temperature and drift upward as $\beta J$ grows, as expected when a single pooled coefficient absorbs $N$-dependent loop corrections and still-higher terms. This pooled fit should not be interpreted as a sharp measurement of either correction. Its useful message is more modest: the parameter-free tensors already capture nearly all of the operator-resolved agreement. At $\beta J=2$, fitting both amplitudes raises $R^2$ only from $0.977$ to $0.994$, and even at $\beta J=3$ the improvement is $0.932$ to $0.976$. Most of the explanatory power therefore lies in identifying the structures rather than tuning their amplitudes.

\begin{table}[t]
\centering
\begin{tabular}{c cc ccc}
\toprule
$\beta J$ & $c_{\rm lead}/(-4)$ & $c_2/(+4)$ & $R^2$ (leading) & $R^2$ ($+$deg-2, no fit) & $R^2$ ($+$deg-2, fitted) \\
\midrule
$0.5$ & $1.012$ & $1.023$ & $0.992$ & $1.000$ & $1.000$ \\
$1.0$ & $1.044$ & $1.079$ & $0.967$ & $0.998$ & $1.000$ \\
$1.5$ & $1.092$ & $1.152$ & $0.924$ & $0.990$ & $0.998$ \\
$2.0$ & $1.137$ & $1.240$ & $0.876$ & $0.977$ & $0.994$ \\
$3.0$ & $1.234$ & $1.403$ & $0.756$ & $0.932$ & $0.976$ \\
\bottomrule
\end{tabular}
\caption{How much does a little fitting buy? One global amplitude per structure (leading $J_X$, degree-two $W^{(2)}T^{(2)}$) fitted jointly across all operators and sizes ($N = 10$--$20$ pooled, $2880$ points per row, $1440$ at $\beta J = 1.5, 3$), quoted as ratios to the rule values $-4$ and $+4$. The first two $R^2$ columns are parameter-free identity-line statistics; only the final column uses the two fitted amplitudes. The fit gives a modest improvement because the rule-level structures already carry nearly all of the agreement.}
\label{tab:w4fit}
\end{table}

\subsection[Temperature dependence at weight 4]{Temperature dependence at weight $4$}
\label{app:w4-temperature}

The main tests vary $N$ at fixed temperature and establish the expected
finite-size convergence.  Here we instead hold the largest accessible size,
$N=24$, fixed and ask how the complete weight-$4$ prediction behaves toward
the high- and low-temperature ends of the range.  We use $160$ disorder
realizations and $48$ sampled quartets per realization at
$\beta J=0.5,2,4$.  A single diagonalization supplies all three temperatures,
and the same realizations and operator stream are used throughout.  As a
pipeline control, all $7680$ values at $\beta J=2$ reproduce the independent
large-$N$ push dataset row by row, with identical keys
and numerical tolerances $10^{-12}$ absolute and $10^{-10}$ relative to allow
for independent-eigensolver roundoff.

\begin{figure}[h]
\centering
\includegraphics[width=\textwidth]{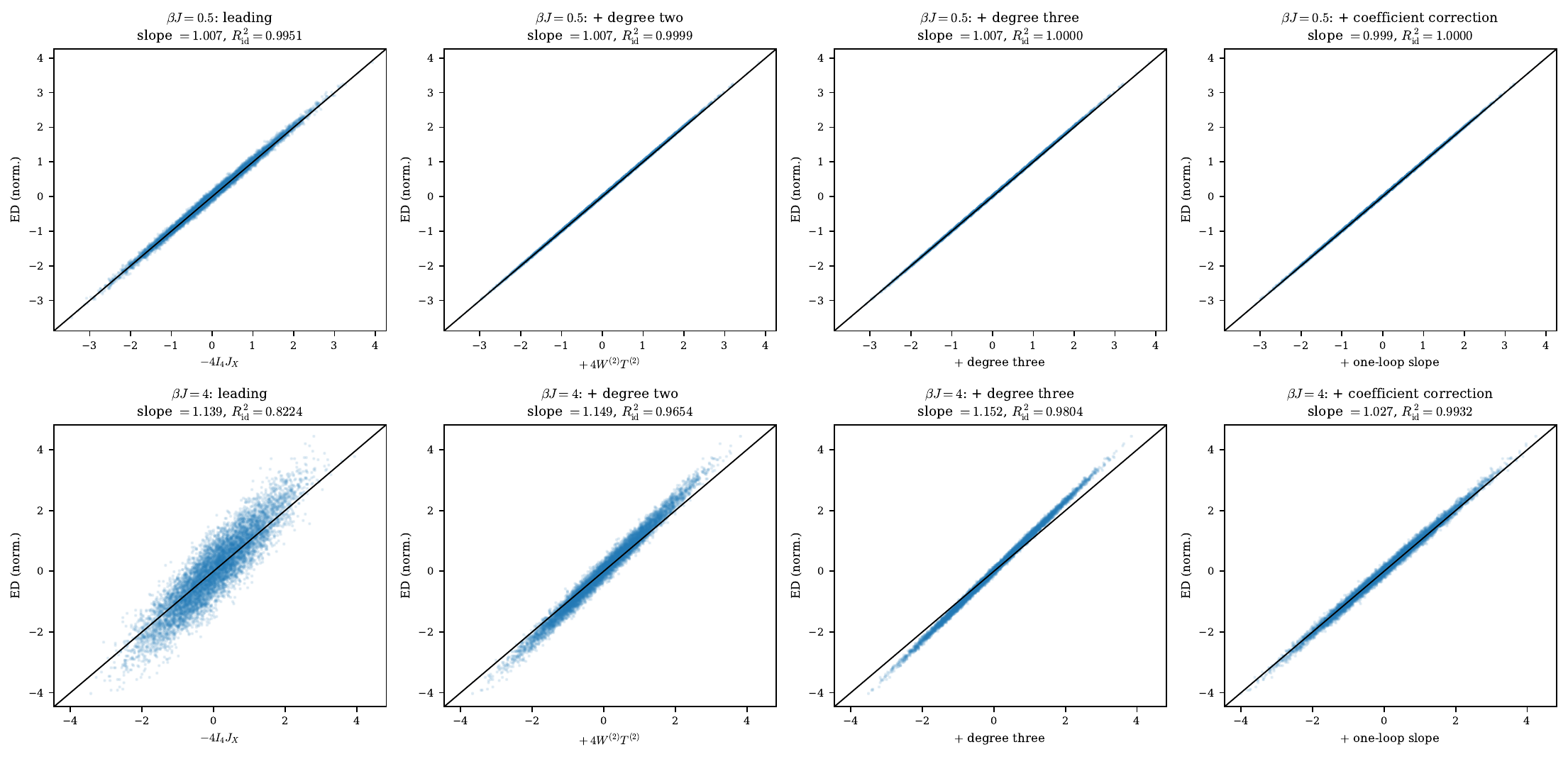}
\caption{Fixed-$N$ temperature stress test for the weight-$4$ prediction
($N=24$, $160$ seeds, $48$ quartets per seed;
\texttt{numerics-v3/case\_w4\_temperature/analyze.py}).  Rows are
$\beta J=0.5$ and $4$; columns successively add the degree-two tensor, the
Wick-ordered degree-three tensors, and the parameter-free one-loop coefficient
correction.  Axes are normalized by the standard deviation of the leading
prediction at each temperature.  Every panel reports the through-origin slope
and the identity-line $R^2_{\rm id}$; the diagonal is not fitted.  At
$\beta J=4$ the leading scatter is large, but the predicted fluctuation
structures collapse it systematically and the final coefficient correction
brings the cloud close to the identity line.}
\label{fig:w4-temperature-stress}
\end{figure}

For each temperature we evaluate, without fitting, the leading term, the
degree-two and Wick-ordered degree-three fluctuation tensors, followed by the
one-loop coefficient correction in Eq.~\eqref{eq:w4slopecorr}.
Figure~\ref{fig:w4-temperature-stress} shows the two endpoint temperatures.
The high-temperature cloud is already
nearly one-dimensional at leading order: $R^2_{\rm id}=0.9951$, rising to
$0.999940$ after degree two and $0.999999$ for the complete prediction.  Its
remaining rms residual is only $1.1\times10^{-3}$ of the rms leading signal.

At $\beta J=4$ the leading description is visibly less complete: its relative
scatter is $0.507$ and $R^2_{\rm id}=0.8224$.  Nevertheless, the failure is
highly structured.  Degree two raises $R^2_{\rm id}$ to $0.9654$, degree three
raises it to $0.9804$, and the complete prediction reaches $0.9932$, leaving
an rms residual $0.103$ times the leading signal.  The predicted fluctuation
has correlation $0.9944$ with the observed one, although its fitted amplitude
is $1.211$, exposing the expected growth of higher-order coefficient effects.

The slope test is sharper still; Table~\ref{tab:w4-temperature-stress} collects the diagnostics at all three temperatures.  At $\beta J=4$ ED gives
$s_{\rm ED}=1.1389(52)$, in close agreement with the complete one-loop value
$1.1381$, whereas the unresummed bare-rung estimate $1.2515$ is decisively too
large.  The energy anchor, which uses every coupling rather than the sampled
operators, resolves a small residual difference: $s_H=1.14232(59)$, about
three percent of the predicted slope correction above the one-loop result.
Thus the low-temperature endpoint is better than the raw scatter might
suggest: the diagrammatic directions and the resummed mean coefficient remain
accurate, while the first clear limitation appears in the amplitudes of the
subleading structures.

\begin{table}[h]
\centering
\small
\setlength{\tabcolsep}{4pt}
\begin{tabular}{c c c c c c c c c}
\hline
$\beta J$ & $s_{\rm ED}$ & $s_{\rm 1\mbox{-}loop}$ & $s_{\rm bare}$
& $\rho_{\rm fluc}$ & $a_{\rm fluc}$
& $R^2_{\rm lead}$ & $R^2_{\rm full}$ & $s_H$ \\
\hline
$0.5$ & $1.0066(7)$  & $1.0077$ & $1.0080$ & $0.9999$ & $1.011$ & $0.9951$ & $0.999999$ & $1.00691(10)$ \\
$2$   & $1.0704(29)$ & $1.0724$ & $1.0986$ & $0.9994$ & $1.110$ & $0.9325$ & $0.9993$   & $1.07189(37)$ \\
$4$   & $1.1389(52)$ & $1.1381$ & $1.2515$ & $0.9944$ & $1.211$ & $0.8224$ & $0.9932$   & $1.14232(59)$ \\
\hline
\end{tabular}
\caption{Diagnostics for the fixed-$N$ stress test.  Here $s_{\rm ED}$ is
the through-origin slope of ED against the leading prediction, with bootstrap
errors over disorder realizations; $s_{\rm 1\mbox{-}loop}=1+\delta_{\rm slope}$
and $s_{\rm bare}=1+\delta_{\rm rung}$ are parameter-free.  After subtracting
the one-loop-corrected leading term, $\rho_{\rm fluc}$ is the correlation of
the residual with the predicted degree-two plus degree-three fluctuation and
$a_{\rm fluc}$ is its fitted amplitude (a diagnostic only).  The final column
is the independent energy-anchor slope
$s_H=\langle H\rangle_\beta/[-I_4\sum_XJ_X^2]$.}
\label{tab:w4-temperature-stress}
\end{table}

\section[First Corrections at Weight 8]{First Corrections at Weight $8$}
\label{app:w8}

The weight-8 tests of Section~\ref{sec:leading-w8} left one loose end: the operator-to-operator scatter about the leading prediction \eqref{eq:w8pred} decreases with $N$ much more slowly than at weight 4, and a naive linear extrapolation in $1/\sqrt{N}$ leaves a sizeable intercept (Figure~\ref{fig:w8scatter}). This appendix identifies the structures responsible, computes the first correction, and shows that the slow decrease is a property of the first-order budget itself: evaluated exactly, the predicted scatter is flat across the sizes studied and reaches its asymptotic $N^{-1/2}$ form only for $N$ in the hundreds, so the intercept of a linear extrapolation is an artifact of pre-asymptotic sizes rather than evidence of an unexplained structure.

\subsection{The two structures at first order}
\label{app:w8-structures}

The counting differs from weight 4 in one important way. At $W=4$ the leading diagram has one vertex, so the first fluctuation correction ($T^{(2)}$, two vertices) is one vertex beyond leading and the degree-three skeletons sit two beyond. At $W=8$ the leading diagram already has two vertices, so \emph{every} three-vertex structure enters at the first correction order, relative size $N^{-1/2}$ by the power counting \eqref{eq:powercount}. Enumerating the allocations of the eight external Majoranas over three quartic vertices, only two connected classes survive:
\begin{itemize}
    \item \textbf{Bubble} (allocation $4|2|2$, all blocks even): one vertex absorbs a full external quartet $A$, exactly as in the leading diagram, while the other quartet $B$ splits into two pairs on two vertices joined by two internal lines --- the fluctuation tensor $T^{(2)}_B$ of \eqref{eq:corr-t2} substituted for one leading vertex. The kernel factorizes into the two pieces already in hand, $I_4(\beta)\, W^{(2)}(\beta)$.
    \item \textbf{Chain} (allocation $3|2|3$, odd end blocks): two vertices each absorb an external \emph{triple} and connect through a middle vertex carrying the remaining external pair, with two single internal lines. Its kernel is
    \beq
        K_{\rm ch8}(\beta) = \int_0^\beta d\tau_1\, d\tau_2\, d\tau_3\;
        G(\tau_1)^3\, G(\tau_1 - \tau_3)\, G(\tau_3)^2\, G(\tau_3 - \tau_2)\, G(\tau_2)^3 ,
        \label{eq:w8chainkernel}
    \eeq
    with the external legs running to the operator at $\tau = 0$ and the internal lines continued antiperiodically.
\end{itemize}
\begin{center}
\begin{tikzpicture}[line width=0.8pt, every node/.style={font=\small}]
  \begin{scope}[xshift=-3.6cm]
    \coordinate (va) at (-1.5,0);
    \foreach \ang in {120,160,200,240} {\draw (va) -- ++(\ang:1.05);}
    \fill (va) circle (2.2pt);
    \node[below=10pt] at (va) {$J_A$};
    \coordinate (v1) at (0.3,0); \coordinate (v2) at (1.7,0);
    \draw (v1) -- ++(120:1.0); \draw (v1) -- ++(240:1.0);
    \draw (v2) -- ++(60:1.0); \draw (v2) -- ++(-60:1.0);
    \draw (v1) .. controls (0.75,0.3) and (1.25,0.3) .. (v2);
    \draw (v1) .. controls (0.75,-0.3) and (1.25,-0.3) .. (v2);
    \fill (v1) circle (2.2pt); \fill (v2) circle (2.2pt);
    \node[below=12pt] at (1.0,0) {bubble ($4|2|2$)};
  \end{scope}
  \begin{scope}[xshift=3.4cm]
    \coordinate (c1) at (-1.6,0); \coordinate (cm) at (0,0); \coordinate (c2) at (1.6,0);
    \foreach \ang in {105,140,175} {\draw (c1) -- ++(\ang:1.0);}
    \foreach \ang in {75,40,5} {\draw (c2) -- ++(\ang:1.0);}
    \draw (cm) -- ++(70:0.95); \draw (cm) -- ++(110:0.95);
    \draw (c1) -- (cm); \draw (cm) -- (c2);
    \fill (c1) circle (2.2pt); \fill (cm) circle (2.2pt); \fill (c2) circle (2.2pt);
    \node[below=12pt] at (cm) {chain ($3|2|3$)};
  \end{scope}
\end{tikzpicture}
\end{center}
The remaining allocation $4|3|1$ has no consistent internal pairing (the single-leg vertex would need three internal lines against one available partner), and allocations involving a vertex with no external legs collapse onto the melon already inside $G$. The two coupling polynomials are
\begin{align}
P_{{\rm bub},X}
 &= \sum_{\substack{A\sqcup B=X\\|A|=|B|=4}}
    \eta(A,B) J_A T^{(2)}_B, \nonumber\\
P_{{\rm chain},X}
 &= \sum_{\substack{C\sqcup D\sqcup E=X\\|C|=|E|=3,\ |D|=2}}^{\prime}
    \eta(C,D,E)\sum_{u,v=1}^N
    J_{c_1c_2c_3u}J_{u d_1d_2v}J_{e_1e_2e_3v}.
\label{eq:w8polynomials}
\end{align}
Each external block is written in increasing order and $\eta$ is the parity of its displayed concatenation relative to sorted $X$. The bubble sum distinguishes the lone quartet $A$ from the quartet $B$ carrying $T^{(2)}_B$. The prime on the chain sum retains each unordered pair of end triples once: exchanging $C,E$ and relabeling $u,v$ leaves the signed term unchanged. In particular, the last coupling is $J_{Ev}$, not $J_{vE}=-J_{Ev}$. Couplings with repeated indices vanish by antisymmetry. The corrected weight-8 prediction through first order is

\beq
    \xi_X = 16\,I_4^2 L_X
    -16\,I_4 W^{(2)}P_{{\rm bub},X}
    -16\,K_{\rm ch8}P_{{\rm chain},X}+\cdots,
    \qquad L_X=\sum_P\eta(P)J_AJ_B.
    \label{eq:w8corrected}
\eeq
The coefficient $-16$ follows from \eqref{eq:coeffrule} at $(W,r)=(8,3)$.

\subsection{Exact validation at third order}
\label{app:w8-cubic}

Every sign and coefficient in \eqref{eq:w8corrected} can be checked against an exact trace. Because $\langle \mu_X H \rangle_0 = 0$ at weight 8, the $\beta^3$ term of the Taylor expansion \eqref{eq:taylor} is exactly $-\langle \mu_X H^3\rangle_0 / 6$. Regressing the dense-trace moment $\langle \mu_X H^3 \rangle_0$ (computed at $N = 12$, all $495$ octets, three disorder realizations) onto the three structures gives
\beq
    \big\langle \mu_X H^3 \big\rangle_0
    = 0 \cdot L_X + \tfrac{3}{32}\, P_{{\rm bub},X} - \tfrac{1}{32}\, P_{{\rm chain},X} ,
\eeq
with coefficient of determination $R^2 = 1.000000$: the two structures \emph{exhaust} the cubic moment exactly, the vanishing $L_X$ coefficient confirming that no coincident-coupling piece survives at this order. The bubble coefficient equals $6 \cdot 16 \cdot [I_4 W^{(2)}]_{\beta^3} = 3/32$ as required, and the chain coefficient fixes the orientation of \eqref{eq:w8chainkernel}: the high-temperature limit is $K_{\rm ch8} \to -\beta^3/3072$, and --- in contrast to the odd-block topologies at weight 4, which carry one extra sign relative to the naive antiperiodic evaluation --- the weight-8 chain requires no additional orientation sign.

\subsection{The scatter budget}
\label{app:w8-budget}

Both cubic structures are Wick-orthogonal to the quadratic $L_X$, so their ensemble projection onto the leading tensor vanishes and they feed only the operator-to-operator scatter. Orthogonality constrains the ensemble projection; a finite-sample regression slope need not be exactly unchanged. Their exact Gaussian Gram entries are
\begin{align}
    \operatorname{var}(L) &= 35\,\sigma_J^4, \nonumber\\
    \operatorname{var}(P_{\rm bub})
      &= \left[210\binom{N-4}{2}+2520\right]\sigma_J^6, \nonumber\\
    \operatorname{var}(P_{\rm chain})
      &= 280(N^2-5N+30)\,\sigma_J^6, \nonumber\\
    \operatorname{cov}(P_{\rm bub},P_{\rm chain})
      &= 840(2N-7)\,\sigma_J^6.
    \label{eq:w8variances}
\end{align}
Every monomial in either cubic polynomial contains three distinct couplings. After collecting identical monomials across external block assignments, unequal monomials are orthogonal and each has variance $\sigma_J^6$; the Gram entries are therefore the sums of products of their integer coefficients. The count can be organized by whether zero, one, or two internal flavors lie outside $X$, giving polynomials of degree at most two in $N$.

For comparison, retaining only pairings within one block assignment gives $210\binom{N-4}{2}$ for the bubble and $280[(N-5)^2-(N-8)]$ for the chain, in units of $\sigma_J^6$. The assignment multiplicities are $35\cdot2\cdot3=210$ and $\binom82\binom63/2=280$, respectively. For a fixed chain assignment each internal index has $N-5$ possible values, with $N-8$ forbidden coincidences. Cross-assignment pairings add $2520$ to the bubble count and $1680N-840$ to the chain count. The latter is a relative-$1/N$ contribution, essential at accessible sizes.

The two cubic structures are not orthogonal to each other: their correlation is $0.969$ at $N=10$ and $0.453$ at $N=24$. This illustrates the basis non-uniqueness in Appendix~\ref{sec:rules-chaos}: covariant structures within a chaos sector can have a nontrivial Gram matrix. Since $K_{\rm ch8}<0$ and the covariance is positive, their contributions cancel substantially in the physical combination. The predicted scatter includes this covariance:
\beq
    r_{\rm pred}^2
    = \frac{ \big(I_4 W^{(2)}\big)^2 \operatorname{var}(P_{\rm bub})
           + K_{\rm ch8}^2 \operatorname{var}(P_{\rm chain})
           + 2\, I_4 W^{(2)} K_{\rm ch8}\, \operatorname{cov}(P_{\rm bub}, P_{\rm chain})}
           { I_4^4\, \operatorname{var}(L) } .
    \label{eq:w8predscatter}
\eeq
The Gram entries in \eqref{eq:w8variances} make \eqref{eq:w8predscatter} deterministic, with no ensemble Monte Carlo error. The cancellation is large: at $N=10$, $\beta J=0.5$ the separate terms give $r\approx0.206$ in quadrature, while the full covariance gives $r_{\rm pred}=0.0680$, close to the measured $0.0675$. At $\beta J=2$ the exact predictions are $0.3278$ at $N=16$ and $0.3266$ at $N=24$, and in between they are flat and slightly non-monotonic: $0.3323$, $0.3325$, and $0.3303$ at $N=18,20,22$. The individual variances fall as $1/N$, but the bubble--chain correlation that cancels them decays slowly, from $0.68$ at $N=16$ to $0.45$ at $N=24$, and the two effects compensate at these sizes. Asymptotically the covariance is subleading and $r_{\rm pred}\sqrt N\to\big[18\,(W^{(2)}/I_4)^2+48\,(K_{\rm ch8}/I_4^2)^2\big]^{1/2}\approx2.1$ at $\beta J=2$, but at $N=24$ the product is still $1.60$, so the $N^{-1/2}$ regime is reached only for $N$ in the hundreds. The bubble dominates the surviving piece at larger $\beta J$, where the kernel ratios are $W^{(2)}/I_4=0.439$ and $|K_{\rm ch8}|/I_4^2=0.139$ at $\beta J=2$; in the decoupled large-$N$ limit the bubble alone gives $r_{\rm bub}\to\sqrt{2}\,\times$ the weight-4 scatter, counting the two quartets that fluctuate independently.

\subsection{Comparison with exact diagonalization}
\label{app:w8-ed}

\begin{figure}[t]
\begin{center}
\includegraphics[width=0.98\textwidth]{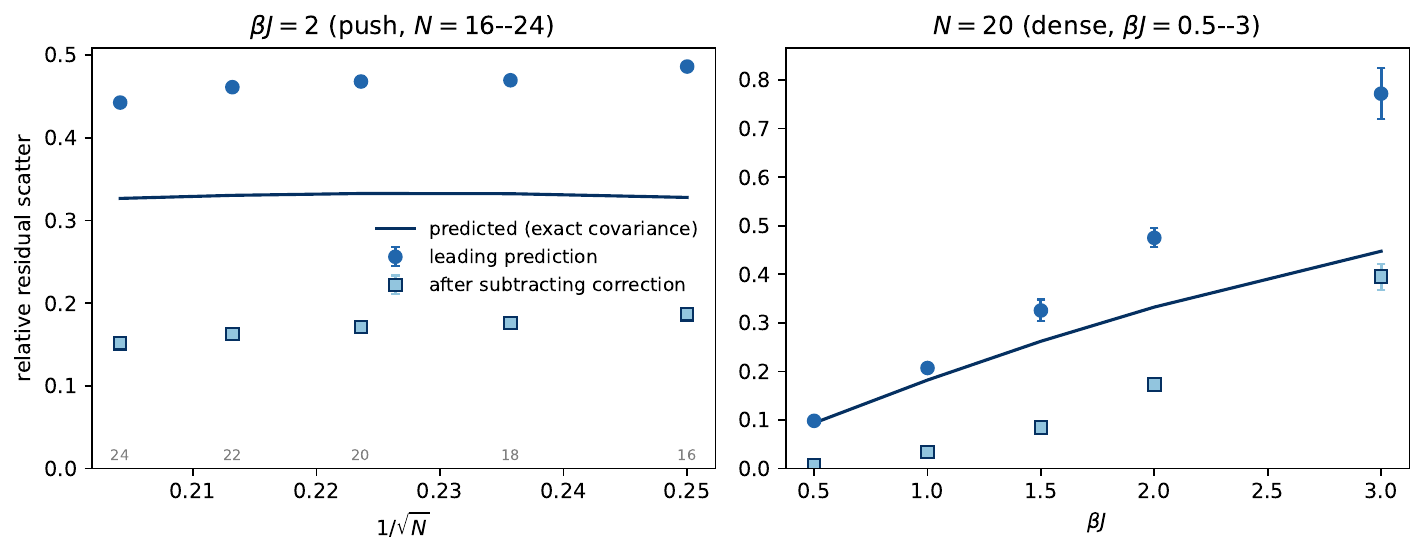}
\end{center}
\caption{The weight-8 scatter budget. Left: scatter about the fitted leading slope at $\beta J=2$ (push dataset, $N=16$--$24$, $160$ seeds $\times$ $48$ octets) versus $1/\sqrt N$: measured scatter (circles), prediction from the exact Gaussian covariance \eqref{eq:w8predscatter} (line; flat across these sizes, see text), and residual after subtracting the parameter-free cubic correction (squares). The subtraction statistic \eqref{eq:w8subscatter} refits only the leading coefficient and retains the leading normalization. Point error bars are standard errors from $2000$ disorder-seed block-bootstrap resamples. Right: the same quantities versus $\beta J$ at $N=20$. Subtraction reduces the scatter by $2.6$--$2.9$ at $\beta J=2$ in the left panel and by $13.6$ at $\beta J=0.5$ in the right panel.}
\label{fig:w8fluct}
\end{figure}

Figure~\ref{fig:w8fluct} confronts the budget with the exact-diagonalization data of Section~\ref{sec:leading-w8}, applying \eqref{eq:w8corrected} operator by operator on the stored realizations. To specify the subtraction test, write $x_i^{(0)}=16I_4^2L_{X_i}$ for the leading prediction, $c_i$ for its parameter-free cubic correction, and $y_i$ for ED. We plot
\beq
s_{\rm sub}=\frac{\sum_i x_i^{(0)}(y_i-c_i)}{\sum_i(x_i^{(0)})^2},
\qquad
r_{\rm sub}=\left[\frac{\sum_i(y_i-c_i-s_{\rm sub}x_i^{(0)})^2}
{\sum_i(x_i^{(0)})^2}\right]^{1/2}.
\label{eq:w8subscatter}
\eeq
Thus the correction is subtracted at its predicted amplitude; only the remaining leading coefficient is fitted, and both pre- and post-subtraction scatters use the same leading normalization.

The direction of the correction is well captured: across the studied sizes and temperatures, the residual about the fitted leading slope has correlation $0.921$--$0.999$ with the computed correction, with the lowest correlation at $N=20$, $\beta J=3$. Its magnitude retains a finite-$N$ excess. At $\beta J=0.5$ the measured scatters exceed the exact budget by up to about $10\%$, at the edge of the statistical errors; for example, at $N=16$ the prediction is $0.0919$ against $0.1002(41)$ measured, two standard errors apart, where the uncertainty is a seed-block bootstrap standard error. At $\beta J=2$, the fitted amplitude of the correction falls from $1.43$ at $N=16$ to $1.31$ at $N=24$, similar to the kernel-dressing excess seen elsewhere in the paper.

Parameter-free subtraction still substantially reduces the scatter. In the push dataset at $\beta J=2$ ($160$ seeds, $48$ octets each), it falls from $0.4860(49)$ to $0.1865(21)$ at $N=16$ and from $0.4425(55)$ to $0.1515(19)$ at $N=24$, an improvement of $2.6$--$2.9$. At $N=20$, $\beta J=0.5$, it falls from $0.0983(41)$ to $0.00722(33)$, a factor of $13.6$. The leading-projection slope changes little in these samples, but its ensemble orthogonality does not require an exactly unchanged finite-sample slope. We have not isolated the asymptotic $N$ scaling of the residual.

The slow decrease and apparent intercept of Figure~\ref{fig:w8scatter} are thus explained. The first-order budget itself is flat across $N=16$--$24$, so a linear extrapolation in $1/\sqrt N$ over this window has no asymptotic meaning and its intercept is not evidence of a missing structure; the measured decrease from $0.486$ to $0.442$ is carried by the falling amplitude excess, $1.43\to1.31$, rather than by the $N^{-1/2}$ of the budget. The two computed structures account for most of the direction of the deviations and for the bulk of their magnitude. The remaining amplitude excess and residual scatter are the expected kernel-dressing and $r=4$ effects, but we have not isolated their $N$ scaling at these sizes. The leading slope excess of Figure~\ref{fig:w8slope} requires a coherent kernel correction, now acting on both vertices together with a connected cross-vertex rung. Its computation parallels Section~\ref{sec:corrections-w4slope}; we do not carry it out here.

\section{Numerical Methods}
\label{app:numerics}

Three numerical components enter the comparisons of this paper: the melonic saddle, which supplies every kernel; finite-$N$ exact diagonalization, which supplies the data the predictions are tested against; and, for the case study of Section~\ref{sec:corrections-w4slope}, the one-loop factors $\kappa_L$ and $\kappa_S$. All are implemented in the \texttt{numerics-v3} directory of the repository accompanying this paper, with one script per figure and a test suite pinning the checks quoted below. The directory also contains an exact-version \texttt{requirements.txt}, fixed seeds for every stochastic calculation, provenance headers in generated CSV files, and a top-level \texttt{run\_all.py}; its \texttt{--quick} mode performs a reduced smoke reproduction, while the default regenerates the production artifacts, including Appendix~\ref{app:w8}.

\paragraph{Artifact map.} Table~\ref{tab:numerical-artifacts} locates the
producers used in the comparisons above. Paths are relative to
\texttt{numerics-v3}; the section READMEs give the complete CSV and figure map.
The default production push uses $160$ disorder seeds, and the full workflow
includes the tagged MQ fidelity campaigns at $N=12,14$ and the additional
$24$ seeds at $N=10$. Quick-mode data have a separate suffix and figures go
to a preview directory. The weight-eight budget uses exact Gaussian Gram
entries; Monte Carlo moment estimation remains an optional cross-check.

\begin{table}[t]
\centering
\small
\begin{tabular}{@{}lp{0.69\textwidth}@{}}
\toprule
Artifact & Producing scripts \\
\midrule
Figure~\ref{fig:w4intro} & \texttt{case\_w4/fig\_w4\_intro.py} \\
Figure~\ref{fig:w4order3} & \texttt{case\_w4/fig\_w4\_order3.py} \\
Table~\ref{tab:w4pred} & \texttt{case\_w4/case\_study\_w4.py}, using \texttt{syk/oneloop.py} \\
Figure~\ref{fig:w4slopecheck} & \texttt{case\_w4/fig\_case\_slope.py} \\
Figure~\ref{fig:w4anchor} & \texttt{case\_w4/run\_anchor.py}, \texttt{case\_w4/fig\_case\_anchor.py} \\
Table~\ref{tab:w4fit} & \texttt{case\_w4/table\_w4\_fit.py} \\
Figure~\ref{fig:w8fluct} & \texttt{case\_w8/scan\_w8\_fluct.py}, \texttt{case\_w8/fig\_w8\_fluct.py} \\
MQ fidelity ensembles & \texttt{mq/scan\_fidelity.py}; tagged campaign arguments are in \texttt{run\_all.py} \\
\bottomrule
\end{tabular}
\caption{Numerical producers for the detailed correction tests. All production figure scripts write to \texttt{figures-v3}.}
\label{tab:numerical-artifacts}
\end{table}

\subsection{Statistical diagnostics}
\label{app:numerics-statistics}

We use three distinct diagnostics and keep their roles separate. Given a
parameter-free prediction $x_i$ and ED value $y_i$, the through-origin slope
and relative residual scatter are
\beq
 s=\frac{\sum_i x_i y_i}{\sum_i x_i^2},\qquad
 r_{\rm rel}=
 \left[\frac{\sum_i(y_i-sx_i)^2}{\sum_i x_i^2}\right]^{1/2}.
 \label{eq:numerical-diagnostics}
\eeq
The slope is a fitted diagnostic, whereas $r_{\rm rel}$ measures the residual
about that fitted line. The weight-eight subtraction test uses the explicitly
defined $r_{\rm sub}$ in \eqref{eq:w8subscatter}: its correction is fixed
and its denominator remains the root-mean-square leading prediction. For genuinely parameter-free comparisons we instead
quote the identity-line coefficient of determination
\beq
 R^2_{\rm id}=1-
 \frac{\sum_i(y_i-x_i)^2}{\sum_i(y_i-\bar y)^2}.
 \label{eq:identity-r2}
\eeq
This centered-denominator convention is used throughout the code and paper;
$R^2_{\rm id}$ can be negative when the identity-line prediction is worse
than the sample mean. It should not be confused with the $R^2$ of the fitted
line $y=sx$.

In pooled scatter plots spanning several $N$, both coordinates at each $N$
are divided by the standard deviation of the leading prediction at that
$N$ before pooling. This removes the known overall $N$ scaling without
altering the within-$N$ through-origin slope; pooled diagnostics consequently
weight the normalized observations rather than the raw amplitudes. All
operators measured in one disorder realization share the same couplings and
are correlated. For operator-resolved slopes and scatters, error bars
therefore use a block bootstrap over disorder seeds, retaining every operator
from a resampled seed as one block. The main size scans use $2000$ bootstrap
replicates, the one-loop slope check uses $400$, and the fixed-$N$ temperature
stress test uses $500$. These error bars are bootstrap standard errors. The
energy anchor has one aggregate number per seed and uses the ordinary standard
error of their mean. Extrapolation lines in the main size scans are weighted
by inverse squared bootstrap standard errors, and the $R^2$ quoted for such a
line uses the same weights. These
extrapolations and their reported intercepts are descriptive finite-size
diagnostics, not formal confidence statements about the $N\to\infty$ limit.
ED and SD discretization errors are far below the displayed statistical
errors: parity-block ED agrees with the independent dense implementation to
machine precision, and the SD convergence check is described next.

\subsection{Schwinger--Dyson solver}
\label{app:numerics-sd}

The saddle equations \eqref{eq:SD} are solved on a midpoint imaginary-time grid, $\tau_j = (j + \tfrac12)\,\beta/M$, with the transform to fermionic Matsubara frequencies $\omega_n = 2\pi(n + \tfrac12)/\beta$ implemented as a phase-decorated fast Fourier transform; on the midpoint grid the forward and inverse transforms are exact inverses of one another. One detail matters for accuracy: the free propagator $G_0(i\omega_n) = 1/(-i\omega_n)$ decays slowly in frequency, and transforming it naively produces Gibbs ringing at the edges of the interval. We therefore transform only the difference $G(i\omega_n) - 1/(-i\omega_n)$, which decays like $1/\omega_n^2$, and add back its exact image $G_0(\tau) = \tfrac12$ on $(0, \beta)$. With this tail handling the kernel $I_4$ is converged to ${\sim}10^{-7}$ relative accuracy already at $M = 512$; production runs use $M = 1024$.

The iteration is plain damped fixed-point mixing, $G \to (1 - \alpha) G + \alpha\, G_{\rm new}$ with $\alpha = 0.15$, run to a max-norm residual below $10^{-11}$. Over the temperature range used in this paper ($\beta J \le 4$, and the Maldacena--Qi matrix generalization of Section~\ref{sec:mq}) no convergence difficulties arise; at much larger $\beta J$ the fixed-point iteration slows and smaller mixing (or a weighted update) is the standard remedy. Because of the exact-leg-count normalization \eqref{eq:H}, the solver is always run at the physical coupling $J$ of the instance: there is no separate finite-$N$ calibration step. The temperature kernels ($I_4$, $W^{(2)}$, the degree-three and bilocal kernels) are then evaluated on the grid by direct quadrature and FFT convolutions.

\subsection{Exact diagonalization}
\label{app:numerics-ed}

The Majorana algebra is realized by Jordan--Wigner matrices on $N/2$ qubits, $\chi_i = \gamma_i/\sqrt{2}$, and the strings $\mu_X$ carry the phase convention \eqref{eq:mu}. Since $p$ is even, $H$ commutes with fermion parity, and every equal-time observable in this paper is an even string; the Hamiltonian and all probes are therefore built directly in the two occupation-parity blocks, each of dimension $2^{N/2 - 1}$. A thermal one-point function needs only the diagonal matrix elements of $\mu_X$ in the energy bases of the two blocks, combined with the shared partition function, so a full realization at one $N$ costs two dense diagonalizations: at $N = 24$, two $2048 \times 2048$ problems, which is what makes the largest sizes of Figures~\ref{fig:w4slope}--\ref{fig:w8scatter} affordable. The full-space dense path is retained for the bilocal correlators of Section~\ref{sec:time-dep-w2imag}, which require off-diagonal matrix elements, and as a reference implementation: the parity-block spectra and one-point values are verified against it to machine precision.

\subsection{One-loop factors}
\label{app:numerics-oneloop}

The full derivation is given in Appendix~\ref{app:w4-oneloop}. The ladder factor $\kappa_L(\beta)$ is evaluated as the Neumann series of the Bethe--Salpeter kernel on the saddle: each application of $K$ is a pair of dense $M \times M$ matrix multiplications, the series converges geometrically for the couplings studied (its tail ratio estimates the spectral radius of $K$), and the result is Richardson-extrapolated in $1/M$. The self-energy factor $\kappa_S(\beta)$ is obtained from the one-loop tadpole of the collective $(g, \sigma)$ fluctuations: the quadratic form of Appendix~\ref{sec:rules} is assembled as an exact Hessian over antisymmetric bilocals on the grid, inverted directly, and contracted with the cubic vertices; the shift $6 J^2\, \partial_{J^2} G_*$ from the counterterm vertex \eqref{eq:deltaI} is added by central differencing. Two exact discrete identities tie the construction together and are enforced in the tests: the $\sigma\sigma$ block of the Hessian equals the free antisymmetrized four-point kernel, and the $gg$ block of its inverse equals the resummed ladder $(1 - K)^{-1} F_0$ --- the same object that defines $\kappa_L$ --- so the two channels are computed within one consistent discretization. The values in Table~\ref{tab:w4pred} carry these Richardson extrapolations.

\subsection{Real-time continuation}
\label{app:numerics-spectral}

The real-time kernels of Section~\ref{sec:time-dep} require the melonic propagator off the imaginary axis, where the Matsubara series diverges; the continuation goes through the spectral function. We solve the SD equations directly in real frequency by the standard weak-coupling spectral iteration: given $\rho(\omega)$, the Wightman functions $G^{\gtrless}(t)$ are fast Fourier transforms of $\rho$ weighted by the Fermi factors, the self-energy is local in real time, $\Sigma^{\gtrless}(t) = J^2 [G^{\gtrless}(t)]^3$, the retarded self-energy follows from the one-sided transform of $\Sigma^> + \Sigma^<$, and Dyson's equation returns the updated $\rho = \mathrm{Im}\, G^R/\pi$; damped mixing converges rapidly at the temperatures used here. The kernel continuation is then algebraic: $F = G^3$ is local in time, $K(z) = G(z)^2 F(z)$ is the unique continuation of the Matsubara product, and on the symmetric line
\beq
    K(\beta/2 + it) = \int d\omega\; \frac{\rho_K(\omega)\, e^{-i\omega t}}{2\cosh(\beta\omega/2)},
    \qquad
    \rho_K = \frac{1}{\pi}\, \mathrm{Im}\big[ G^R(\omega)^2 F^R(\omega) \big],
\eeq
with the $1/\cosh$ providing exponential damping. The tests enforce the sum rule $\int \rho = 1$, agreement of the reconstructed $G(\tau)$ and $K(\tau)$ with the imaginary-time solver at the $10^{-3}$ level, the exact symmetry of $K(\tau)$ about $\beta/2$, and the reality of $K(\beta/2 + it)$.

\bibliographystyle{unsrtnat}
\bibliography{refs}

\end{document}